\documentclass[12pt,dvipsnames]{article}
 
\usepackage{iftex}
\ifLuaTeX
  \usepackage{polyglossia} 
  \usepackage{fontspec} 
  \setmainlanguage{english}
  \usepackage[nosort]{cite}
\else 
  \usepackage[utf8]{inputenc}
  \usepackage[T1]{fontenc}
  \usepackage[nosort]{cite}
  \usepackage[english]{babel} 
\fi

\usepackage[top=2cm,bottom=2cm,outer=2.5cm,inner=2.5cm,marginparwidth=2.5cm]{geometry}
\usepackage{fvextra} 
\usepackage{authblk} 

\ifLuaTeX
  \usepackage[unicode]{hyperref}
\else
  \usepackage{hyperref}
\fi

\usepackage{amsmath}
\usepackage{amsfonts}
\usepackage{amssymb}
\usepackage{dsfont} 
\usepackage{braket} 

\usepackage{lmodern} 
\usepackage{caption}
\usepackage{fancyhdr} 
\usepackage{abstract} 

\usepackage{graphicx} 
\usepackage{array} 
\usepackage{tikz}
\usepackage{pgf}
\usepackage{pgfplots} 

\hypersetup{
pdfstartview={FitH},
pdftitle={},
pdfauthor={},
pdfsubject={},
pdfcreator={},
pdfkeywords={},
colorlinks=true,
citecolor=MidnightBlue,
linkcolor=MidnightBlue,
urlcolor=MidnightBlue
}

\numberwithin{equation}{section}
\newcommand\frontmatter{%
  \clearpage
  \pagenumbering{roman}
}

\newcommand\mainmatter{%
  \clearpage
  \pagenumbering{arabic}
}

\newcommand{\vev}[1]{\left\langle #1 \right\rangle}
\newcommand{\norm}[1]{\left| #1 \right|}
\newcommand{\diff}{\mathrm{d}}

\DeclareMathOperator{\str}{str}
\DeclareMathOperator{\disc}{Disc}
\DeclareMathOperator{\ddisc}{dDisc}
\DeclareMathOperator{\res}{Res}

\def\bR {\mathbb{R}}

\def\cB{{\mathcal{B}}}
\def\cD{{\mathcal{D}}}
\def\cF{{\mathcal{F}}}
\def\cG{{\mathcal{G}}}
\def\cN{{\mathcal{N}}}
\def\cL{{\mathcal{L}}}
\def\cO{{\mathcal{O}}}
\def\O2{{\Phi}}
\def\Vop{{\mathsf{V}}}

\newcommand{\su}{\mathfrak{su}}
\renewcommand{\sl}{\mathfrak{sl}}

\newcommand{\osp}{\mathfrak{osp}}
\newcommand{\sof}{\mathfrak{so}}

\definecolor{mDarkBrown}{HTML}{604c38}
\definecolor{mDarkTeal}{HTML}{23373b}
\definecolor{mLightBrown}{HTML}{EB811B}
\definecolor{mLightGreen}{HTML}{14B03D}

\title{
Bootstrapping string dynamics in the 6d $\cN = (2,0)$ theories}
\author[1]{Carlo Meneghelli}
\author[2,3]{Maxime Tr\'epanier}
\affil[1]{\it Dipartimento SMFI, Universit\`a di Parma, \protect\\
Viale G.P.~Usberti 7/A, 43121, Parma, PR, Italy}
\affil[2]{\it Department of Mathematics, King's College London,\protect\\
  London, WC2R 2LS, United Kingdom}
\affil[3]{\it Perimeter Institute for Theoretical Physics,\protect\\Waterloo, ON, Canada, N2L 2Y5\protect\\[1em]
 {\tt  \mbox{carlo.meneghelli$\bullet$unipr.it,
 trepanier.maxime$\bullet$gmail.com}}}
\date{}

\begin{document}

\frontmatter
\maketitle
\thispagestyle{empty}
    
\begin{abstract}
  We present two complementary approaches to calculating the 2-point function of
  stress tensors in the presence of a 1/2 BPS surface defect of the 6d
  $\cN = (2,0)$ theories. First, we use analytical bootstrap techniques at large
  $N$ to obtain the first nontrivial correction to this correlator, from which
  we extract the defect CFT (dCFT) data characterising the 2d dCFT of the 1/2
  BPS plane. Along the way we derive a supersymmetric inversion formula, obtain
  the relevant superconformal blocks and check that crossing symmetry is
  satisfied. Notably our result features a holomorphic function whose appearance
  is related to the chiral algebra construction of Beem, Rastelli and van Rees.
  Second, we use that chiral algebra description to obtain exact results for the
  BPS sector of the dCFT, valid at any $N$ and for any choice of surface
  operator. These results provide a window into the dynamics of strings of the
  mysterious 6d theories.
\end{abstract}

\mainmatter

\tableofcontents

\section{Introduction and summary}

Calculating observables in the 6d $\cN = (2,0)$ superconformal field theories is
a challenging problem. Due to their lack of a lagrangian description, the
theories remain largely mysterious and outside the reach of conventional field
theoretic methods.

The most established way to study these theories is through
holography.  At large $N$, the $A_{N-1}$ and $D_N$ $\cN = (2,0)$ theories are
dual to 11d supergravity on an $AdS_7 \times S^4$~\cite{maldacena:1997re} and
$AdS_7 \times S^4/\mathbb{Z}_2$~\cite{Witten:1998xy,Aharony:1998rm} background
respectively, where the radius of $AdS_7$ (in Planck
units) is related to $N$ as $R_{AdS}/l_P = (8 \pi N)^{1/3}$.
This supergravity description is useful and leads to concrete predictions;
unfortunately it is also impractical beyond the large $N$ limit: Subleading
corrections probe high-energy corrections to 11d supergravity coming from
M-theory, and we currently have no way to determine these systematically. It is
therefore imperative to find new ways to calculate observables beyond the large
$N$ limit.

The more modern approach to calculating observables is to rely on the
methods of the conformal
bootstrap~\cite{Ferrara:1973yt,Ferrara:1973vz,Polyakov:1974gs,Rattazzi:2008pe}
and the chiral algebra subsector~\cite{Beem:2013sza,Beem:2014kka}. In the context of the
$\cN = (2,0)$ theories, the chiral algebra description was used to calculate
protected CFT data and obtain information about the spectrum of BPS
operators~\cite{Beem:2014kka,Chester:2018dga}.
The bootstrap constraints on the 4-point function of
stress tensor supermultiplets were studied first numerically
in~\cite{Beem:2015aoa} and analytically at large $N$
in~\cite{Rastelli:2017ymc,Zhou:2017zaw,Heslop:2017sco,Chester:2018dga,Alday:2020tgi}
(see also~\cite{Arutyunov:2002ff,Heslop:2004du,Alday:2020lbp,Lemos:2021azv,Kantor:2022epi}).
These works have led to remarkable progress in understanding both the (2,0)
theories at large $N$ and, through holography and the flat space limit of
Mellin amplitudes~\cite{Penedones:2010ue}, scattering amplitudes in M-theory.

In this paper we take a first step to generalise this approach to include
correlators involving surface
operators~\cite{Witten:1995zh,Strominger:1995ac,ganor:1996nf,Howe:1997ue}.
Surface operators are particularly interesting because they play a role
analogous to the Wilson lines of gauge theories and capture interesting physical
properties of these theories not accessible to local operators, such as
higher-form
symmetries~\cite{Gaiotto:2014kfa,DelZotto:2015isa,Bhardwaj:2020phs,Apruzzi:2021mlh}
and measuring the string
potential~\cite{maldacena:1998im,Drukker:2021vyx,Drukker:2022beq}.  In addition
to what they compute, they are useful because they provide a wealth of new
observables, such as their expectation value and correlators with other
operators, and thus provide a larger playground to study the 6d theories.

More precisely, we study the 2-point function of the stress tensor
superprimaries in the presence of a 1/2 BPS defect $V$ defined over a
plane in $\bR^6$.  The set of such defects $V$ is expected to be equal to the set of finite dimensional
representations of the $ADE$ group entering the classification of 6d
$\cN = (2,0)$ theories~\cite{DHoker:2008rje,bachas:2013vza}, and in the
following we keep the choice of representation arbitrary.

The bootstrap approach to this kind of correlator was first developed in the context
of boundary CFTs~\cite{Liendo:2012hy} and later generalised to defect
CFTs~\cite{Billo:2016cpy} and super-CFTs~\cite{Liendo:2016ymz}. At large $N$, an effective approach to calculating
correlators based on the defect version~\cite{Lemos:2017vnx} of the inversion formula~\cite{Caron-Huot:2017vep} was outlined
in~\cite{Barrat:2021yvp} for the Wilson line in $\cN = 4$ SYM (see
also~\cite{Gimenez-Grau:2021wiv,Gimenez-Grau:2022ebb,Bianchi:2022sbz}).
Here we adapt the strategy of~\cite{Barrat:2021yvp} to obtain the first nontrivial correction to our
correlator at large $N$. 

In the rest of this introduction we present a summary of our results.

\subsection{Summary}

The superprimaries of the stress tensor multiplet are scalars transforming in
the symmetric traceless representation of $\sof(5)$ R-symmetry, and we denote
them by $\O2^{I_1 I_2}$, with $I = 1, \dots, 5$ R-symmetry indices. Their
conformal dimension is protected by supersymmetry and fixed to $\Delta = 4$.
It's convenient to introduce a polarisation vector $u$ to avoid carrying
indices, and we define
\begin{align}
  \O2(x,u) \equiv
  \O2^{I_1 I_2}(x) u_{I_1} u_{I_2}\,, \qquad
  I = 1, \dots, 5\,.
\end{align}
Note that we can enforce tracelessness by requiring $u$ to be a null vector, $u^2 = 0$.
A review of this embedding space formalism can be found in~\cite{Costa:2011mg}.

In the absence of a defect, the 2-point functions of these operators are completely fixed by conformal
symmetry up to the choice of normalisation for the operators, but the presence
of the defect $V$ breaks the conformal symmetry $\sof(2,6) \to \sof(2,2) \times
\sof(4)$ and the R-symmetry $\sof(5) \to \sof(4)$, leading to 3 independent
cross-ratios $z, \bar{z}, \omega$. Writing $x^\perp$ for the coordinates
perpendicular to the plane and taking $n$ to be the unit vector specifying the
embedding of $\sof(4) \subset \sof(5)$, we define these cross-ratios by
\begin{align}
  \frac{z+\bar{z}}{2\sqrt{z \bar{z}}} =
  \frac{x_1^{\perp} \cdot x_2^\perp}{\norm{x_1^\perp} \norm{x_2^\perp}}\,,
  \qquad
  \frac{(1-z)(1-\bar{z})}{\sqrt{z \bar{z}}} =
  \frac{x_{12}^2}{\norm{x_1^\perp} \norm{x_2^\perp}}\,,
  \qquad
  \frac{(1-\omega)^2}{\omega} =
  \frac{u_1 \cdot u_2}{(u_1 \cdot n)(u_2 \cdot n)}\,.
 \label{eqn:crossratios}
\end{align}
Here and below we use the short-hand notation $x_{12} \equiv x_1 - x_2$, and
when restricting to coordinates along the plane we use the notation
$x^\parallel$. This choice of cross-ratios is convenient and admits a geometric
interpretation reviewed in section~\ref{sec:kinematics}, see
figure~\ref{fig:crossratios} there.

The 2-point function is then constrained by kinematics to take the form
\begin{align}
  \vev{\O2(x_1,u_1) \O2(x_2,u_2) V} = 
  \frac{(u_1 \cdot n)^2 (u_2 \cdot n)^2}{|x_1^\perp|^{4} |x_2^\perp|^{4}}
  \cF(z,\bar{z},\omega)\,.
  \label{eqn:2pointfunction}
\end{align}
As we derive in section~\ref{sec:kinematics} the function $\cF$ is not arbitrary and must
satisfy additional constraints from supersymmetry known as superconformal Ward
identities
\begin{align}
  (\partial_z + \partial_\omega) \cF(z,\bar{z},\omega)|_{z=\omega} = 0\,,
  \qquad
  (\partial_{\bar{z}} + \partial_\omega) \cF(z,\bar{z},\omega)|_{\bar{z}=\omega} = 0\,.
  \label{eqn:SCWI}
\end{align}
These constraints can be understood as Cauchy-Riemann equations for the
functions $\cF(z,\bar{z},z)$ and $\cF(z,\bar{z},\bar{z})$. They can be
solved explicitly in terms of 2 functions
$F(z,\bar{z}), \zeta(z)$
\begin{equation}
  \begin{aligned}
    \cF(z,\bar{z},\omega) &=
    \frac{(z-\omega)(\bar{z}-\omega)
    (z-\omega^{-1})(\bar{z}-\omega^{-1})}{(z-1)^2(\bar{z}-1)^2} F(z,\bar{z})\\
    &+ \frac{z(\bar{z}-\omega)(\bar{z}-\omega^{-1})(\omega-1)^2}
    {\omega (\bar{z}-z)(\bar{z} - z^{-1}) (z-1)^2 } \bar{\zeta}(\bar{z})
    + \frac{\bar{z}(z-\omega)(z-\omega^{-1})(\omega-1)^2}{\omega
      (z-\bar{z})(z - \bar{z}^{-1}) (\bar{z}-1)^2} \zeta(z)\,.
  \end{aligned}
  \label{eqn:solWIk2}
\end{equation}
Our goal is then to calculate $F$ and $\zeta$ at large $N$.

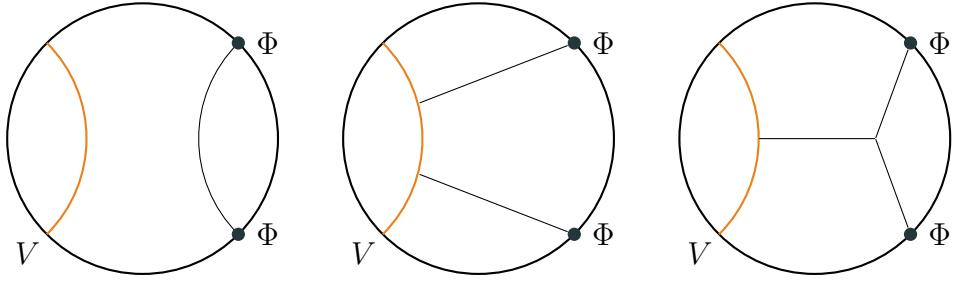
\begin{figure}[tb]
  \begin{tikzpicture}[scale=1.2]
    \draw[thick] (1.5,0) arc (0:360:1.5);
    \draw[thick,mLightBrown] ({1.5*cos(-135)},{1.5*sin(-135)}) arc (-45:45:1.5);
    \draw ({1.5*cos(-45)},{1.5*sin(-45)}) arc (-135:-225:1.5);
    \node[circle,fill=mDarkTeal,inner sep=0pt,minimum size=5pt,label=right:{$\O2$}]
    (bulkop) at ({1.5*cos(-45)},{1.5*sin(-45)}) {};
    \node[circle,fill=mDarkTeal,inner sep=0pt,minimum size=5pt,label=right:{$\O2$}]
    (bulkop2) at ({1.5*cos(45)},{1.5*sin(45)}) {};
    \node (defect) at ({1.8*cos(-135)},{1.8*sin(-135)}) {$V$};
  \end{tikzpicture}
  \hspace*{1em}
  \begin{tikzpicture}[scale=1.2]
    \draw[thick] (1.5,0) arc (0:360:1.5);
    \draw[thick,mLightBrown] ({1.5*cos(-135)},{1.5*sin(-135)}) arc (-45:45:1.5);
    \node[circle,fill=mDarkTeal,inner sep=0pt,minimum size=5pt,label=right:{$\O2$}]
    (bulkop) at ({1.5*cos(-45)},{1.5*sin(45)}) {};
    \node[circle,fill=mDarkTeal,inner sep=0pt,minimum size=5pt,label=right:{$\O2$}]
    (bulkop2) at ({1.5*cos(45)},{1.5*sin(-45)}) {};
    \node (defect) at ({1.8*cos(-135)},{1.8*sin(-135)}) {$V$};
    \node[inner sep=0pt] (defecti1) at ({1.5*(-2*cos(45)+cos(15))},{1.5*sin(15)}) {};
    \node[inner sep=0pt] (defecti2) at ({1.5*(-2*cos(45)+cos(15))},{1.5*sin(-15)}) {};
    \draw (bulkop) -- (defecti1);
    \draw (bulkop2) -- (defecti2);
  \end{tikzpicture}
  \hspace*{1em}
  \begin{tikzpicture}[scale=1.2]
    \draw[thick] (1.5,0) arc (0:360:1.5);
    \draw[thick,mLightBrown] ({1.5*cos(-135)},{1.5*sin(-135)}) arc (-45:45:1.5);
    \node[circle,fill=mDarkTeal,inner sep=0pt,minimum size=5pt,label=right:{$\O2$}]
    (bulkop) at ({1.5*cos(-45)},{1.5*sin(45)}) {};
    \node[circle,fill=mDarkTeal,inner sep=0pt,minimum size=5pt,label=right:{$\O2$}]
    (bulkop2) at ({1.5*cos(45)},{1.5*sin(-45)}) {};
    \node (defect) at ({1.8*cos(-135)},{1.8*sin(-135)}) {$V$};
    \node[inner sep=0pt,minimum size=0pt] (int12) at ({1.5*(2*cos(45)-cos(15))},0) {};
    \node[inner sep=0pt,minimum size=0pt] (defecti) at ({1.5*(-2*cos(45)+1)},0) {};
    \draw (bulkop) -- (int12) -- (bulkop2);
    \draw (int12) -- (defecti);
  \end{tikzpicture}
  \centering
  \caption{Leading Witten diagrams at large $N$. In these pictures the orange
  curve represents the minimal M2-brane dual to the surface operator $V$, and the
  black lines denote propagators in $AdS$.}
  \label{fig:witten}
\end{figure}

The structure of the correlator at large $N$ is easy to understand from
supergravity.  The stress tensor superprimaries are dual to Kaluza-Klein modes
on the $S^4$~\cite{maldacena:1997re,Witten:1998qj}, and the 1/2 BPS plane in the
fundamental representation of $A_{N-1}$ is dual to an M2-brane extended along an
$AdS_3 \subset AdS_7 \times S^4$~\cite{maldacena:1998im}; for the symmetric
representation with $M$ indices we take $M$ coincident
M2-branes~\cite{DHoker:2008rje,bachas:2013vza}.  The leading contributions to
the correlator are given by the Witten diagrams presented in
figure~\ref{fig:witten}, and we can evaluate their respective order in $N$ and
$M$ by dimensional analysis. Propagators contribute as $G_{11}^{-1} \sim
N^{-3}$, and conversely vertices contribute as $N^3$. Interactions with the
M2-branes are proportional to the M2-brane tension $T_{M2} \sim N$, so for $M$
M2-branes we expect a factor $MN$.  Schematically we then expect
\begin{align}
    \vev{\O2(x_1, u_1) \O2(x_2, u_2) V} =
    \frac{(u_1 \cdot u_2)^2}{|x_{12}|^8} + \frac{M^2}{N} \frac{(u_1 \cdot n)^2
    (u_2 \cdot n)^2}{|x_1^\perp|^4 |x_2^\perp|^4} + \frac{M}{N^2}
    (\dots)
    + O(N^{-3})
    \label{eqn:sugraexpansion}
\end{align}
The relative factors of $M^2/N$ and $M/N^2$ are in direct correspondence with
the Witten diagrams.

The first two terms correspond to disconnected correlators and are known
exactly. The first is the free propagator, its coefficient
is one by normalisation. The second is the square of the one-point function,
which can be expressed in terms of anomaly coefficients $c,d$ (see
Appendix~\ref{sec:calca2} for a derivation)
\begin{align}
  \vev{\O2(x, u) V} = a_2 \frac{(u \cdot n)^2}{|x^\perp|^4}\,, \qquad
  a_2 = -\frac{d}{\sqrt{c/2}}\,.
  \label{eqn:a2}
\end{align}
These anomaly coefficients are defined for any $ADE$ group $\mathfrak{g}$
specifying the $\cN = (2,0)$ theory and any choice of representation for the
surface operator $V$, defined by its highest weight $\Lambda$.  A conjecture for
their exact values was proposed for $c$ in~\cite{Beem:2014kka} and $d$
in~\cite{Jensen:2018rxu}; both pass many consistency checks. Writing
$d_\mathfrak{g}$, $h^\vee_\mathfrak{g}$ and $r_\mathfrak{g}$ for the dimension,
the dual Coxeter number and the rank of $\mathfrak{g}$, and $\rho = \frac{1}{2}
\sum_{\alpha>0} \alpha$ for the sum of positive roots (the Weyl vector), they
are given by
\begin{align}
  c = 4 d_\mathfrak{g} h^\vee_\mathfrak{g} + r_\mathfrak{g}\,, \qquad
  d = \frac{1}{2}\left( \Lambda, \Lambda \right) + 2 \left( \Lambda, \rho \right)\,.
  \label{eqn:anomalyexact}
\end{align}
In particular for $\su(n)$ and for a symmetric representation of rank $M$, these
coefficients are
\begin{align}
  c = 4N^3 - 3N -1\,,
  \qquad
  d = M (N-1) \left( 1 + \frac{M}{2N} \right)\,.
  \label{eqn:anomalycoeffs}
\end{align}
The coefficient $c$ agrees with the supergravity calculations
of~\cite{Henningson:1998gx,Intriligator:2000eq,Tseytlin:2000sf,Beccaria:2014qea}.
The coefficient $d$ was initially obtained from holographic entanglement
entropy~\cite{Jensen:2018rxu}, and also agrees with a calculation from the superconformal
index~\cite{Chalabi:2020iie} and explicit supergravity calculations
for $M=1$ at large $N$~\cite{graham:1999pm,Drukker:2020dcz,Drukker:2020swu}.
With these results, we get that in the large $N$ limit $a_2^2 \sim M^2/N$, which
matches~\eqref{eqn:sugraexpansion}.

The term proportional to $M/N^2$ in~\eqref{eqn:sugraexpansion} multiplies a nontrivial function of the cross-ratios, and encodes the interaction between
$\O2$ and $V$ via the exchange of a stress tensor multiplet. Its
coefficient is also known exactly and follows from the superconformal block
decomposition of $\cF$
\begin{align}
  \cF(z,\bar{z},\omega) = 
  \frac{(z \bar{z})^2 (1-\omega)^4}{(1-z)^4 (1-\bar{z})^4 \omega^2}
  \sum_{\cO_l} \lambda_{22l} a_l \cG_l(z,\bar{z},\omega)\,.
  \label{eqn:bulkblockdecomp}
\end{align}
In this decomposition, the functions $\cG$ are the superconformal blocks
associated with the exchange of a given bulk supermultiplet, and
the coefficients $\lambda_{22l}$, $a_l$ are respectively the structure
constants appearing in the 3-point function of local operators and 1-point
functions in the presence of $V$. For the stress tensor, $\lambda_{222}$ is
known exactly and is given by
\begin{align}
  \lambda_{222} = \sqrt{\frac{8}{c}}\,.
  \label{eqn:lambda222}
\end{align}
In the large $N$ limit we also reproduce $a_2 \lambda_{222} \sim M/N^2$.

The expectations from supergravity then translate into the expansion
\begin{equation}
\begin{aligned}
 \zeta(z) &= \frac{z^2}{(1-z)^4} + \frac{2d^2}{c} - \frac{4d}{c} \zeta^{(1)}(z) +
  O(c^{-2})\\
  F(z,\bar{z}) &=\,\,\,\,\,\,0\,\,\,\,\,\,+\, \frac{2d^2}{c} -\frac{4d}{c} F^{(1)}(z, \bar{z}) + O(c^{-2})\,.
\end{aligned}
  \label{eqn:Fsolexpansion}
\end{equation}

In section~\ref{sec:inversion}, we adapt the strategy presented
in~\cite{Barrat:2021yvp} to calculate $\zeta, F$ from bootstrap techniques.
The observation is as follows. The function $\cF$ can
have a branch cut at $\bar{z} = 1$ (physically this corresponds to having null
separated bulk operators, refer to figure~\ref{fig:crossratios}), and its discontinuity
along that branch cut can be used to reconstruct the correlator via the
dispersion relation derived in~\cite{Barrat:2022psm,Bianchi:2022ppi}.  From the block
decomposition of $\cF$~\eqref{eqn:bulkblockdecomp} one can show that the only
blocks contributing to the discontinuity are either those with low enough twist
$\Delta - \ell < 8$ or long multiplets with anomalous dimensions. Long multiplets
arise from double-trace operators (in the large $N$ limit they have the
schematic form $\O2_k \square^n \partial^{\ell} \O2_k$, with $\O2_k$ a 1/2 BPS
operator of dimension $2k$) and are expected to have conformal dimensions
\begin{align}
  \Delta = 4k + 2n + \ell + \frac{\gamma}{c} + \dots\,,
  \qquad
  k \ge 2\,,
  \label{eqn:doubletracespectrum}
\end{align}
so at large $c$ their contribution to the discontinuity is subleading.
We can conclude that to order $c^{-1}$, the only superblocks that contribute
to the discontinuity are the exchange of the identity and stress tensor
multiplet, which correspond to diagrams 1 and 3 of figure~\ref{fig:witten}. A
straightforward strategy is then to obtain their respective superblocks,
calculate the discontinuity and reconstruct the correlator from its
discontinuity.

Unfortunately this strategy is incomplete for two reasons.
First, the dispersion relation of~\cite{Barrat:2022psm,Bianchi:2022ppi} may not
reconstruct the full correlator, for instance it misses the disconnected diagram
in the middle of figure~\ref{fig:witten}. Second, it is also not manifestly
supersymmetric, and indeed to satisfy the superconformal Ward
identities~\eqref{eqn:SCWI}, the result obtained this way must be supplemented
by an infinite number of conformal blocks, see for
instance~\cite{Barrat:2021yvp}.

In section~\ref{sec:inversion} we resolve both of these issues by deriving a
manifestly supersymmetric inversion formula~\eqref{eqn:susyinversionB}
(see~\cite{Lemos:2021azv} for a similar idea applied to 4-point functions of
local operators). As we review below, in addition to the decomposition into bulk
superconformal blocks~\eqref{eqn:bulkblockdecomp}, $\cF$ admits a decomposition
in defect superconformal blocks $\hat{\cG}$
\begin{align}
  \cF(z,\bar{z},\omega) = \sum_{\hat{\cO}_l} b_{kl}^2 \hat\cG_{l}(z,\bar{z},\omega)\,.
  \label{eqn:defectblockdecomp}
\end{align}
The supersymmetric inversion formula calculates the coefficients $b_{kl}^2$
entering this decomposition directly from the discontinuity of $\cF$.
Resumming these superblocks we are guaranteed to obtain a supersymmetric
correlator, and furthermore we observe that it calculates all but two
superblocks contributing to $\cF$: the defect identity (corresponding to the
middle diagram in figure~\ref{fig:witten}) and the displacement
supermultiplet. The contribution from the defect identity is simply the
$d^2/c$ term in~\eqref{eqn:Fsolexpansion}, and the contribution from the
displacement multiplet is fixed by kinematics~\cite{Drukker:2020atp} and
proportional to $d/c$.

Implementing this strategy we obtain the subleading terms in~\eqref{eqn:Fsolexpansion}
\begin{equation}
  \begin{aligned}
    \zeta^{(1)}(z) &= \frac{z}{(1-z)^2}\,,\\
    F^{(1)}(z,\bar{z}) &=
    \frac{ z \bar{z}}{(1-z \bar{z})^6}
    \left[ 
      2 \left( 1 + z\bar{z} + (z \bar{z})^2 \right) \left( 1 + 18 z \bar{z} + (z \bar{z})^2 \right)
      - (z + \bar{z}) \left( 1+z \bar{z} \right) \left( 1 + 28 z \bar{z} + (z
      \bar{z})^2 \right)
     \right] \\
    &+ 6 \frac{(z \bar{z})^2 \log{z \bar{z}}}{(1-z \bar{z})^7} \left[
      \left( 1 + z \bar{z} \right) \left( 3 + 4 z \bar{z} + 3 (z \bar{z})^2 \right)
    - 2 \left( z + \bar{z} \right) (1 + 3 z \bar{z} + (z \bar{z})^2)
    \right]\,.
  \end{aligned}
  \label{eqn:corrresult}
\end{equation}
We expect subleading corrections to $\cF$ in $d/c$ and $1/c$, so this result is
valid for any choice of representation for $V$, as long as $1 \ll d \ll c$. 

\begin{figure}[tb]
  \centering
  \begin{tikzpicture}
    \draw[thick,mLightBrown] (-2,0) -- (2,0);
    \draw[dashed] (-1,1.5) -- (-1,0.05) -- (1,0.05) -- (1,1.5);
    \node[circle,fill=mDarkTeal,inner sep=0pt,minimum size=5pt,label=left:{$\O2$}]
    (bulkop) at (-1,1.5) {};
    \node[circle,fill=mDarkTeal,inner sep=0pt,minimum size=5pt,label=right:{$\O2$}]
    (bulkop) at (1,1.5) {};
    \node at (3,1) {$=$};
    \draw[thick,mLightBrown] (4,0) -- (8,0);
    \draw[dashed] (5,1.5) -- (6,1) -- (7,1.5);
    \draw[dashed] (6,1) -- (6,0);
    \node[circle,fill=mDarkTeal,inner sep=0pt,minimum size=5pt,label=left:{$\O2$}]
    (bulkop) at (5,1.5) {};
    \node[circle,fill=mDarkTeal,inner sep=0pt,minimum size=5pt,label=right:{$\O2$}]
    (bulkop) at (7,1.5) {};
  \end{tikzpicture}
  \caption{Crossing symmetry constraint for a 2-point function of bulk operators
  with a defect. On the left, the correlator is evaluated using the defect
  channel decomposition~\eqref{eqn:defectblockdecomp}, on the right with the
  bulk channel decomposition~\eqref{eqn:bulkblockdecomp}.}
  \label{fig:crossing}
\end{figure}
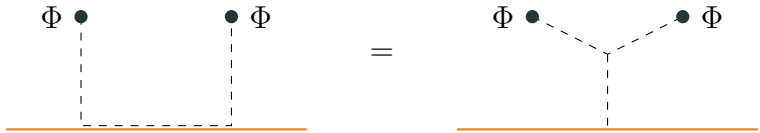

A nontrivial check of our result is that the correlator we obtain satisfies
crossing symmetry, see figure~\ref{fig:crossing}.
The terms corresponding to the exchange of bulk and defect identity are known to
be crossing symmetric by themselves, but in addition we check that
$\zeta^{(1)}, F^{(1)}$ lead to a crossing symmetric correlator.

Finally our result leads to one more surprise. A fundamental difference between
the case of the Wilson line in $\cN = 4$ SYM studied in~\cite{Barrat:2021yvp}
and surface operators in the (2,0) theories is the lack of results from
supersymmetric localization. Yet, we show in section~\ref{sec:chiral} that we can
obtain exact results in 6d as well. As was shown in~\cite{Beem:2014kka}, a
protected subsector of the 6d theories obey the structure of a chiral algebra,
and for the $A_{N-1}$ 6d theories the corresponding chiral algebra is expected
to be the $\mathcal{W}_N$-algebras.  In parallel to the analogous constructions
in 4d~\cite{Beem:2013sza,Cordova:2017mhb}, we show in section~\ref{sec:chiral}
that surface operators are identified with a module of the
$\mathcal{W}_N$-algebras, and using that description we obtain exact results for
the protected defect CFT (dCFT) data. In particular, we are able to calculate
the holomorphic function $\zeta(z)$ and find that the bootstrap result obtained
above is exact
\begin{align}
  \zeta(z) = 
  \frac{z^2}{(1-z)^4} + \frac{2d^2}{c} - \frac{4d}{c} \frac{z}{(1-z)^2}\,.
  \label{eqn:zetafull}
\end{align}
This is a nontrivial check of our bootstrap result, and
unlike~\eqref{eqn:Fsolexpansion}, this expression has no further corrections in
$c,d$, and we conjecture that it is valid for any $ADE$ theory and any choice of
representation for $V$.

The rest of this paper is organised as follows. In section~\ref{sec:kinematics}
we review the definition of the dCFT data, study the kinematics of the 2-point
function, derive the superconformal Ward identities and study the two
superconformal blocks decompositions of $\cF$. In section~\ref{sec:chiral} we
use the chiral algebra description to obtain protected dCFT data and calculate
$\zeta(z)$~\eqref{eqn:zetafull}. In section~\ref{sec:inversion} we present a
supersymmetric inversion formula and use it to obtain the dCFT data associated
with the defect channel and our main result~\eqref{eqn:corrresult}.  Finally, in
section~\ref{sec:crossing} we check crossing symmetry and obtain dCFT data
associated with the bulk channel.

We also include three appendices. Appendix~\ref{sec:bulkblocks} contains a
review of the conformal blocks in the bulk channel, along with a derivation of
the superconformal blocks relevant to this paper.
Appendix~\ref{sec:defectblocks} does the same for defect channel blocks.
Appendix~\ref{sec:calca2} presents the calculation of $a_2$ in terms of the
anomaly coefficients for our choice of normalisation.

\section{Kinematics}
\label{sec:kinematics}

We begin by reviewing the constraints from kinematics on correlators, the
definition of defect CFT data and the two superconformal blocks decomposition
of the 2-point function~\eqref{eqn:2pointfunction}.

\subsection{Review of defect CFT data}

A CFT can be defined by its spectrum of operators along with the structure constants
appearing in their 3-point functions. For the $A_{N-1}$ $\cN=(2,0)$ theories,
the spectrum is expected to contain a set of 1/2 BPS operators $\O2_k$ ($k = 2, \dots, N$)
transforming in the symmetric traceless representation of $\sof(5)$ with $k$
indices ($\O2_{k=2} \equiv \O2$ is the superprimary of the stress tensor multiplet). For these
operators the 3-point functions take the form
\begin{equation}
  \begin{aligned}
  &\vev{\O2_{k_1}(x_1,u_1) \O2_{k_2}(x_2,u_2)
  \O2_{k_3}(x_3, u_3)}\\
  &\qquad\qquad = \lambda_{k_1 k_2 k_3} 
  \left( \frac{-2u_1 \cdot u_2}{x_{12}^4} \right)^{\frac{k_{123}}{2}}
  \left( \frac{-2u_1 \cdot u_3}{x_{13}^4} \right)^{\frac{k_{132}}{2}}
  \left( \frac{-2u_2 \cdot u_3}{x_{23}^4} \right)^{\frac{k_{231}}{2}}\,,
  \end{aligned}
  \label{eqn:3pts}
\end{equation}
where we use the shorthand notation $k_{ijk} \equiv k_i+k_j-k_k$, and the
factors $-2$ are introduced for later convenience.
The constants $\lambda$ are not fixed by symmetry and are pieces of CFT data
defining the theory.

The value for these structure constants $\lambda$ depends on the choice of
normalisation of the operators, and for definiteness in this paper we take
the 2-point function to be
\begin{align}
  \vev{\O2_k(x_1,u_2) \O2_k(x_2,u_2)}
  =
  \left(\frac{2u_1 \cdot u_2}{x_{12}^4}\right)^k\,.
  \label{eqn:2pts}
\end{align}

In addition to this usual CFT data characterising the algebra of local
operators, surface operators enrich the theory by a new set of defect CFT data
that characterise correlators involving the surface operators.

The 1/2 BPS operators $\O2_k$ can acquire an expectation value in the presence
of a surface operators $V$ and are constrained by the residual conformal
symmetry to take the form
\begin{align}
  \vev{\O2_k(x,u) V} = a_{k} \frac{(u \cdot n)^k}{|x^\perp|^{2k}}\,.
  \label{eqn:1pts}
\end{align}
The coefficients $a_k$ are independent pieces of dCFT data.

Finally, the dCFT contains defect operators $\hat{\cO}$ that can be inserted on
the defect $V$.
These defect operators sit in multiplets of the $\osp(4^*|2) \oplus \osp(4^*|2)$
algebra preserved by the plane $V$, which includes as a bosonic subalgebra the
$\sof(2,2)$ group of rigid 2d conformal symmetries along the plane, the $\sof(4)$
rotations of the space transverse to the plane and the residual $\sof(4)$
R-symmetry. For defect operators of conformal dimension $\hat{\Delta}$ and in
representations of transverse spin $s$ and R-symmetry spin $r$, we
again introduce an index-free notation by contracting the operators with
polarisation vectors $u_\parallel^i$ and $v^m$ (with $i,m = 1,2,3,4$ and
$u, v$ s.t. $u_\parallel^2 = v^2 = 0$)
\begin{align}
  \hat{\cO}_{\hat{\Delta},s,r}(x_\parallel,u_\parallel,v) \equiv
  \hat{\cO}_{m_1 m_2 \dots m_{s}}^{i_1 i_2 \dots i_{r}}(x_\parallel)
  u^{i_1}_\parallel u^{i_2}_\parallel \dots
  v^{m_1} v^{m_2} \dots 
  \label{eqn:defnOl}
\end{align}
As above, for definiteness we assume the normalisation
\begin{align}
  \vev{V[\hat{\cO}_{\hat{\Delta},s,r}(x_1,u_1,v_1)
  \hat{\cO}_{\hat{\Delta},s,r}(x_2,u_2,v_2)]}
  =
  \frac{(2u_1 \cdot u_2)^{r} (2 v_1 \cdot v_2)^{s}}{x_{12}^{2\hat{\Delta}}}\,.
  \label{eqn:2ptsnormspin}
\end{align}
Their correlators with bulk operators is fixed by kinematics up to a
coefficient $b$
\begin{align}
  \vev{\O2_k(x_1^\parallel,x_1^\perp,u_1)
  V[\hat{\cO}_{\hat{\Delta},s,r}(x_2^\parallel,u_2^\parallel,v_2)]}
  =
  b_{k,\{\hat{\Delta},s,r\}} \frac{(u_1 \cdot n)^{k-r} (-2u_1 \cdot
  u_2^\parallel)^{r} (x^\perp_1 \cdot v_2)^{s}}
  {|x_1^\perp|^{2k-\hat{\Delta}+s} \left[ (x_1^\perp)^2 + (x_{12}^\parallel)^2
  \right]^{\hat{\Delta}}}\,.
  \label{eqn:bulkdefect2pts}
\end{align}

There are two distinguished defect operators. In the case of the defect
identity ($\hat{\Delta} = s = r = 0$), the
correlators~\eqref{eqn:bulkdefect2pts} reduce to a 1-point
function~\eqref{eqn:1pts} and $b_{k,\left\{ 0,0,0 \right\} } \equiv a_k$. The
other universal operator is the displacement supermultiplet, which arises from
the broken symmetries in the presence of $V$. We denote it by $B[1]$
anticipating the notation for defect supermultiplets (see
section~\ref{sec:defectchannel}), and its superprimary by $\hat{\cO}_{B[1]}$.
This supermultiplet is studied in details in~\cite{Drukker:2020atp}, and the
coefficient $b_{2,B[1]}$ was shown to be fixed by Ward identities to
\begin{align}
  \vev{\O2(x_1^\parallel,x_1^\perp,u_1) V[\hat{\cO}_{B[1]}(x_2^\parallel,u_2^\parallel)]}
  =
  b_{2,B[1]}
  \frac{(u_1 \cdot n) (-2u_1 \cdot u_2^\parallel)}
  {|x_1^\perp|^{2} \left[ (x_1^\perp)^2 + (x_{12}^\parallel)^2
  \right]^{2}}\,, \qquad
  b_{2,B[1]} = 2 \sqrt{\frac{d}{c}}\,.
  \label{eqn:disp}
\end{align}

\subsection{Kinematics for 2-point functions}
\label{sec:kinematics2pts}

We now turn to the correlators involving two bulk operators at points $x_1$ and
$x_2$ and a defect $V$. The kinematics for these correlators are studied
in~\cite{Billo:2016cpy} and reviewed below.  Unlike the previous correlators,
these correlators are not completely fixed by kinematics constraints and involve
an arbitrary function of 3 cross-ratios $z, \bar{z}, \omega$.

To understand their origin, it is convenient to use the symmetries of the
correlator to fix the position of the operators. Using the translations and the special conformal
transformations preserved by the defect, one can show that there is a frame in which both
$x^\parallel_{1,2} = 0$, so in that frame the only invariants under the residual
symmetry are 2 linearly independent combinations built out of\footnote{Transverse rotations and dilatations preserve this frame and act on $x_1^\perp$, $x_2^\perp$.}
\begin{align}
  \frac{x_1^\perp \cdot x_2^\perp}{|x_1^\perp||x_2^\perp|}\,, \qquad
  \frac{|x_2^\perp|}{|x_1^\perp|}\,, \qquad
  \frac{(x_{12}^\perp)^2}{|x_1^\perp||x_2^\perp|}\,.
  \label{eqn:confinvbos}
\end{align}
We can recover the full conformal invariants by undoing the frame fixing, and it
is convenient to take the 2 cross-ratios $z, \bar{z}$ to
be~\eqref{eqn:crossratios} as defined in~\cite{Lemos:2017vnx}.

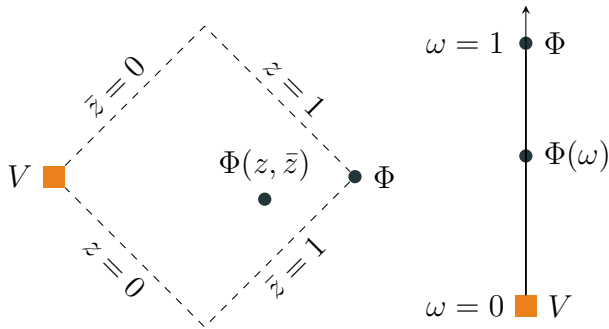
\begin{figure}[tb]
  \centering
  \begin{tikzpicture}
    \node[fill=mLightBrown,inner sep=0pt,minimum size=8pt,label=left:{$V$}] (defect) at (-2,0) {};
    \node[circle,fill=mDarkTeal,inner sep=0pt,minimum size=5pt,label=right:{$\O2$}] (bulkop) at (2,0) {};
    \node[circle,fill=mDarkTeal,inner sep=0pt,minimum
    size=5pt,label={$\O2(z,\bar{z})$}]
    (bulkop2) at (0.8,-.3) {};
    \node[rotate=45] at (-1.2,1.2) {$\bar{z}=0$};
    \node[rotate=-45] at (1.2,1.2) {$z=1$};
    \node[rotate=-45] at (-1.2,-1.2) {$z=0$};
    \node[rotate=45] at (1.2,-1.2) {$\bar{z}=1$};
    \draw[dashed] (defect) -- (0,2) -- (bulkop) -- (0,-2) -- (defect);
  \end{tikzpicture}
  \begin{tikzpicture}
    \node[fill=mLightBrown,inner sep=0pt,minimum size=8pt,label=right:{$V$}] (defect) at
    (0,-2) {};
    \node[label=left:{$\omega=0$}] at (0,-2) {};
    \node[circle,fill=mDarkTeal,inner sep=0pt,minimum
    size=5pt,label=right:{$\O2$}] (bulkop) at (0,1.5) {};
    \node[label=left:{$\omega=1$}] at (0,1.5) {};
    \node[circle,fill=mDarkTeal,inner sep=0pt,minimum size=5pt,label=right:{$\O2(\omega)$}]
    (bulkop2) at (0,0) {};
    \draw (defect) -- (bulkop) -- (defect);
    \draw[-stealth] (defect) -- (0,2);
  \end{tikzpicture}
  \caption{Interpretation for the cross-ratios $z, \bar{z}, \omega$ in
  lorentzian kinematics. Upon choosing an appropriate conformal frame, one can
  always bring the correlator~\eqref{eqn:2pointfunction} to the following kinematic
  configuration. On the left, we draw a plane transverse to $V$ and
  containing both bulk operators. One can set $V$ at the origin at fix the
  location of one $\O2$; $z, \bar{z}$ are then the coordinates of the
  remaining $\O2$. The dotted lines are the locations of the lightcones. On
  the right, we draw the transverse direction to $V$ in R-symmetry space. We can
  again fix $V$ at the origin and set the location of one $\O2$. The position
  of the other $\O2$ is $\omega$.}
  \label{fig:crossratios}
\end{figure}

These have a nice interpretation. If we further fix $x_1^\perp =
(1,0,0,0)$ and $x_2^\perp = (x,y,0,0)$, then it's easy to show
that~\eqref{eqn:crossratios} is solved by $z = x+iy$ and $\bar{z} = x-iy$, so
in that frame $z, \bar{z}$ are interpreted as the location of one of the
operators. In particular in Lorentzian kinematics both $z, \bar{z}$ are real and
are interpreted as lightcone coordinates, see figure~\ref{fig:crossratios}.

In addition to these spacetime cross-ratios, the
correlator~\eqref{eqn:2pointfunction} admits many R-symmetry tensor structures,
and they can be packaged as a sum over an R-symmetry cross-ratio $\sigma$
\begin{align}
  \sum_{j=0}^2 F_{j}(z,\bar{z}) \sigma^j\,, \qquad
  \sigma = \frac{u_1 \cdot u_2}{(u_1 \cdot n)(u_2 \cdot n)}\,.
  \label{eqn:Fsigmadecomposition}
\end{align}
We can interpret this geometrically as follows. Polarisation vectors $u$ take
values in the projective space $\mathbb{P}^4$. The space of $u$ subject to the
tracelessness condition $u^2=0$ can then parametrised by coordinates $y$ in $\bR^3$
\begin{align}
  u = [y^i: \frac{1}{2}(1-y^2):\frac{i}{2}(1+y^2)]\,.
  \label{eqn:projectiveu}
\end{align}

Surface operators preserve an $\sof(4) \subset \sof(5)$, so they pick a
direction $n$ in projective space. We can choose $n$ to be a direction in
$\bR^3$, so that $V$ splits $y$ into $\left\{ y^\parallel, y^\perp \right\}$. As
for the spacetime part we can act with preserved R-symmetry to set a frame where
$y_{1,2}^\parallel = 0$, leaving only the invariant
\begin{align}
  \omega = \frac{|y^\perp_2|}{|y^\perp_1|}\,.
  \label{eqn:sigmaomega}
\end{align}
Comparing with~\eqref{eqn:Fsigmadecomposition} we can trade $\sigma$ for
$\omega$ using the relation
\begin{align}
  \sigma = \frac{(1-\omega)^2}{\omega}\,.
\end{align}

\subsection{Superconformal Ward identities}

In addition to the constraints from conformal and R-symmetry, the
correlator~\eqref{eqn:2pointfunction} obeys constraints from supersymmetry. A
nice way to derive them is to promote $\O2$ to superfields taking values in
superspace, so that the correlator also encodes the 2-point functions of all
operators in the stress tensor multiplet.

In addition to the coordinates $x \subset \bR^6$ and $y \subset \bR^3$ (arising
from realising geometrically the R-symmetry, see the discussion
around~\eqref{eqn:projectiveu}), we can introduce 8 Grassmann parameters
$\theta_{\alpha a}$, where the spinor indices run over $\alpha = 1,\dots,4$ and
$a = 1,2$. Together, $x, \theta, y$ organise in the supercoordinates $X$ of analytic
superspace~\cite{Dolan:2004mu,Heslop:2004du}
\begin{align}
  X =
  \left( \begin{array}{c|c}
    x_{\alpha\beta} & \theta_{\alpha b}\\ \hline
    \theta_{a\beta} & y_{ab}\\
  \end{array} \right)\
  \in GL(4|2)\,.
\end{align}
In this equation $x$ is an antisymmetric matrix and $y$ is symmetric. One can
check that this supermatrix $X$ satisfies ``graded antisymmetry'' with respect
to the supermatrix $\Sigma$, which is the constraint ($X^{st}$ is the
supertranspose)
\begin{align}
  X^{st} = - X \Sigma\,, \qquad
  \Sigma = 
  \left( \begin{array}{c|c}
    \mathds{1}_4 & 0\\ \hline
    0 & -\mathds{1}_2\\
  \end{array} \right)\,.
\end{align}
With this property, $X$ can be shown to parametrise the superspace
$OSp(8^*|4)/G$, with $G$ the stabiliser of $\O2$ (see~\cite{Heslop:2004du} for
more details).

Surface operators also have a simple description in this superspace similar to the one presented in \cite{Liendo:2016ymz} for defects in four dimensional $\mathcal{N}=4$ SCFTs. The operator $V$
picks a plane in $\bR^6$ and a direction in $\bR^3$, for definiteness we can
take them along $x^1, x^2$ and $y^1$. Then $V$ naturally defines the supermatrix
\begin{align}
  \Pi =
  \left( \begin{array}{c|c}
    i \gamma_{12} & 0\\ \hline
    0 & \rho_1\\
  \end{array} \right)\,. 
\end{align}
This decomposes $X$ into $X_\parallel$ and $X_\perp$ according to (anti)symmetry
under
\begin{align}
  X = X_\parallel + X_\perp\,, \qquad
  (\Pi X_\parallel)^{st} = - \Sigma \Pi X_\parallel\,, \qquad
  (\Pi X_\perp)^{st} = \Sigma \Pi X_\perp\,.
\end{align}

We are then interested in uplifting the 2-point
function~\eqref{eqn:2pointfunction} to superspace. The correlator is constrained
to be a function of superconformal invariants, and as for the case of 4-point
functions of $\O2$ discussed in~\cite{Dolan:2004mu}, a simple counting argument
shows that there are no
superconformal invariants built out of fermionic coordinates $\theta$'s only.
This is because our 2-point function depends on 16 fermionic coordinates
($\theta$ for both $X_1$ and $X_2$), and correspondingly the defect preserves 16
supercharges. So there are enough preserved supersymmetries to fix a frame where
all the $\theta$'s vanish. In turn this implies that the 2-point function of all
the superdescendants is uniquely fixed in terms of the function $\cF$ introduced
in~\eqref{eqn:2pointfunction} by finding the appropriate superspace extension to
the cross-ratios $z, \bar{z}, \omega$.

To find the superspace cross-ratios, we can use the previous strategy and again
fix a frame where $X_1^\parallel = X_2^\parallel = 0$.  For some appropriate
choice of basis for the gamma matrices, $X_\perp$ takes the form
\begin{align}
  X_\perp =
  \left( \begin{array}{cc|cc}
    0 & x_\perp & 0 & \theta_{\perp}\\
    -x_\perp^T  & 0 & \eta_{\perp} & 0\\ \hline
    0 & \eta_{\perp}^T & 0 & y_\perp\\
    \theta_{\perp}^T & 0 & y_\perp & 0\\
  \end{array} \right)\,. 
\end{align}
The cross-ratios are then given by the superconformal invariants built out of 2
points, which are the eigenvalues of $X_1 X_2^{-1}$. Let us call them $Z,
\bar{Z}$ and $\Omega$. If we set $\theta, \eta = 0$ it is simple to check that they
match the definitions for $z, \bar{z}, \omega$ introduced
in~\eqref{eqn:crossratios}.

When $\theta, \eta \neq 0$ these eigenvalues receive corrections. To find them, we
consider the 3 identities
\begin{align}
  \frac{1}{2} \str\left( X_1 X_2^{-1} \right)^n = Z^n + \bar{Z}^n - \Omega^n\,, \qquad
  n = 1,2,3\,.
\end{align}
For our purposes it is sufficient to take $X_2$ purely bosonic.
Solving these equations to first order in $\theta, \eta$ we find
(with $\pi_\pm$ some projector satisfying $\pi_+ + \pi_- = 1$; in the frame of
section~\ref{sec:kinematics2pts} they are $\pi_+ = 
\left(\begin{smallmatrix}
  1 & 0 \\
  0 & 0\\
\end{smallmatrix}\right)$ and $\pi_- = \left(\begin{smallmatrix}
  0 & 0 \\
  0 & 1\\
\end{smallmatrix}\right)$)
\begin{equation}
\begin{gathered}
  Z = z - \frac{(\eta^T \pi_+ \theta)}{z-\omega} + \dots\,,
  \qquad
  \bar{Z} = \bar{z} - \frac{(\eta^T \pi_- \theta)}{\bar{z} - \omega} + \dots\,,\\
  \Omega = \omega - \frac{\eta^T \pi_+ \theta}{z - \omega} - \frac{\eta^T \pi_-
  \theta}{\bar{z} - \omega}
  + \dots
\end{gathered}
\end{equation}
Expanding $\cF(Z,\bar{Z},\Omega)$ in fermionic coordinates, we then get
\begin{equation}
\begin{aligned}
  F(Z,\bar{Z},\Omega) &-
  F(z,\bar{z},\omega) =\\
  &-\left( \partial_Z F|_{z} + \partial_\Omega F|_{\omega} \right)
  \frac{(\eta^T \pi_+ \theta)}{z - \omega}
  -
  \left( \partial_{\bar{Z}} F|_{\bar{z}} + \partial_\Omega F|_{\omega} \right)
  \frac{(\eta^T \pi_- \theta)}{\bar{z} - w}
  + \dots
\end{aligned}
\end{equation}
This has unphysical poles at $z = \omega$ and $\bar{z} = \omega$. Indeed the
correlator should be well-defined at these points, and requiring the absence of
singularities leads to the superconformal Ward identities\footnote{
This strategy has been originally applied in \cite{Dolan:2004mu}
and later in a variety on examples, see e.g.~\cite{Liendo:2015cgi,Liendo:2016ymz,Lemos:2016xke,Liendo:2018ukf}
}~\eqref{eqn:SCWI}.

\subsection{Superconformal block expansion: bulk channel}
\label{sec:bulkchannelexpansion}

Consider evaluating the correlator~\eqref{eqn:2pointfunction} using the OPE of
the two bulk operators $\O2 \O2$. The bulk operators that appear in this OPE
are symmetric traceless tensors of dimension $\Delta$, spin $\ell$ and
R-symmetry spin $R$. As we review in Appendix~\ref{sec:bulkblocks}, for each of
them we can calculate a conformal block that encodes their contribution to
$\cF$. These conformal blocks consist of two parts.
First, the spacetime dependence is encoded in a conformal block
$g_{\Delta,\ell}^{st}(z,\bar{z})$ expressed in terms of the conformal blocks
$g^{a,b}_{\Delta,\ell}$ for 4-point functions of local operators in
4d~\cite{Dolan:2000ut}
\begin{align}
  g^{st}_{\Delta,\ell}(z,\bar{z}) = 
  \frac{(1-z)(1-\bar{z})}{1 - z \bar{z}}
  g^{0,0}_{\Delta-1,\ell+1}\left( 1-z, 1-\bar{z} \right)\,.
  \label{eqn:confblockso8sol}
\end{align}
The expression for $g^{(a,b)}_{\Delta,\ell}(z,\bar{z})$ is given
in~\eqref{eqn:gab4d}.  The surprising appearance of 4d conformal blocks in the
context of surface operators in 6d is part of a set of relations
between conformal blocks uncovered and explained in~\cite{Isachenkov:2018pef}.

Second, the R-symmetry dependence takes the form of a
conformal block $h_R(\omega)$ and is given in terms of the Legendre polynomials
$P_R(x)$
\begin{align}
  h_R(\omega) =
  \frac{R!(R+1)!}{(2R+1)!}
  \left( \frac{1-\omega}{1+\omega} \right)
  P_{R+1}\left( \frac{1+\omega}{1-\omega} \right)\,.
  \label{eqn:confblockso5sol}
\end{align}
Together they contribute to $\cF$ as
\begin{align}
  g_{\Delta,\ell,R}(z,\bar{z},\omega)
  &= g^{st}_{\Delta,\ell}(z,\bar{z}) h_R(\omega)\,.
  \label{eqn:confblockbulkch}
\end{align}
The derivation of these conformal blocks along with their appropriate
normalisation is reviewed in Appendix~\ref{sec:bulkblocks}. 

In a SCFT, operators organise into supermultiplets, and correspondingly the
contributions from conformal blocks organise into superconformal blocks.
For instance, the stress tensor multiplet (denoted $\cD[2,0]$) contains the
operators listed in table~\ref{tab:stresstensormult}.
The only operators that can contribute to the exchange are the superprimary
$\O2^{IJ}$ and the stress tensor $T^{\mu\nu}$, so the superconformal block is a
linear combination
\begin{align}
  \cG_{\cD[2,0]} = g_{4,0,2} + \alpha g_{6,2,0}\,.
  \label{eqn:sblockD20}
\end{align}
Imposing the superconformal Ward identity~\eqref{eqn:SCWI} fixes the parameter
$\alpha = -\frac{3}{700}$, and plugging the explicit conformal blocks we get
\begin{align}
  \cG_{\cD[2,0]}(z,\bar{z}, \omega) =
  -12 \left[1 + \frac{1(1+z)(z-\omega)(z\omega
    -1)(\bar{z}-1)^2}{2(1-z)(z-\bar{z})(z \bar{z}-1)(\omega-1)^2} \log(z)
    - (z \leftrightarrow \bar{z}) \right]\,.
  \label{eqn:stresstensorsblock}
\end{align}
We can read the contribution to $F,\zeta$~\eqref{eqn:solWIk2} by taking the limits
$\omega \to \bar{z}$ and $\omega \to 1$
\begin{align}
  \zeta_{\cD[2,0]}(z) =  -\frac{12z^2}{(1-z)^4} \left[ 1 + \frac{1}{2} \frac{1+z}{1-z} \log
    z
  \right]\,, \qquad
  F_{\cD[2,0]} = 0\,.
  \label{eqn:D20sblockcontrib}
\end{align}

\begin{table}
  \centering
  \begin{tabular}{c|c}
    Primary & Representation\\
    \hline
    $T^{\mu\nu}$ & $[2,0,0]_6^{(0,0)}$\\
    $J^\mu$ & $[1,1,0]_{11/2}^{(0,1)}$\\
    $j^\mu$ & $[1,0,0]_5^{(0,2)}$\\
    $H^{\mu\nu\rho}$ & $[0,2,0]_5^{(0,1)}$\\
    $\chi^I$ & $[0,1,0]_{9/2}^{(1,1)}$\\
    $\O2^{IJ}$ & $[0,0,0]_4^{(2,0)}$\\
  \end{tabular}
  \caption{Content of the stress tensor multiplet derived in~\cite{Howe:1983fr}.
  We specify the representations by their Dynkin labels as
  $[j_1,j_2,j_3]^{(k_1,k_2)}_\Delta$. Here $[j_1,j_2,j_3]$ denotes the Dynkin
labels for $\sof(6)$ Lorentz symmetry; the spin $l$ representations are
$[0,l,0]$. $(k_1,k_2)$ denote the $\sof(5)$ Dynkin labels; the spin $R$
representations are $(R,0)$.
}
  \label{tab:stresstensormult}
\end{table}

The superblocks that can contribute to the 2-point
function of 1/2 BPS operators $\O2_k$ are constrained by selection rules. These
were derived
in~\cite{Eden:2001wg,Arutyunov:2002ff,Ferrara:2001uj,Heslop:2004du}, see
also~\cite{Beem:2015aoa}. For two stress tensor superprimaries ($k=2$) they read
\begin{equation}
\begin{aligned}
  \cD[2,0] \times \cD[2,0] =&\ 
  \mathds{1} + \cD[4,0] + \cD[2,0] + \cD[0,4]\\
  &+ \sum_{\ell=0,2,\dots} \cB[2,0]_\ell + \cB[0,2]_{\ell+1}\\
  &+ \sum \cL[0,0]_{\Delta,\ell}\,.
  \label{eqn:selrule2ptsbulk}
\end{aligned}
\end{equation}
For the supermultiplets appearing on the right we follow the notation
from~\cite{Beem:2015aoa}. The first letter indicates the shortening condition,
while $[k_1,k_2]$ specify the $\sof(5)$ representation of the superprimary.

The representations appearing on the right are all those allowed by
representation theory ($\cB[0,0]_\ell$ multiplets are also allowed on the grounds
of representation theory but they contain higher spin conserved currents).
Out of these only $\mathds{1}$, $\cD[2,0], \cD[4,0], \cB[2,0]_\ell$ and $\cL[0,0]$
have symmetric traceless tensors that can acquire an expectation value and
contribute to the correlator. We derive their superblocks in
Appendix~\ref{sec:bulkblocks} and collect the results in Table~\ref{tab:1pts}.

\begin{table}[h]
  \centering
  \renewcommand{\arraystretch}{1.3}
  \begin{tabular}{c|c|c|c|c|c}
    Multiplet & $\Delta$ & $R$ & $\cG(z,\bar{z},\omega)$ & $\zeta(z)$ & $F(z,\bar{z})$\\ \hline
    $\mathds{1}$ & 0 & 0 & 1 & $\frac{z^2}{(1-z)^4}$ & 0\\
    $\cD[2,0]$  & 4 & 2 & \eqref{eqn:sblockD20} &
    $\frac{z^2}{(1-z)^4} k_{4}(1-z)$ & $0$ \\
    $\cD[4,0]$  & 8 & 4 &  \eqref{eqn:sblockD40} &
    $\frac{z^2}{(1-z)^4} k_{8}(1-z)$ &
    $\frac{(z \bar{z})^2}{(1-z)^4(1-\bar{z})^4} g^{st}_{8,0}(z,\bar{z})$\\
    $\cB[2,0]_\ell$  & $8+\ell$ & 2 &  \eqref{eqn:sblockB20} &
    $- \frac{z^2}{(1-z)^4} k_{2(\ell+6)}(1-z)$ &
    $-\frac{(z \bar{z})^2}{(1-z)^4 (1-\bar{z})^4} g^{st}_{\ell+10,\ell+2}(z,\bar{z})$\\
    $\cL[0,0]_{\Delta, \ell}$ & $\Delta > 6 + \ell$ & 0 & \eqref{eqn:sblockL00}, \eqref{eqn:sblockL00l} &
    $0$ & $\frac{(z \bar{z})^2}{(1-z)^4 (1-\bar{z})^4} g^{st}_{\Delta+4,\ell}(z,\bar{z})$ \\
  \end{tabular}
  \renewcommand{\arraystretch}{1}
  \caption{List of supermultiplets contributing to the correlator, along with
    $\Delta,R$ for the superprimary. The conformal blocks entering each
    superblock are detailed in the Appendix~\ref{sec:tabsblockbulk}. We also
    list the contributions of each superblock to $\zeta$ and $F$, with
    $k_{2h}(z)$ defined in~\eqref{eqn:kbdef} and $g_{\Delta,\ell}^{st}(z,\bar{z})$
    defined in~\eqref{eqn:confblockso8sol}.
    }
    \label{tab:1pts}
\end{table}

We note that the relation between $g^{st}_{\Delta,\ell}$ and the 4d conformal blocks
$g^{(a,b)}_{\Delta,\ell}$ also extends to a relation between the superconformal
blocks of table~\ref{tab:1pts} and the superconformal blocks of local operators
in 4d $\cN = 2$ theories~\cite{Gimenez-Grau:2020jrx}. The relation to $\cN = 2$
blocks in particular is natural because the latter depend on a single R-symmetry
cross-ratio.  Comparing our Table~\ref{tab:1pts} with their Table 4, we see that
the superconformal blocks share the same structure.  It would be interesting to
understand how the relations of~\cite{Isachenkov:2018pef} extend more generally
to superconformal blocks for arbitrary defects and bulk operators.

\subsection{Superconformal block expansion: defect channel}
\label{sec:defectchannel}

A second way to evaluate the 2-point function~\eqref{eqn:2pointfunction} is to
use the defect operator expansion (dOE). This expresses bulk
operators in terms of defect operator insertions $\hat{\cO}$ on the defect $V$.
To see which defect operators can enter the dOE, recall that defect operators
$\hat{\cO}$ transform under the symmetry $\sof(2,2) \times \sof(4) \times
\sof(4)$ preserved by the plane $V$. Their representations are labelled as $[s,
r]_h [\bar{s}, \bar{r}]_{\bar{h}}$, where $s, r, \bar{s}, \bar{r}$ are $\su(2)$ 
labels and $h, \bar{h}$ are $\sl(2)$  labels. Their conformal dimension is
$\hat{\Delta} = h + \bar{h}$, and their 2d spin is $h - \bar{h}$.

Consider then the correlator
\begin{align}
\label{O2Ohat}
  \vev{\O2_k(x_1,y_1) V[\hat{\cO}(x_2^\parallel,y_2^\parallel)]}\,,
\end{align}
where $\O2_k$ is the 1/2-BPS primary and $\hat{\cO}$ is a defect
operator in an arbitrary representation (we suppress the additional indices).
First, notice that we can choose a frame where $x_1^\parallel = x_2^\parallel =
0$, idem for $y^\parallel$, so there is no quantity that carries 2d spin. This
means that the only defect operators $\hat{\cO}$ that can have a nonzero correlator \eqref{O2Ohat} are
those with
\begin{align}
\label{hishbar}
  h = \bar{h} = \frac{\hat{\Delta}}{2}\,, \qquad
  r = \bar{r}\,.
\end{align}
A second constraint is that the only quantity that transforms
under the $\sof(4)$ acting on the transverse space is $x_1^\perp$. Therefore the
only representations that can arise are the symmetric traceless, which are those
for which
\begin{align}
\label{sissbar}
  s = \bar{s}\,.
\end{align}
We label these defect operators as $\hat{\cO}_{\hat{\Delta},s,r}$,
and their correlator with bulk operators $\O2_k$ are given
in~\eqref{eqn:bulkdefect2pts}.

From these selection rules we can write a dOE for $\O2_k V$
\begin{align}
  \O2_k(x,y) V =
  \sum_{\{ \hat{\Delta},s,r \}} b_{k,\{\hat{\Delta},s,r\}}
  \frac{|y^\perp|^{k-r}}{|x^\perp|^{2k-\hat{\Delta}}}
  C_{k,\{ \hat{\Delta},s,r \}}
  V[\hat{\cO}_{\hat{\Delta},s,r}(x^\parallel, y^\parallel,v)]\,,
  \label{eqn:doe}
\end{align}
where $C_{k,\{ \hat{\Delta},s,r \}}$ are differential operators encoding the
contribution from the descendants.

Acting twice with the dOE inside the correlator~\eqref{eqn:2pointfunction} gives
a sum of 2-point functions of defect operators $\hat{\cO}_{\hat{\Delta},s,r}$.
As we review in Appendix~\ref{sec:defectblocks}, for each of them we can
calculate a conformal block $\hat{g}$ that encodes their contribution to $\cF$,
we find
\begin{align}
  \hat{g}_{\hat{\Delta},s,r}(z,\bar{z},\omega) =
  (-1)^r
  \frac{(z \bar{z})^{\frac{\Delta-s}{2}}}{(1- z \bar{z})}
  \frac{z^{s+1}-\bar{z}^{s+1}}{z - \bar{z}}
  \frac{\omega^{r+1}-\omega^{-(r+1)}}{\omega - \omega^{-1}}\,.
  \label{eqn:defectchannelconfbloc}
\end{align}

As for the bulk channels, defect operators also organise into supermultiplets,
this time of the $\osp(4^*|2) \oplus \osp(4^*|2)$ superalgebra. The corresponding
unitary supermultiplets were obtained
in~\cite{Gunaydin:1990ag,Agmon:2020pde,Drukker:2020atp}. The supermultiplets
that can appear in the dOE are those which contain defect operators with at most
$r \le 2$, in order to respect the decomposition~\eqref{eqn:Fsigmadecomposition}
of $\cF$. The list of superconformal blocks that correspond to these selection
rules and can contribute to the correlator are given in table~\ref{tab:defectsblocks}.

\begin{table}[h]
  \centering
  \renewcommand{\arraystretch}{1.5}
  \begin{tabular}{c|c|c|c|c|c}
    Multiplet & $\hat\Delta$ & $r$ & $\hat\cG$ & $\zeta(z)$ & $F(z,\bar{z})$
    \\ \hline
    $\mathds{1}$ & 0 & 0 & 1 & $1$ & $1$\\
    $B[1]$ & $2$ & $1$ & \eqref{eqn:sblockBr} & $-z$
    & $-z \bar{z} \left[ 1 + \frac{(1-z)(1-\bar{z})}{1 - z \bar{z}} \right]$ \\
    $B[2]$ & $4$ & $2$ & \eqref{eqn:sblockBr} & $z^2$
    & $(z \bar{z})^{2} \left[ 1 + \frac{2 (1-z)(1-\bar{z})}{1 - z \bar{z}} \right]$ \\
    $A[0]_s$  & $2+s$ & 0 & \eqref{eqn:sblockA0} & $-z^{s+2}$
    & $\frac{ \bar{z} z^{s+2} (1-z)^2 - z \bar{z}^{s+2} (1-\bar{z})^2}{(z-\bar{z})(1-z \bar{z})}$ \\
    $A[1]_s$  & $4+s$ & $1$ & \eqref{eqn:sblockA1} & $z^{s+3}$
    & $-\frac{\bar{z}^2 (\bar{z}-2) z^{s+3} (1-z)^2 - z^2 (z-2) \bar{z}^{s+3} (1-\bar{z})^2}{(z-\bar{z})(1-z \bar{z})}$ \\
    $L[0]_s$  & $\hat{\Delta} > 2 + s$ & 0 & \eqref{eqn:sblockL} & 0
    & $ (z \bar{z})^{\frac{\hat{\Delta}+s}{2}} \frac{(z-1)^2 (\bar{z}-1)^2}{1 - z \bar{z}} \frac{z^{s+1} -
  \bar{z}^{s+1}}{z - \bar{z}}$\\
  \end{tabular}
  \renewcommand{\arraystretch}{1}
  \caption{List of defect supermultiplets contributing to the correlator.
    $\hat{\Delta},r$ denote the conformal dimension and R-symmetry charge of the
    superprimary of the supermultiplet. For each superconformal block, its
    content in terms of conformal blocks is written in
    Appendix~\ref{sec:defectblocktable}, and we write its contribution to $\zeta, F$.
  }
    \label{tab:defectsblocks}
\end{table}

Note that at the unitarity bound $\hat{\Delta} \to 2+s$, the long multiplets decompose according to the
recombination rules~\cite{Agmon:2020pde}
\begin{equation}
\begin{aligned}
  L[0]_{s=0}|_{\Delta \to 2} &=
  A[0]_0 + B[2]\,,\\
  L[0]_{s \ge 1}|_{\Delta \to 2+s} &= 
  A[0]_s + A[1]_{s-1}\,.
\end{aligned}
\end{equation}
Correspondingly, we find that the superconformal blocks obey the identities
\begin{equation}
\begin{aligned}
  \hat{\cG}_{L[0]_0}|_{\Delta = 2}
  &=
  \hat{\cG}_{A[0]_0} + \hat{\cG}_{B[2]}\,,\\
  \hat{\cG}_{L[0]_s}|_{\Delta = 2+s}
  &=
  \hat{\cG}_{A[0]_s} +
  \hat{\cG}_{A[1]_{s-1}}\,.
  \label{eqn:recombination}
\end{aligned}
\end{equation}
These recombination rules suggest that the multiplet $B[2]$ is the analytic
continuation of $A[1]_s$ to $s=-1$. Indeed, we find that the superconformal
blocks satisfy $\hat{\cG}_{B[2]} = \hat{\cG}_{A[1]_{-1}}$, and furthermore
$\hat{\cG}_{B[1]} = -\hat{\cG}_{A[1]_{-2}}$, $\hat{\cG}_{\mathds{1}} =
-\hat{\cG}_{A[0]_{-2}}$.

\section{Chiral algebra}
\label{sec:chiral}

We now show how to obtain exact dCFT data and calculate $\zeta(z)$ directly from
the chiral algebra.

\subsection{Reminder of the chiral algebra map for local operators}
\label{Walgreminder}

As shown in~\cite{Beem:2014kka}, to any 6d $\cN = (2,0)$ superconformal field theory
one can associate a chiral algebra (or VOA) by passing to the cohomology of a
certain nilpotent supercharge.  Under this cohomological reduction $\chi$,
representations of $\mathfrak{osp}(8^*|4)$ are mapped into representations of
$\mathfrak{sl}(2)$. For the representations appearing in this work, this map
gives (see table 1 in~\cite{Beem:2014kka} for the complete list)
\begin{equation}
\label{cohomologicalreductionMultiplets}
\chi \quad : \quad
\mathcal{D}[k,0]\,\mapsto\, v_k\,,
\qquad
\mathcal{B}[k_1,k_2]_\ell\,\mapsto\, v_{k_1+k_2+4+\ell}\,,
\qquad
\mathcal{L}[k_1,k_2]_{\Delta,\ell}\,\mapsto\, 0\,,
\end{equation}
where $v_h$ denotes the highest weight representation of $\mathfrak{sl}(2)$ with
highest weight $h$. This fact has a manifestation at the level of the conformal
blocks in the bulk channel, see the fifth column of Table \ref{tab:1pts} where
one recognizes $k_{2h}(1-z)$ as the $\mathfrak{sl}(2)$ conformal blocks for the
exchange of an operator of weight $h$.  It is clear that the map
\eqref{cohomologicalreductionMultiplets} cannot be inverted as different
$\mathfrak{osp}(8^*|4)$ representations map to the same representation
$v_h$.

There is more structure to the map $\chi$, and also the OPEs of the 6d theory
reduce to OPEs for operators associated with the representations $v_h$, giving
rise to a VOA. In the case of the $(2,0)$ theories of type $A_{N-1}$,
the associated VOA is (conjecturally) the well-known $\mathcal{W}_N$ algebra, where the
central charge takes the value $c$ given in~\eqref{eqn:anomalycoeffs}
(see~\cite{watts1997} for a pedagogical introduction to $\mathcal{W}$-algebras
and~\cite{Bouwknegt:1992wg} for a review).%
\footnote{
  This $\mathcal{W}_N$ algebra can be constructed from the corresponding current
  algebra using the Drinfel'd-Sokolov construction. $\mathcal{W}$-algebras
  also exist for the $D_N$ and $E_{6,7,8}$ current algebras, and the proposal
  of~\cite{Beem:2014kka} extends to these theories as well.
}

The $\mathcal{W}_N$ algebra is generated by primary operators $W_p(z)$ ($p=2,3,\dots, N$) of
conformal weight $h=p$, where $W_2(z)=T(z)$ is the stress tensor. Under the map
$\chi$, the half-BPS operators that generate the chiral ring map to the
generators of the $\mathcal{W}$-algebra as
\begin{equation}
  \chi : \O2_k(x,y)\,\mapsto W_k(z)\,,
\end{equation}
up to a factor coming from the relative normalisation of $\O2_k$ and $W_k$.

At large $N$, the operators $\O2_k$ have a simple interpretation in holography
as the ``single-trace'' Kaluza-Klein modes on $S^4$, and the spectrum of local
operators is freely generated by these modes along with their derivatives.
Correspondingly, at large $N$ the Hilbert space $\mathcal{H}$ of the VOA is
freely generated by $W_k(z)$ and their derivatives, as can be seen from the
(unrefined) superconformal
index~\cite{Kim:2013nva,Beem:2014kka,Bullimore:2014upa}
\begin{equation}
  \mathcal{I}_{A_{N-1}}(q) =
  \text{Tr}_\mathcal{H}\, q^{L_0} =
  q^{-\frac{c}{24}}
  \text{PE} \left(
  \frac{q^2 + q^3 + \dots + q^N}{1-q}
  \right)\,,
\label{indexnodefect}
\end{equation}
with $L_0$ the generator of the Virasoro algebra.

The structure constants appearing in the 3-point functions of
$\O2_k$~\eqref{eqn:3pts} are captured by the VOA and can be calculated exactly
at any $N$.  Restricting the points $x_i$ to lie on a plane (arbitrary but
fixed, referred to as the \emph{chiral algebra plane}) with coordinate
$(z_i,\bar{z}_i)$ and taking the R-symmetry variables $y_i$ to lie on a line
with coordinate $\omega_i$, one finds that upon setting $\omega_i = \bar{z}_i$
the correlator~\eqref{eqn:3pts} reduces to
\begin{align}
  \vev{ \O2_{k_1}(z_1, \bar{z}_1; \bar{z}_1)
  \O2_{k_2}(z_2, \bar{z}_2; \bar{z}_2)
  \O2_{k_3}(z_3, \bar{z}_3; \bar{z}_3) }
  = 
  \frac{\lambda_{k_1 k_2 k_3}}
  {(z_{12})^{k_{123}}(z_{13})^{k_{132}}(z_{23})^{k_{231}}}\,,
\end{align}
with the short-hand notations $z_{ij} = z_i - z_j$, $k_{ijk} = k_i + k_j -
k_k$, and $\O2_{p}(z,\bar{z};\bar{z})=
\O2_{p}(x^\parallel=0,x^\perp=\text{diag}(z,\bar{z}),
y^\parallel=0,y^\perp=\bar{z})$.
The twisted operators $\O2_k(z,\bar{z},\bar{z})$ sit in the cohomology
used to define the VOA, and using the map $\chi$ the structure constants
$\lambda_{k_1 k_2 k_3}$ can be identified with the corresponding structure
constants of the VOA. As pointed out in~\cite{Beem:2014kka}, using results from
\cite{Campoleoni:2011hg} one can reproduce the supergravity results of
\cite{Bastianelli:1999en} (see also~\cite{Corrado:1999pi}) at leading order at
large $N$, which read
\begin{align}
  \lambda_{k_1 k_2 k_3} =
  \frac{2^{k_1+k_2+k_3-2}}{(\pi N)^{3/2}} 
  \Gamma\left( \frac{k_1+k_2+k_3}{2} \right)
  \frac{
    \Gamma\left( \frac{k_{123}+1}{2} \right)
    \Gamma\left( \frac{k_{231}+1}{2} \right)
    \Gamma\left( \frac{k_{312}+1}{2} \right)}{
      \sqrt{\Gamma( 2k_1-1) \Gamma(2k_2-1) \Gamma(2k_3-1)}
    } + \dots\,.
    \label{eqn:structureconstantslargeN}
\end{align}

\paragraph{An example:}
As an example let us recall how to compute the 4-point function of the stress
tensor $T$ from the singular part of the OPE:
\begin{equation}
\label{TTOPE}
T(z)T(w) =
\frac{c/2}{(z-w)^4}+
\frac{2T(w)}{(z-w)^2}+
\frac{\partial T(w)}{(z-w)}
+ \dots
\end{equation}
We consider the 4-point function
\begin{equation}
\langle T(z_1)T(z_2)T(z_3) T(z_4)\rangle\,,
\end{equation}
and think of it as a function of $z_2$ with the other point fixed. This is a
meromorphic function, whose poles arise when $z_2$ approaches any of the other
three points and its residues are related to lower point correlators of $T$
by~\eqref{TTOPE}. This information is enough to completely determine the
function, and following this strategy one finds that
\begin{equation}
\label{Tfourpoint}
\frac{\langle T(0)T(z)T(1) T(\infty)\rangle}{\langle T(0)T(z)\rangle \langle T(z)T(1)\rangle}\,=\,
1+z^4+\left(\frac{z}{z-1}\right)^4
+\frac{8}{c}
\left(\frac{z}{z-1}\right)^2(z^2-z+1)\,.
\end{equation}
This function can be expanded in $\mathfrak{sl}(2)$ blocks as
\begin{equation}
\label{TTTTandOPE}
\frac{\langle T(0)T(z)T(1) T(\infty)\rangle}{\langle T(0)T(z)\rangle \langle T(z)T(1)\rangle}\,=\,
1+ \frac{8}{c}\, \,k_{2}(z)+
\sum_{n=0}^{\infty}\, \lambda^2_n\,\,
k_{2h_n}(z)\,,
\end{equation}
where $k_{2h}(z)=z^h\,{}_2F_1(h,h,2h;z)$, $h_n=2n+4$
and the structure constants are
\begin{equation}
\label{lamndasquaredinTTTT}
  \lambda^2_n\,=
  \frac{(2n+6)!(2n+3)!(2n+1)(n+1)}{18(4n+5)!}
  + \frac{8}{c} \frac{(2n+3)!(2n+2)!(4n^2+14n+11)}{(4n+5)!}\,.
\end{equation}
It is instructive to identify the operators that are exchanged in \eqref{TTTTandOPE}
and write
\begin{equation}
\label{TTfullOPE}
T(z_1)T(z_2)=\frac{c/2}{z_{12}^4}
+ \frac{2}{z_{12}^{2}}\mathcal{D}_{2,2;2}(z_{12},\partial_2)T(z_2)
+\sum_{n=0}^{\infty} 
\frac{\lambda_{TT}{}^{[TT]_n}}{z_{12}^{4-h_n}}\mathcal{D}_{2,2;h_n}(z_{12},\partial_2)[TT]_n(z_2)\,,
\end{equation}
where $\mathcal{D}_{2,2;h_n}$ are differential operators encoding the
contributions from descendants; their explicit expression can be found in
e.g.~Appendix A of \cite{Bonetti:2018fqz} to which we also refer for further
explanations.  The expansion \eqref{TTfullOPE} is a completion of \eqref{TTOPE}
to include all the regular terms where 
\begin{equation}
\label{TTcompositedef}
  [TT]_n(z):=\sum_{\ell=0}^{2n}
  (-1)^{\ell}
  {2n\choose\ell}
  \frac{(2n+3)!}{(\ell+3)!(2n-\ell+3)!} \text{NO}[\partial^{\ell}T,\partial^{2n-\ell} T]-\frac{1}{4(n+1)(2n+5)}\partial^{2n+2} T\,,
\end{equation}
where $\text{NO}$ denotes normal ordering.
With these conventions we can compute the norm and OPE coefficients to find
\begin{equation}
\label{TTnormandlambda}
\langle [TT]_n(z_1)\,[TT]_m(z_2)\rangle=\frac{\delta_{n,m}\,g_n}{(z_1-z_2)^{2(2n+4)}}\,,
\quad
\lambda_{TT}{}^{[TT]_n}=\frac{(2n+1)(n+1) (2n+6)!}{(4n+5)!}\,,
\end{equation}
and the coefficients $\lambda^2_n$ in \eqref{lamndasquaredinTTTT} are recovered as 
\begin{equation}
\lambda^2_n\,=\frac{1}{(c/2)^2}\,\left(\lambda_{TT}{}^{[TT]_n}\right)^2\,g_n\,.
\end{equation}

\subsection{Adding a surface defect}

The VOA also captures surface operators $V$ orthogonal to
the chiral algebra plane, since the $\osp(4^*|2) \oplus \osp(4^*|2)$ subalgebra
preserved by $V$ contains the supercharge used to define the cohomology.
This is very similar to the case of surface defects in four dimensional
$\mathcal{N}=2$ SCFTs considered in \cite{Cordova:2017mhb},
\cite{Bianchi:2019sxz} (see also \cite{Argyres:2022npi}):
The defect $V$ intersects the chiral algebra plane at $z=0$ and
$z = \infty$, and under the map $\chi$ corresponds to inserting two vertex
operators $\Vop_\Lambda(0)$ and $\bar{\Vop}_\Lambda(\infty)$ at these points, which
defines a module of the associated W-algebra.

The fate of defect operators $V[\hat{\cO}]$ under the chiral algebra map
presents some important differences compared to the case of the bulk operators,
which are related to the fact that these operators are bound to $V$ and
therefore cannot be translated away from the origin (or infinity).  While
primary bulk operators are mapped to $\mathfrak{sl}(2)$ primaries, defect
primary operators can be mapped to $\mathfrak{sl}(2)$ descendants. The simplest
but very important example is given by the displacement operator which, as we
will see below, is mapped to $\partial \Vop_{\Lambda}(0)$ (or equivalently
$L_{-1} \ket{\Vop_{\Lambda}} $).

The cohomological reduction carries over to representations of
$\mathfrak{osp}(4^*|2)\oplus \mathfrak{osp}(4^*|2)$. In this work we consider
only representations with same left/right quantum numbers (they satisfy
\eqref{hishbar} and \eqref{sissbar}), and for these the map $\chi$ gives
\begin{equation}
\chi \quad :\quad
B[r] \mapsto s_r\,,
\qquad
A[r]_s \mapsto s_{r+s+2}\,,
\qquad
L[r]_s \mapsto 0\,,
\label{eqn:chiralmapdefect}
\end{equation}
where $s_d$ denotes a one dimensional representation of $\mathbb{C}^*$ with
weight $d$. The weights $d$ can be read from table \ref{tab:defectsblocks}.

\paragraph{ $\mathcal{W}$ modules.}
Modules for the W-algebras introduced in section \ref{Walgreminder} have been
extensively studied in the literature. The ones relevant to surface operators
$V$ are a special class of the so-called completely degenerate
representations, see e.g.~\cite{Drukker:2010jp} (also
\cite{Fateev:1987vh,Chalabi:2020iie,Afkhami-Jeddi:2017idc,Linshaw:2017tvv}),
and are labelled by a single highest weight $\Lambda$.%
\footnote{The value of the central charge for the W-algebra in this work
  corresponds to setting $b=1$ in \cite{Drukker:2010jp}.}
The module is constructed by acting with the negative modes of the W-algebra
generators on the highest weight vector $\ket{\Vop_{\Lambda}}$, which satisfies
\begin{equation}
L_n \ket{\Vop_{\Lambda}}=W^{(p)}_n\ket{\Vop_{\Lambda}}=0\,,
\qquad  n>0\,,\,\,\,\,\,\,\,\,\, p=3,4,\dots,N\,,
\end{equation}
and is labelled by its weights under the W-algebra generators
\begin{equation}
L_0 \ket{\Vop_{\Lambda}}=
\Delta(\Lambda)
\ket{\Vop_{\Lambda}}
\quad 
W^{(p)}_0\ket{\Vop_{\Lambda}}=\omega_p(\Lambda)\,\ket{\Vop_{\Lambda}}\,,
\qquad  p=3,4,\dots,N\,.
\end{equation}
The eigenvalue of $L_0$ in particular is the conformal weight and is given by
\begin{equation}
\label{DeltaofLambda}
  \Delta(\Lambda)=-\frac{1}{2} (\Lambda,\Lambda) - 2\,(\rho,\Lambda) =
  -d(\Lambda)\,,
\end{equation}
with $d$ the anomaly coefficient introduced in~\eqref{eqn:anomalyexact}. The
appearance of the coefficient $d$ supports the identification of
$\ket{\Vop_\Lambda}$ as the module associated to the surface operator $V$ with
representation $\Lambda$, and remarkably indicates that the entire dependence on
$d$ is captured by the chiral algebra.

The expressions for $\omega_p(\Lambda)$ are obtained from a free field
realization of the $\mathcal{W}_N$ algebra in~\cite{Fateev:1987zh}.%
\footnote{This free field realisation appears naturally
since the $\mathcal{W}_N$ algebra is the chiral sector of Toda field
theory which in turn can be obtained by compactifying the (2,0) theories on a 4-sphere, see~\cite{Alday:2009qq,Cordova:2016cmu}.}
For example, with their choice of normalisation for $W_3$, $w_3(\Lambda)$ reads 
\begin{equation}
\begin{aligned}
  w_3(\Lambda) &= \sum_{1 \le i_1 < i_2 < i_3 \le N}
  (\Lambda+2\rho,\epsilon_{i_1}) (\Lambda+2\rho,\epsilon_{i_2})
  (\Lambda+2\rho,\epsilon_{i_3})\\
  & \qquad + 2(N-1) \sum_{1 \le i_1 < i_2 \le N} (\Lambda+2\rho,\epsilon_{i_1}) (\Lambda+2\rho,\epsilon_{i_2})
  + 8\binom{n+2}{4}\,,
\end{aligned}
\end{equation}
where $\epsilon_i$ are the weights of the fundamental representation of
$\su(N)$.
Taking for example the totally symmetric representation $\Lambda = [M,0,\dots]$,
the expression above reduces to 
\begin{align}
  w_3(M) = \frac{(M-N)(N-1)}{3N^2} \left( M^2 (N-2) + M N (N-5) - 3 N^2 (N+1)
  \right)\,.
\end{align}

Let us look more closely at the structure of these modules. 
Their character is captured by the surface index computed
in~\cite{Bullimore:2014upa}\footnote{
The character of fully degenerate module and the surface index can be computed independently. The fact that they are equal provides strong evidence for the conjectured identification of the corresponding modules, see \cite{Bullimore:2014upa}.}  (up to a factor $q^{-\frac{c}{24}}$).  It can be
written as a plethystic exponential
\begin{align}
  \mathcal{I}_\Lambda =
  q^{\Delta(\Lambda) - \frac{c}{24}}
  \text{PE} \left(
  \frac{(N-1)q}{1-q} - \sum_{\alpha>0} q^{(\rho + \Lambda,\alpha)}
  \right)\,,
  \qquad
  \text{PE} \left(f(q)\right) \equiv \text{exp}
  \left(\sum_{k=1}^{\infty}\,\frac{f(q^k)}{k}\right)\,,
\end{align}
where the sum runs over the positive roots.
Setting $\Lambda = 0$ we recover the superconformal index without defects given in \eqref{indexnodefect}. 
For the totally symmetric representations $\Lambda = [M, 0, \dots]$, the index above reduces to
\begin{align}
  \mathcal{I}_\Lambda =
  q^{\Delta(\Lambda) - \frac{c}{24}}
  \text{PE} \left(
  \frac{q^{N+M} - q^{M+1} + \sum_{k = 1}^{N-1} q^k}{1-q}
  \right)\,.
\end{align}
The structure of these modules at large $N$ (for any $M$) is encoded in this formula
\begin{equation}
 \text{PE} \left(
  \frac{q^{N+M} - q^{M+1} + \sum_{k = 1}^{N-1} q^k}{1-q}
  \right)=
   \text{PE} \left(
  \frac{\sum_{k = 1}^{N} q^k}{1-q}
  \right)\, \prod_{\ell=1}^{M}\frac{1}{1-q^\ell} +
  O(q^{N})\,.
\end{equation}
By looking at the index for more general $\Lambda$, we notice a puzzling
feature: if we count how many states in the module have the same $L_0$ quantum
number as the state associated to the displacement operator, namely
$\Delta(\Lambda)+1$, we find that this number coincides with the number of
non-zero entries in the weight $\Lambda$.  This feature is not new and something
similar happens for surface defects in four dimensions,
see~\cite{Bianchi:2019sxz}. We postpone the analysis of these states from the
point of view of the surface defect to the future.

The module structure can be translated in the language of OPEs
\begin{align}
\label{TVOPE}
T(z) \Vop_{\Lambda}(0)&\sim \frac{\Delta(\Lambda)}{z^2} \,\Vop_{\Lambda}(0)+\frac{2}{z} \,\,\partial \Vop_{\Lambda}(0)\,,
\\
\label{WVOPE}
W_k(z) \Vop_{\Lambda}(0)&\sim \frac{\omega_k(\Lambda)}{z^k} \,\Vop_{\Lambda}(0)+ \frac{1}{z^{k-1}} \,
\left(\frac{k\, \omega_k(\Lambda)}{2\,\Delta(\Lambda)} \partial\Vop_{\Lambda}(0)+\Vop^{(k)}_{\Lambda}(0)\right)+\dots 
\end{align}
In this way we can make direct contact with the (cohomological reduction of the)
defect OPE of the $\Phi_k$ half-BPS operators.
We will now use these OPEs to determine certain quantities exactly in the 6d
theory with the surface defect.

A correlator of the 1/2-BPS bulk operators $\Phi_k$ (with appropriate twist
$\omega = \bar{z}$) in the presence of the surface defect $V$ is given by
\begin{equation}
\label{correbulkOPwithdefectinWalg}
  \vev{\O2_{p_1}(z_1,\bar{z}_1;\bar{z}_1)
  \dots \O2_{p_n}(z_n,\bar{z}_n;\bar{z}_n)\,V_{\Lambda}}
  \propto
    \frac{\vev{\bar{\Vop}_{\Lambda}(\infty) W_{p_1}(z_1)
  \dots W_{p_n}(z_n)\Vop_{\Lambda}(0)}}
    {\vev{\bar{\Vop}_{\Lambda}(\infty) \Vop_{\Lambda}(0)}}\,,
\end{equation}
where we have used ``$\propto$'' instead of ``$=$'' due to different
normalization conventions in the two sides (recall that on the left
$\vev{V_\Lambda} = 1$ so the normalisation is omitted). On the left we write $V_\Lambda$ to
emphasize the dependence on the representation $\Lambda$. On the right
$\bar{\Vop}_{\Lambda}$ denotes the conjugate of $\Vop_{\Lambda}$. 

Several generalizations of the correlator \eqref{correbulkOPwithdefectinWalg}
can be computed in the chiral algebra. For example, we can insert (twisted
translated) defect operators at zero or infinity, which corresponds to replacing
$\Vop_{\Lambda}(0)$ and $\bar{\Vop}_{\Lambda}(\infty)$ in the right hand side
with the appropriate descendants.

The simplest quantity of the type \eqref{correbulkOPwithdefectinWalg} is the one
point function of the half-BPS operators $\O2_k$ starting with $k=2$ which is
the stress tensor. In this case, the OPEs \eqref{TVOPE} immediately allow to
compute
  \begin{align}
    \frac{\vev{\bar{\Vop}_{\Lambda}(\infty) T(z) \Vop_{\Lambda}(0)}}
    {\vev{\bar{\Vop}_{\Lambda}(\infty) \Vop_{\Lambda}(0)}}
    =
    \frac{\Delta(\Lambda)}{z^2}\,.
  \end{align}
Dividing by $\sqrt{c/2}$ to account for the normalisation of the stress
tensor and recalling the identity $ \Delta(\Lambda)= -d(\Lambda)$ pointed
out above, we recover precisely the coefficient $a_2$~\eqref{eqn:a2}!

The next quantity that can be determined using the connection to the W-algebra
is the 2-point function of the stress tensor superprimary with the
displacement operator given in~\eqref{eqn:disp}. Setting $x_{1,2}^\parallel =
y_{1,2}^\parallel = 0$, $x^\perp = (z,\bar{z})$, $y^\perp = \bar{z}$, we
get a holomorphic correlator
\begin{align}
  \vev{\O2(z,\bar{z};\bar{z}) V[\hat{\cO}_{B[1]}(0,0)]}
  =
  \frac{b_{2,B[1]}}{z^3}\,.
\end{align}
In the chiral algebra, the insertion of the  displacement operator at the origin
is equivalent to replace $\Vop_{\Lambda}(0)$ in
\eqref{correbulkOPwithdefectinWalg} with its descendant
$\partial\Vop_{\Lambda}(0)$.  Using the OPEs \eqref{TVOPE} one easily computes 
  \begin{align}
    \frac{\vev{\bar{\Vop}_{\Lambda}(\infty) T(z) \partial \Vop_{\Lambda}(0)}}{\vev{\bar{\Vop}_{\Lambda}(\infty) \Vop_{\Lambda}(0)}}
    =
    \frac{2\Delta(\Lambda)}{z^3}\,.
  \end{align}
To reproduce the coefficient $b_{2,B[1]}$ given in~\eqref{eqn:disp} we have to
divide this expression by the square root of the norm of $T$, namely
$\sqrt{c/2}$, and the square root of the norm of $\partial \Vop_\Lambda(0)$,
which is $\sqrt{-2\Delta(\Lambda)}$.%
\footnote{
  The norm is computed by recalling that $\partial \Vop_{\Lambda}(0)\sim
  L_{-1}|\Vop_{\Lambda}\rangle$. By inspecting the defect OPE in 6d we obtain
  that the conjuguate is $- \bra{\Vop_\Lambda} L_{+1}$. We obtain the norm by
  using the commutation relation $[L_{+1},L_{-1}]=2L_0$ and the properties
  $|\Vop_{\Lambda}\rangle$: $L_{+1}|\Vop_{\Lambda}\rangle=0$ and
  $L_{0}|\Vop_{\Lambda}\rangle=\Delta\,|\Vop_{\Lambda}\rangle$.
}
The result is $2\sqrt{-\Delta(\Lambda)/c} $ which matches \eqref{eqn:disp}.

The next quantity we consider is the 2-point function of $ \O2$ in the
presence of the defect. In this case, the protected part $\zeta(z)$ of the
correlator $\langle \O2 \O2 V\rangle$ is captured by the 4-point function
\begin{equation}
\label{ChiralALgebraVTTV}
\frac{\langle \bar{\Vop}_{\Lambda}(0)T(z)T(1) \Vop_{\Lambda}(\infty)\rangle}{\langle \bar{\Vop}_{-\Lambda}(0)\Vop_{\Lambda}(\infty)\rangle \langle T(z)T(1)\rangle}\,=\,
1+ \frac{\Delta(\Lambda)^2}{c}2 Z^2
+ \frac{\Delta(\Lambda)}{c}4 Z\,,
\qquad
Z:=\frac{(z-1)^2}{z}\,.
\end{equation}
More precisely the quantity \eqref{ChiralALgebraVTTV} is equal to $Z^2\zeta(z)$.
The expression on the right is obtained using similar methods as the one used to
compute \eqref{Tfourpoint}, namely by reconstructing the function from its poles
in $z$ and the OPE of $T(z)$.

We can identify the protected CFT data by expanding~\eqref{ChiralALgebraVTTV} in
conformal blocks. In the bulk channel $z \sim 1$, the function can be expanded
in $\sl(2)$ blocks $k_{2h}(z)$
\begin{equation}
\label{ZetainbulkOPEs}
    Z^2\zeta(z)=
    1+\frac{4 \Delta(\Lambda)}{c}\,k_4(1-z)
    +\sum_{n=0}^{\infty}\,\alpha_n\,k_{2h_n}(1-z)\,,
\end{equation}
where $h_n=4+2n$ and
\begin{equation}
    \alpha_n=\frac{(2n+2)!(2n+3)!}{(2n+3)!(4n+5)}\left(
    (2n+5)\,\frac{\Delta(\Lambda)^2}{c}
    +\frac{1}{n+1}\frac{\Delta(\Lambda)}{c}
    \right)\,.
\end{equation}
The operators exchanged in the $TT$ OPEs (singular and regular) entering \eqref{ZetainbulkOPEs} are the same as the one contributing to \eqref{TTTTandOPE} and \eqref{TTfullOPE}. 
Using the explicit expressions for the bilinears in the stress tensor in  \eqref{TTcompositedef} and the OPEs \eqref{TVOPE} one finds
\begin{equation}
[TT]_n(z)\Vop_{\Lambda}(0)\sim\,\frac{u_n}{z^{h_n}}\,\Vop_{\Lambda}(0)+\dots\,,
\qquad
u_n=
\frac{(2n)!\left((n+1)(2n+5)\Delta(\Lambda)^2+\Delta(\Lambda) \right)}{(n+2)(n+3)(2n+5)}\,.
\end{equation}
Inserting the explicit expression of $\lambda_{TT}{}^{[TT]_n}$ given in \eqref{TTnormandlambda} we can check that 
\begin{equation}
\alpha_n=\frac{\lambda_{TT}{}^{[TT]_n}\, u_n}{c/2}\,,
\end{equation}
as it should.
By comparing to the content of table \ref{tab:1pts} we can read off the ``$a\lambda$'' OPE coefficients to be
\begin{equation}
(a\lambda)_{\mathcal{D}[2,0]}=\frac{4 \Delta(\Lambda)}{c}\,,
\qquad
(a\lambda)_{\mathcal{D}[4,0]}=\alpha_0\,,
\qquad
(a\lambda)_{\mathcal{B}[2,0]_{\ell=2n-2}}=-\alpha_n\,.
\label{eqn:chiralonepts}
\end{equation}
Notice that for any central charge and for any $\Delta(\Lambda)$ the operator transforming in the $\mathcal{D}[4,0]$ representation being exchanged here is a composite operator which is orthogonal to $\Phi_4$. 

To obtain OPE data associated to the defect operator expansion (dOE) we need to
expand the function $\zeta(z)$ in defect channel blocks (which are simply the
monomials $z^k$) for $z$ close to zero
\begin{equation}
\zeta(z)=2\frac{\Delta(\Lambda)^2}{c} +\sum_{n=1}^{\infty}\,\beta_n
\,z^n\,,
\qquad
\beta_n=n\left(\frac{1}{6}(n^2-1)+4\frac{\Delta(\Lambda)}{c} \right)\,.
\end{equation}
We can identify the coefficients $b^2$ in 6d from
table~\ref{tab:defectsblocks} and obtain
\begin{equation}
b^2_{\mathbb{I}}=2\frac{\Delta(\Lambda)^2}{c}\,,
\qquad
b^2_{\hat{B}[1]}=-\beta_{1}\,,
\qquad
b^2_{\hat{B}[2]}-b^2_{\hat{A}[0]_0}=\beta_{2}\,,
\qquad
b^2_{\hat{A}[1]_s}-b^2_{\hat{A}[0]_{s+1}}=\beta_{s+3}\,,
\label{eqn:chiralbcoeff}
\end{equation}
where $s=0,1,\dots$.
Notice that there is an ambiguity here since the
chiral algebra map~\eqref{eqn:chiralmapdefect} is not invertible and the blocks
$z^n$ may originate from multiple defect blocks in 6d. However in
section~\ref{sec:inversion}, we show that the
supermultiplets of type $A[0]_s$ do not appear in this OPE.
With this additional input we can identify the coefficients $b^2$ in 6d uniquely from the W-algebra.
It would be interesting to understand if there is a deeper reason for the absence of these $A[0]_s$ multiplets.
We leave the detailed study of the W-algebra, its modules and other exact computations, such as the protected part of correlators $\vev{\O2_{k_1}
\O2_{k_2} V_\Lambda}$, for future work.

\section{Obtaining the correlator}
\label{sec:inversion}

The superconformal block expansion~\eqref{eqn:defectblockdecomp} constructs
the function $\cF$ in terms of the coefficients $b_{kl}^2$ and the spectrum of
supermultiplets contributing to the correlator. The inverse relation is known
as the (lorentzian) inversion formula~\cite{Caron-Huot:2017vep,Lemos:2017vnx}
and expresses the dCFT data in terms of the discontinuity of $\cF$ at its branch
cut $\bar{z} = 1$, defined as%
\footnote{More precisely the correlator is a distribution and the discontinuity
should be understood in that sense. In particular rational functions can have
discontinuities.}
\begin{align}
  \disc \cF(z,\bar{z},\omega) =
  \cF(z,\bar{z},\omega) - \cF(z,\bar{z},\omega)|_{(1-\bar{z}) \to e^{2\pi i}
  (1-\bar{z})}\,.
\end{align}

The inversion formula of~\cite{Lemos:2017vnx} applies to defects of general
dimension (and codimension) and doesn't take into account R-symmetry or
supersymmetry, which are specific to each setup. To address the first we can apply
the formula to each R-symmetry channel $\cF|_r(z,\bar{z})$, which is the
restriction of $\cF$ to the contributions of defect conformal
blocks~\eqref{eqn:defectchannelconfbloc} of a fixed representation $r$ (for our
correlator, $r$ can take values $0,1,2$).

For every $r$, the inversion formula defines a function $B_r(\hat{\Delta},s)$
constructed in such a way that it has poles at isolated values of $\hat{\Delta}
= \hat{\Delta}_{m,s,r}$ whenever a block with these quantum numbers contributes
to $\cF|_r$. At each pole the residue gives the OPE coefficient
\begin{align}
  b_{\hat{\Delta}_{m,s,r},s,r}^2 = -\res_{\hat{\Delta} = \hat{\Delta}_{m,s,r}}
  B_r(\hat{\Delta}, s)\,.
\end{align}

This gives the conformal block decomposition of $\cF$. There is a simple
observation made in~\cite{Alday:2017vkk,Lemos:2021azv} to reorganise these
conformal blocks into superconformal blocks, thus ensuring that supersymmetry is
preserved.  From the explicit expression for defect channel superblocks (see
Appendix~\ref{sec:defectblocktable}) we observe that R-symmetry blocks with $r=2$ and given
labels $\hat{\Delta}, s$ appear only once in any given superconformal block, except for the superblocks associated to the identity, $B[1]$ and $A[0]_s$ operators, for which the $r=2$ component is absent. This means
that, from the conformal block decomposition of $\cF|_{r=2}$ we can read the
superconformal block decomposition of $\cF$, up to the OPE coefficients of identity, $B[1]$ and $A[0]_s$ operators (more on that below). Explicitly, we define the function
$\mathsf{B}(\hat{\Delta},s)$ from the inversion formula~\cite{Lemos:2017vnx}
applied to $\cF|_{r=2}$ as
\begin{align}
  \mathsf{B}(\hat{\Delta},s)=
  B_{r=2}(\hat{\Delta}+2,s)\,=
  \int\limits_0^1 \frac{\diff z}{2} z^{-\frac{\hat{\Delta}-s}{2}-2}
  \int\limits_1^{1/z} \frac{\diff \bar{z}}{2\pi i}
  & (1-z \bar{z})(\bar{z}-z)
  \bar{z}^{-\frac{\hat{\Delta}+s}{2}-3}
  \disc \cF|_{r=2}(z,\bar{z})\,.
  \label{eqn:susyinversionB}
\end{align}
The shift in $\hat{\Delta}$ is included so that for long
supermultiplets~\eqref{eqn:sblockL}, $\mathsf{B}(\hat{\Delta},s)$  has poles at the
location of the superprimary. For short and semishort multiplets the shift is
different and can be treated separately, for instance the $A_s[1]$ multiplets
correspond to poles of  $\mathsf{B}(\hat{\Delta},s+1)$ at $\hat{\Delta}=s+3$, see~\eqref{eqn:sblockA1}.
$\cF|_{r=2}$ can be
expressed directly in terms of $\zeta, F$ by extracting the part
of~\eqref{eqn:solWIk2} proportional to the R-symmetry block with $r=2$ (the
Chebyshev polynomial of second kind $U_2(\frac{\omega + \omega^{-1}}{2})$),
which gives
\begin{align}
  \cF|_{r=2} =
  \frac{z \bar{z}}{(1-z)^2(1-\bar{z})^2} F(z,\bar{z}) + \frac{z \bar{z}}{(z -
    \bar{z})(1-z \bar{z})} \left( \frac{\bar{z}}{(1-\bar{z})^2} \zeta(z) -
    \frac{z}{(1-z)^2} \bar{\zeta}(\bar{z}) \right)\,.
\end{align}

Following~\cite{Caron-Huot:2017vep,Lemos:2017vnx}, we can evaluate the
discontinuity of $\cF$ by relying on the bulk channel decomposition of $\cF$,
which converges in the region of integration.
Consider then calculating the contribution to the discontinuity from a single
bulk superblock and applying the inversion formula to it.
There is a contribution to the discontinuity if the block gives rise to a branch cut
for $F$ at $\bar{z} = 1$, or as shown in~\cite{Lemos:2017vnx,Barrat:2021yvp} if it gives
rise to a pole in $\cF$.

Comparing with Table~\ref{tab:1pts} and using the $\bar{z} = 1$ expansion of
$g^{st}$
\begin{align}
  g_{\Delta,\ell}^{st}(z,\bar{z}) =
  (1-\bar{z})^{\frac{\Delta - \ell}{2}} k_{\Delta+l}(1-z) + \dots
  \label{eqn:discFzbar}
\end{align}
it's easy to see that branch cuts can only arise from long operators with anomalous
dimensions. At large $N$ the exchanged operators are ``double-traces'' and their anomalous dimension is suppressed by a factor $c^{-1}$, see~\eqref{eqn:doubletracespectrum}. Since
long multiplets enter the correlator at order $d^2/c$, they would contribute to branch cuts at
order $d^2/c^2$, so do not contribute to the order we consider in our calculation.
This suppression of double-trace operators is a general feature of the inversion
formula at large $N$ and means the correlator is completely fixed in terms of
the exchange of single-trace operators~\cite{Caron-Huot:2017vep}.

Superblocks may also lead to a pole in $\cF$, either from a pole in
$F(z,\bar{z})$ when the superprimary satisfies $\Delta - \ell < 8$, or from a
pole in $\zeta(z)$. Again comparing to Table~\ref{tab:1pts} we find that the
only superblocks which may contribute to the discontinuity 
by this mechanism are the (bulk) identity and the stress tensor superblock.

This observation was translated into a concrete bootstrap strategy for the
2-point function of stress tensors in the presence of a Wilson line
in~\cite{Barrat:2021yvp}, and we adapt it in the following.
From the discontinuity of the superblocks for the bulk identity and stress
tensor and the inversion formula~\eqref{eqn:susyinversionB}, we calculate the
dCFT data entering the superblock decomposition of $\cF$, up to the defect
identity, $B[1]$ and $A[0]_s$ blocks not captured by the inversion formula. The
contributions from the defect identity and displacement supermultiplet $B[1]$
are fixed by the correlators~\eqref{eqn:a2} and~\eqref{eqn:disp} respectively which encode the choice of $d$ and $c$
so they are unambiguous. The contributions from the $A[0]_s$ multiplets can be
inferred from the chiral algebra once the contribution from the $A[1]_s$
multiplets is known and we find that they are absent; alternatively, we can
check that crossing symmetry is satisfied without adding $A[0]_s$ multiplets.
Resumming all these blocks we obtain the correlator. 

There is one more subtlety. The inversion formula may miss the
contribution of defect multiplets with low-spins, so not reconstruct the full
correlator. Using the variables $z = r w$ and $\bar{z} = r/w$, $\cF$ may have
singularities at $w = 0$. Assuming it is bounded by a power
\begin{align}
  |\cF(z=rw, \bar{z}=r w^{-1})| \lesssim w^{-s_*}\,,\ \text{as}\ w \to 0\,,
  \label{eqn:sstar}
\end{align}
then the convergence of the inversion formula is guaranteed only down
to spin $s > s_*$~\cite{Lemos:2017vnx}. This constraint is easy to understand
from the procedure described above: Poles at $w=0$ naturally arise from an
infinite sum of bulk blocks, whose discontinuity vanishes identically term by
term and thus are not captured by the inversion formula.

We address this issue by checking crossing symmetry in
section~\ref{sec:crossing}, and surprisingly we find that the inversion formula
along with the input from kinematics reconstructs the \textit{full} correlator.

In the following we use the inversion formula~\eqref{eqn:susyinversionB} to
calculate the contributions from the bulk identity and stress tensor multiplet.

To simplify our calculation, it is convenient to massage  the inversion
formula~\eqref{eqn:susyinversionB} as follows. The contributions to $\mathsf{B}$
from $F$ and $\zeta$ can be calculated separatly, with the contribution from $F$
\begin{align}
  \mathsf{B}^F(\hat{\Delta},s) =
  \frac{1}{2} \int\limits_0^1 \diff z
  \frac{z^{-\frac{\hat{\Delta}-s}{2}-1}}{(1-z)^{2}}
  \int\limits_1^{1/z} \frac{\diff \bar{z}}{2\pi i}
  & \frac{(1-z \bar{z})(\bar{z}-z)}{(1-\bar{z})^2} 
  \bar{z}^{-\frac{\hat{\Delta}+s}{2}-2}
  \disc F(z,\bar{z})\,.
\end{align}
In the present case, neither of the bulk identity and the stress tensor
contribute to $F$, so for these $\mathsf{B}^F = 0$.

The contribution from $\zeta$ is more subtle. The function $\zeta(z)$ may only
have poles at $z = 1$ by kinematics and unitarity. The
integral~\eqref{eqn:susyinversionB} can be evaluated from a careful
regularisation of the singularities, but a simpler approach is to express the
discontinuity as a contour integral.  Going back to the Euclidean inversion
formula of~\cite{Lemos:2017vnx}, using the variables $z = r w$ and $\bar{z} =
r/w$, and expressing the Chebyshev polynomials as~\eqref{eqn:chebushevid} we get
\begin{align}
  \frac{1}{2 \pi i} \int\limits_0^1 \diff r (r^{\hat\Delta+1} -
  r^{-(\hat\Delta+1)})
  \oint_{|w|=1} \diff w (w^{s+1}-w^{-(s+1)}) \left[ \frac{\zeta(rw)}{(r-w)^2} -
  \frac{\zeta(r w^{-1})}{(rw - 1)^2} \right]\,.
\end{align}
The contour integral picks up residues inside the disk $|w| = 1$, which can only
be located at the point $w = r$.

The integral contains poles corresponding to exchanged operators and their
shadow blocks. Keeping only poles corresponding to physical operators we get
\begin{equation}
\begin{aligned}
  B^\zeta(\hat{\Delta},s) &=
  -\frac{1}{2 \pi i} \int\limits_0^1 \diff r r^{-(\Delta+1)}
  \oint_{|w|=1} \diff w w^{s+1} \left[ \frac{\zeta(rw)}{(r-w)^2} -
  \frac{\zeta(r w^{-1})}{(rw - 1)^2} \right]\\
  &=
  -\int\limits_0^1 \diff r r^{-(\Delta+1)}
  \res_{w=r} \left[ w^{s+1} \left( \frac{\zeta(rw)}{(r-w)^2} -
  \frac{\zeta(r w^{-1})}{(rw - 1)^2} \right) \right]\,.
\end{aligned}
  \label{eqn:susyinvBzeta}
\end{equation}

\subsection{Inverting the bulk identity}

As a check of the supersymmetric inversion formula~\eqref{eqn:susyinversionB} we
consider the inversion of the bulk identity,
which corresponds to the disconnected correlator $\vev{\O2 \O2} \vev{V}$.
From the table~\ref{tab:1pts}, the
identity block contributes to $\zeta$ and $F$ as
\begin{align}
  \zeta_{\mathds{1}}(z) = \frac{z^2}{(1-z)^4}\,, \qquad
  F_{\mathds{1}}(z,\bar{z}) = 0\,.
  \label{eqn:sblockidentitybulkcontrib}
\end{align}
Plugging in~\eqref{eqn:susyinvBzeta}, we calculate the residue
\begin{equation}
\begin{aligned}
  &\res_{w=r} \left[ w^{s+1} \left( \frac{\zeta(rw)}{(r-w)^2} -
  \frac{\zeta(r w^{-1})}{(rw - 1)^2} \right) \right]\\
  & \quad =
  -\frac{(s+1) r^{s+2}}{6 (1-r^2)^4} \left[ 
  s(s-1) r^4 - 2 (s+3)(s-1) r^2 + (s+3)(s+2)\right]\,.
\end{aligned}
\end{equation}
Including the kernel $r^{-(\Delta+1)}$, the integral of this over $r$ diverges near
$r=1$. This divergence can be understood by expanding the integrand in powers of $r$ using
\begin{align}
  (1-r^2)^{-4} = \sum_{n=0}^\infty \frac{(n+1)(n+2)(n+3)}{6} r^{2n}\,,
\end{align}
and interchanging the sum and integral. In that case the integrals converge term
by term for large enough $\hat\Delta$ and we find
\begin{align}
  \mathsf{B}(\hat\Delta,s)|_{poles} =
  \sum_{m \ge 0}
  \frac{(m+1)(s+1)(m+s+2)(2m+s+3)}{6 (2+2m+s - \hat\Delta)}\,.
\end{align}
From this we read the dimension of long supermultiplets and the corresponding
OPE coefficients
\begin{align}
  (\hat{\Delta}_{m,s})^{(0)} = 2(m+1) + s\,,
  \qquad
  (b_{m,s}^2)^{(0)} =
  \frac{(m+1)(s+1)(m+s+2)(2m+s+3)}{6}\,.
  \label{eqn:sblockidentityOPE}
\end{align}
We added a label ${}^{(0)}$ in anticipation of subleading corrections in
$c$ discussed in section~\ref{sec:exchangestresstensor}.

Notice that when $m=0$, the dimension of operators sit at the unitarity bound
and split into short multiplets according to the recombination
rules~\eqref{eqn:recombination}. We can identify the resulting superblocks by
shifting appropriately the labels according to where the $r=2$ block appear (see
Appendix~\ref{sec:defectblocktable}) to get
\begin{align}
  (b_{0,0}^2)^{(0)} \hat\cG_{B[2]}
  + \sum_{s \ge 0} (b_{0,s+1}^2)^{(0)} \hat\cG_{A[1]_s}
  + \sum_{m>0 \atop s \ge 0} (b_{m,s}^2)^{(0)}
  \hat\cG_{L[0]_{\hat{\Delta}_{m,s},s}}
  =
  \left(\frac{z \bar{z} (1-\omega)^2}{(1-z)^2(1-\bar{z})^2\omega} \right)^2\,.
  \label{eqn:sblockidentity}
\end{align}
In the last step we resummed the superconformal blocks. The result is indeed the
contribution from the bulk identity, it decomposes into $\zeta$ and $F$ exactly
as~\eqref{eqn:sblockidentitybulkcontrib}.
Since the expansion \eqref{eqn:sblockidentity} is unambiguous, this shows that, at this order, there is no exchange of $A[0]_s$ supermultiplets.

Note that the dCFT data~\eqref{eqn:sblockidentityOPE} doesn't depend on the
choice of defect operator $V$, since it corresponds to a disconnected
correlator. So the result simply follows from the branching rules
for the decomposition of bulk operators in terms of representations of
$\osp(4^*|2) \oplus \osp(4^*|2)$. It is clear that this decomposition in
defect blocks always exists, so the bulk identity is always crossing symmetric.

Focusing on the conformal blocks with $r=2$, this dCFT data matches the general
result of~\cite{Lemos:2017vnx} for a defect of dimension $2$ in 6d.

\subsection{Inverting the stress tensor exchange}
\label{sec:exchangestresstensor}

Next we consider the contribution to the correlator arising from the exchange of
a stress tensor supermultiplet. The contribution to $\zeta, F$ from the stress
tensor is given in~\eqref{eqn:D20sblockcontrib}, which we reproduce here for
convenience
\begin{align}
  \zeta_{\cD[2,0]}(z) =  -\frac{12z^2}{(1-z)^4} \left[ 1 + \frac{1}{2} \frac{1+z}{1-z} \log
    z
  \right]\,, \qquad
  F_{\cD[2,0]} = 0\,.
\end{align}
Calculating the residue gives
\begin{align}
  \frac{r^2}{2(1-r^2)^5} \left( s+2 + (38+9s)r^2 + r^4(20-9s) -s r^6 \right)
  - \frac{6 r^4}{(1-r^2)^6} \left( s+3 + 6 r^2 - (s-1) r^4 \right)
  \log{r}\,.
\end{align}
Again expanding $(1-r^2)$ in series around $r=0$ we can perform the integral as
above. The $\log{r}$ terms gives rise to double pole. They should be understood
as the small $c$ expansion of the dCFT data
\begin{align}
\label{b2finalandgamma}
  b_{m,s}^2 = (b_{m,s}^2)^{(0)} + \frac{4d}{c} (b_{m,s}^2)^{(1)} + \dots\,, \qquad
  \hat{\Delta}_{m,s} = 2(m+1) + s + \frac{4d}{c} \gamma_{m,s}^{(1)} + \dots
\end{align}
where the leading terms are simply the OPE data found
in~\eqref{eqn:sblockidentityOPE}.
This leads to an expansion
\begin{align}
  \frac{b_{m,s}^2}{\hat\Delta_{m,s} - \hat\Delta} =
  \frac{(b_{m,s}^2)^{(0))}}{2(m+1)+s - \hat\Delta}
  + \frac{4d}{c} \left( \frac{(b_{m,s}^2)^{(1)}}{2(m+1)+s - \hat\Delta}
  - \frac{(b_{m,s}^2)^{(0)} \gamma_{m,s}^{(1)}}{(2(m+1)+s - \hat\Delta)^2} \right)
  + \dots
\end{align}
Performing the integral and comparing with this expansion we can read the dCFT data
\begin{equation}
\begin{aligned}
  \gamma_{m,s}^{(1)} &=
  -\frac{6m(m+1)(m+2)}{(s+1)(2m+s+3)}\,,\\
  (b_{m,s}^2)^{(1)} &=
  - \left[ (m+1)(s+1) - \frac{1}{2} (m+1)^2 (5m^2 + 2 m (2s+7) + 4s + 6)
  \right]\,,
\end{aligned}
\label{eqn:sblocstresstensorOPE}
\end{equation}
where $m=1,2,\dots$ correspond to long operators while the case $m=0$
corresponds to $A[1]_s$ operators. Note that at this order they do not acquire
an anomalous dimension.

\subsection{The result}

As anticipated in the beginning of this section, the supersymmetric inversion
doesn't capture the contributions from the defect identity and the displacement
supermultiplet; however these blocks are special and their contribution is fixed
by~\eqref{eqn:a2} and~\eqref{eqn:disp} respectively.

Including these and using the dCFT data~\eqref{eqn:sblockidentityOPE}
and~\eqref{eqn:sblocstresstensorOPE}, the full superconformal block
decomposition of the correlator is
\begin{equation}
  \begin{aligned}
    \cF(z,\bar{z},\omega)
    &=
    \frac{2d^2}{c}
    + \frac{4d}{c}
    \hat{\cG}_{B[1]}
    + b_{0,0}^2 \hat{\cG}_{B[2]}
    + \sum_{s \ge 0} b_{0,s+1}^2 \hat{\cG}_{A[1]_{s}}
    +
    \sum_{m \ge 1 \atop s \ge 0}
    b_{m,s}^2 \hat{\cG}_{L[0]_{\hat{\Delta}_{m,s},s}}
    + O(c^{-2})\,,
    \label{eqn:corrsuperblockansatz}
  \end{aligned}
\end{equation}
where the OPE coefficients and spectrum are given in \eqref{b2finalandgamma}.
Notice that all the coefficients $b^2$ in this expansion are positive, as
expected from unitarity. For $m=0$ the coefficients $b$ encode the contribution
of short superblocks, and we find that~\eqref{eqn:sblocstresstensorOPE} agrees
with the chiral algebra calculation~\eqref{eqn:chiralbcoeff}. 

Note that by comparing the coefficients $b_{0,s+1}^2$ of the $A[1]_s$ multiplets
to the chiral algebra result~\eqref{eqn:chiralbcoeff} we can conclude that, at
this order, there are no $A[0]_s$ multiplets exchanged, provided the inversion
formula captures all the $A[1]_s$ multiplets. We cannot exclude the possibility
that the inversion formula misses a finite number of $A[1]_s$ multiplets with $s
< s_*$ for some $s_*$, and correspondingly the correlator could receive
contributions from a finite number of $A[0]_s$ multiplets; however we find that
no such multiplets are required (and are likely excluded) by crossing symmetry.

Resumming these blocks, we find our main result~\eqref{eqn:corrresult}.
The function $\zeta$ we obtain reproduces the result from the chiral
algebra~\eqref{eqn:zetafull}, which suggests that the bootstrap result captures
the full protected sector of the dCFT.

The function $F$ given in~\eqref{eqn:corrresult} has several interesting features. From the definition of
cross-ratios~\eqref{eqn:crossratios} we see that $z \leftrightarrow
\bar{z}$ corresponds to an equivalent kinematic configuration, so is a symmetry
of $F$. Additionally, $(z, \bar{z}) \to (1/z, 1/\bar{z})$ is also a symmetry and corresponds to
exchanging the two bulk operators.

The dependence of $F$ on $z + \bar{z}$ is linear so that $\cF|_{r=2}$ satisfies the expected
Regge behavior~\eqref{eqn:sstar} for $s_* = -1$, which is a consistency check of the inversion
formula. The dependence on $z \bar{z}$ is more complicated, and we note that it
has no branch cut or pole at $z \bar{z} = 1$. Note that the coefficient of the
$\log{z\bar{z}}$ captures information about the anomalous dimensions of defect
operators.

In euclidean signature, $\bar{z}$ is the complex conjugate of $z$, and $F$ is
manifestly real. If we analytically continue the result by taking $z,
\bar{z}$ real, we get a spacelike defect in lorenztian signature (see
figure~\ref{fig:crossratios}). A different analytic continuation is to
take $\sqrt{z \bar{z}} \to i \sqrt{z \bar{z}}$ with $z/\bar{z}$ fixed, which
corresponds to a timelike defect~\cite{Lauria:2017wav}. In that case we have
$\log{z \bar{z}} \to \log{z \bar{z}} + i \pi$, so on this sheet $F$ develops a pole at $z
\bar{z} = 1$, which is where both bulk operators become lightlike separated from
the same point on the defect. The corresponding singularity can be interpreted as
arising from the exchange of ``double-trace'' defect operators,
see~\cite{Maldacena:2015iua,Lauria:2017wav}.

\subsection{An equivalent calculation}

In the calculation of section~\ref{sec:exchangestresstensor} we apply the
inversion formula to the discontinuity arising from the stress tensor block
exchange~\eqref{eqn:D20sblockcontrib}. Since we know the protected sector of the
dCFT from the chiral algebra in section~\ref{sec:chiral} and in particular
the exact $\zeta(z)$~\eqref{eqn:zetafull}, it is natural to write an inversion
formula that already includes the contribution from these protected multiplets.

The only bulk supermultiplets that contribute to the function $\zeta(z)$ are 
the short bulk supermultiplets $\mathds{1}, \cD[2,0], \cD[4,0]$ and $\cB[2,0]_l$
(refer to Table~\ref{tab:1pts}). The coefficients of that expansion are known
exactly from the chiral algebra and are given in~\eqref{eqn:chiralonepts}.
Resumming these superblocks,  we find the contribution to $F$
which we call $F_\text{short}$\footnote{
Such a splitting was first introduced in \cite{Beem:2013qxa} in a similar context.
}
\begin{align}
  F_\text{short}(z,\bar{z}) =
  \frac{2d^2}{c}
  - \frac{4d}{c} \left[
  -\frac{12z \bar{z}}{(1-z)^2(1-\bar{z})^2}
  + \frac{6 z \bar{z}}{(z-\bar{z})(1-z \bar{z})}
  \left(
  -\frac{z (z+1)}{(1-z)^3} \log{z}
  +\frac{\bar{z}(\bar{z}+1)}{(1-\bar{z})^3} \log{\bar{z}}
\right) \right]\,.
\end{align}
The full function $F=F_\text{short}+F_\text{long}$ also receives contributions from long supermultiplets,
which are not protected by supersymmetry.

The inversion formula applied to the exact $\zeta(z)$ and $F_\text{short} +
F_\text{long}$ is equivalent to the original one, and it only depends on the
unknown $F_\text{long}$ encoding the contribution of long multiplets, making it
an excellent starting point for higher order computations.
We can recover the results of section~\ref{sec:exchangestresstensor} by noting
that to order $d/c$, $F_\text{long}$ does not contribute. Plugging $\zeta^{(1)}$ in the
inversion formula, we find that $B^\zeta$ vanishes identically.
$F^{(1)}_\text{short}$ does not have a discontinuity at $\bar{z} = 1$, but it
leads to poles in the inversion formula which contribute to $\mathsf{B}^F$.
To calculate their contributions, one needs to regularise the integrals as
in~\cite{Barrat:2021yvp}, and one can show the result agrees with the dCFT
data~\eqref{eqn:sblocstresstensorOPE}.

\section{Crossing symmetry}
\label{sec:crossing}

A nontrivial check of our result~\eqref{eqn:Fsolexpansion} is that it can be
decomposed both in defect and bulk channels. This suggests that, up to the
missing contributions from the defect identity and displacement supermultiplet,
the inversion formula~\eqref{eqn:susyinversionB} reconstructs the full
correlator at this order in $c$. We note that crossing symmetry is satisfied
independently for all three pieces appearing in~\eqref{eqn:Fsolexpansion},
respectively with coefficients $1, \frac{d^2}{c}$ and $\frac{d}{c}$.

The leading contribution is associated with the exchange of a single bulk
supermultiplet, the bulk identity, and is independent of the choice of defect.
In particular it can be calculated from the trivial defect, and it is known to
be universally crossing symmetric. The corresponding defect channel block
decomposition is obtained in~\eqref{eqn:sblockidentity}.

In this section we verify that crossing symmetry is satisfied also for the two
other terms: We obtain the explicit bulk channel decomposition of $\cF$, given
by
\begin{equation}
\begin{aligned}
  \cF(z,\bar{z},\omega) =
  & \left(\frac{z \bar{z} (1-\omega)^2}{(1-z)^2(1-\bar{z})^2\omega} \right)^2
  \left[ 
    1 - \frac{4d}{c} \cG_{\cD[2,0]} + (a \lambda)_{0,0} \cG_{\cD[4,0]} \right.\\
    & \qquad \left. -\sum_{\ell \ge 0} (a \lambda)_{0,\ell+2}
    \cG_{\cB[2,0]_\ell}
    + \sum_{n,\ell \ge 0} (a \lambda)_{n+2,\ell} \cG_{\cL[0,0]_{\Delta,\ell}}
   \right]\,,
\end{aligned}
\label{eqn:Fbulkdecomp}
\end{equation}
where $\ell$ is even, the dimension of the long superblock is $\Delta = 8 + 2n +
\ell
+ O(c^{-1})$ as in~\eqref{eqn:doubletracespectrum} and the OPE coefficients are
\begin{equation}
\begin{aligned}
  (a \lambda)_{n,\ell} &= 
  \frac{n!(n+1)!(n+\ell+2)!(n+\ell+3)!}{(2n+1)!(2n+2\ell+5)!} \times\\
  & \quad \left[
    \frac{2d^2}{c} \delta_{n,even} (\ell+2)(2n+\ell+5)
  + \frac{d}{2c} (-1)^n (n+4)(n+2)(n+1)(n-1) + O(c^{-2}) \right]\,.
\end{aligned}
\label{eqn:bulkdecompOPE}
\end{equation}
The terms proportional to $\frac{d^2}{c}$ encode the exchange of the defect
identity, and for $n=0$ these coefficients match the chiral algebra
results~\eqref{eqn:chiralonepts} and also the leading lightcone limit
obtained in~\cite{Liendo:2019jpu}.
The terms proportional to $\frac{d}{c}$ include the stress tensor multiplet
$\cD[2,0]$ and are the bulk channel decomposition of~\eqref{eqn:corrresult}.

The appearance of double-trace operators with only even $n$ suggests that the
defect identity enjoys an additional $\mathbb{Z}_2$ symmetry; it would be
interesting to understand it.

In the following we present a supersymmetric inversion formula, this time for
the bulk channel. Applying it to the defect identity we obtain an analytic
derivation of its bulk channel decomposition.  The decomposition of the rest of
the correlator~\eqref{eqn:corrresult} can be obtained in principle from the
inversion formula as well, but here we obtain it directly from the result by
matching an ansatz like~\eqref{eqn:Fbulkdecomp} to high order in $z, \bar{z} \to
1$ using Mathematica.

\subsection{The bulk channel inversion formula}

The OPE coefficients $(a \lambda)$ entering the conformal block decomposition of
$\cF$ can be extracted from the bulk channel inversion formula
of~\cite{Liendo:2019jpu}. Again this formula is general and doesn't explicitly
account for R-symmetry; to apply it we can decompose $\cF$ in representations
of bulk operators labelled by $R$. The inversion formula then defines a
function $C_R(\Delta,l)$ given in terms of the double discontinuity of
$\cF|_R$
\begin{equation}
\begin{aligned}
  \ddisc{\cF|_R(z,\bar{z})} &=
  \cos{\left( \frac{(\Delta_2 - \Delta_1)\pi}{2} \right)} \cF|_R(z,\bar{z})\\
  &\ \ -\frac{1}{2} e^{-i\pi \frac{\Delta_1 + \Delta_2}{2}}
  \cF|_R(z, e^{2\pi i} \bar{z})
  -\frac{1}{2} e^{i\pi \frac{\Delta_1 + \Delta_2}{2}}
  \cF|_R(z, e^{-2\pi i} \bar{z})\,.
\end{aligned}
\end{equation}
$C_R$ is constructed in such a way that it has poles at the
location of exchanged bulk operators $\Delta = \Delta_{n,l}$ and residue
containing the OPE coefficients.

To include supersymmetry we use the fact that each conformal block with
$R=4$ and given $\Delta, \ell$ appears in a single superconformal block (apart from the identity and the stress tensor supermultiplet $\mathcal{D}[2,0]$ when it does not appear), so we
can read the superconformal block decomposition of $\cF$ from the $R=4$ channel, see \ref{sec:tabsblockbulk}.
From the inversion formula of~\cite{Liendo:2019jpu} we can define
\begin{equation}
\begin{aligned}
  \mathsf{C}^t(\Delta,\ell) &=
  C_{R=4}^t(\Delta+4,\ell)\\
  &= -\frac{\kappa_{\Delta+\ell+4}}{2}
  \int\limits_0^1 \diff z \int\limits_0^1 \diff \bar{z}
  \frac{|z-\bar{z}|^2(1-z \bar{z})^2}{(z\bar{z})^{\frac{\Delta_\phi}{2}}
  [(1-z)(1-\bar{z})]^{6-\Delta_\phi}} 
  g^{st}_{5+\ell, \Delta-1}(z, \bar{z})
  \ddisc{\cF|_{R=4}(z,\bar{z})}\,.
\end{aligned}
  \label{eqn:bulksusyinvformula}
\end{equation}
We included a shift $\Delta \to \Delta+4$ so that the function $\mathsf{C}$ has
poles at the location of the superprimary operator, assuming a long
supermultiplet (the corresponding superblock is given in~\eqref{eqn:sblockL00l}).
In this equation, $g^{st}_{\Delta,\ell}(z,\bar{z})$ are the (spacetime part of the) bulk channel
blocks~\eqref{eqn:confblockso8sol} and the coefficient $\kappa_{\beta}$ is
given~\cite{Liendo:2019jpu}%
\footnote{There is a typo there, see also the original definition in~\cite{Caron-Huot:2017vep}.}
\begin{align}
  \kappa_\beta =
  \frac{\Gamma\left( \frac{\beta}{2} \right)^2 \Gamma\left( \frac{\beta}{2}+a \right)
  \Gamma\left( \frac{\beta}{2}-a \right)}{2\pi^2 \Gamma(\beta)
  \Gamma(\beta-1)}\,, \qquad
  a = \frac{\Delta_2 - \Delta_1}{2}\,.
\end{align}

The label ${^t}$ in~\eqref{eqn:bulksusyinvformula} indicates that this captures
the bulk blocks appearing in the ``$t$-channel''. We also need to add the
contributions from the ``$u$-channel'' obtained by exchanging the external bulk
operators (here $\mathsf{C}^u = \mathsf{C}^t$)
\begin{align}
  \mathsf{C}(\Delta,\ell) =
  \mathsf{C}^t(\Delta,\ell) + (-1)^\ell \mathsf{C}^t(\Delta,\ell)\,.
\end{align}

Finally we can express $\cF|_{R=4}$ directly in terms of $\zeta, F$ by
extracting from~\eqref{eqn:solWIk2} the contribution proportional to the
R-symmetry block~\eqref{eqn:confblockso5sol}. We find simply
\begin{align}
  \cF|_{R=4} = F(z,\bar{z})\,.
\end{align}

\subsection{Inverting the defect identity}

As an example of application of the inversion
formula~\eqref{eqn:bulksusyinvformula} we reproduce the defect identity
contribution to the correlator, i.e. the term of order $\frac{d^2}{c}$
in~\eqref{eqn:Fsolexpansion}. This calculation is a simple extension of the one
presented in~\cite{Liendo:2019jpu}.

For an external operator of dimension
$\Delta_\O2$, the double discontinuity of a constant is
\begin{align}
  \ddisc{1} = 2 \sin^2\left( \frac{\pi \Delta_\O2}{2} \right)\,.
\end{align}
In the limit $\Delta_\O2 \to 4$ the double discontinuity vanishes, but
correspondingly the integral in~\eqref{eqn:bulksusyinvformula} diverges such
that $\mathsf{C}^t$ has a well-defined limit.

Since the discontinuity is independent of $z, \bar{z}$, the inversion formula
becomes very simple to evaluate.  Plugging the conformal
blocks~\eqref{eqn:confblockso8sol} in the inversion
formula~\eqref{eqn:bulksusyinvformula}, we obtain (up to a prefactor)
\begin{align}
  \int\limits_0^1 \diff z \diff \bar{z}
  \frac{[(1-z)(1-\bar{z})]^{\Delta_\O2-4}}{(z\bar{z})^{\Delta_\O2/2}}
  (z-\bar{z})(z \bar{z}-1)
  k_{2+\ell-\Delta}(1-z) k_{\ell+\Delta+4}(1-\bar{z})
  - (z \leftrightarrow \bar{z})\,.
\end{align}
Since the integral is symmetric in $z,\bar{z}$, we get a factor of 2 and the
inversion formula reduces to
\begin{equation}
\begin{aligned}
  \mathsf{C}^t(\Delta,\ell) =& \ 
  -\kappa_{\Delta+\ell}
  \sin^2\left( \frac{\pi \Delta_\O2}{2} \right) \times\\
  &\int\limits_0^1 \diff z \diff \bar{z}
  \frac{[(1-z)(1-\bar{z})]^{\Delta_\O2-4}}{(z\bar{z})^{\Delta_\O2/2}}
  (z-\bar{z})(z \bar{z}-1)
  k_{2+\ell-\Delta}(1-z) k_{\ell+\Delta+4}(1-\bar{z})\,.
\end{aligned}
  \label{eqn:inversionformuladefectidentity}
\end{equation}
Introducing the short-hand notation
\begin{align}
  I_{\Delta_\O2,\beta} &\equiv
  \int_0^1 \diff z
  \frac{(1-z)^{\Delta_\O2-2}}{z^{\Delta_\O2/2}}
  k_{\beta}(1-z)\,,
\end{align}
we see that~\eqref{eqn:inversionformuladefectidentity} factorises and the
integral can be written as
\begin{equation}
\begin{aligned}
  \int\limits_0^1 \diff z \diff \bar{z}
  &\frac{[(1-z)(1-\bar{z})]^{\Delta_\O2-4}}{(z\bar{z})^{\Delta_\O2/2}}
  (z-\bar{z})(z \bar{z}-1)
  k_{6+\ell-\Delta}(1-z) k_{\ell+\Delta}(1-\bar{z})\\
  & \quad =
  I_{\Delta_\O2, 2+\ell-\Delta} I_{\Delta_\O2-2,\Delta+\ell+4}
  - I_{\Delta_\O2-2, 2+\ell-\Delta} I_{\Delta_\O2,\Delta+\ell+4}\,.
\end{aligned}
\end{equation}
The integral $I$ can be evaluated. We expand the hypergeometric function in
series, integrate and resum to get
\begin{equation}
  \begin{aligned}
    \frac{\Gamma \left(1-\frac{\Delta_\O2}{2} \right)
    \Gamma \left(\Delta_\O2 + \frac{\beta}{2} - 1 \right)}
    {\Gamma\left( \frac{\Delta_\O2 + \beta}{2} \right)}
    {}_3F_2\left( 
    \begin{matrix}
      \frac{\beta}{2} \,, \frac{\beta}{2} \,, \frac{\beta}{2} + \Delta_\O2 - 1\\
      \beta \,, \frac{\beta+\Delta_\O2}{2} \\
    \end{matrix}
    ; 1 \right)\,.
  \end{aligned}
  \label{eqn:zbintegralid1}
\end{equation}
This can be simplified using Watson's theorem (see (6) of section 4.4
of~\cite{erdelyi1981higher}), which is the identity
\begin{align}
    {}_3F_2\left( 
    \begin{matrix}
      a \,, b \,, c\\
      \frac{a+b+1}{2} \,, 2c \\
    \end{matrix}
    ; 1 \right)
    = \frac{
      \Gamma\left( \frac{1}{2} \right)
      \Gamma\left( c + \frac{1}{2} \right)
      \Gamma\left( \frac{a+b+1}{2} \right)
      \Gamma\left( c + \frac{1-a-b}{2} \right)}
      {\Gamma\left( \frac{a+1}{2} \right)
      \Gamma\left( \frac{b+1}{2} \right)
      \Gamma\left( c + \frac{1-a}{2} \right)
      \Gamma\left( c + \frac{1-b}{2} \right)}\,.
\end{align}
With appropriate values of $a=b$ and $c$ we find
\begin{align}
  I_{\Delta_\O2,\beta} =
  2^{\Delta_\O2+\frac{\beta}{2}-2}
  \frac{\Gamma\left( \frac{\beta+1}{2} \right)
    \Gamma\left( 1-\frac{\Delta_\O2}{2} \right)^2 
    \Gamma\left( \frac{\beta}{4} + \frac{\Delta_\O2-1}{2} \right)}
  {
    \Gamma\left( \frac{\beta+2}{4} \right)^2
    \Gamma\left( \frac{\beta}{4} +\frac{2-\Delta_\O2}{2} \right)
  }\,.
  \label{eqn:integralzbar}
\end{align}

Finally, we can evaluate $\mathsf{C}^t$ by substituting $I_{\Delta_\O2,\beta}$,
simplifying the gamma functions and eliminating the $\sin\left(
\frac{\pi \Delta_\O2}{2} \right)$ by using the following identity for Beta
functions
\begin{align}
  B(x,y) B(x+y,1-y) = \frac{\pi}{x \sin \pi y}\,.
  \label{eqn:betaidentity}
\end{align}
We obtain
\begin{equation}
\begin{aligned}
  &\mathsf{C}^t(\Delta,\ell)|_{poles} =
  \lim_{\Delta_\O2 \to 4}
  2^{2\Delta_\O2-9-\Delta} (\ell+2)(\Delta+1)\ \times\\
  &\qquad
  \left.
  \frac{
  \Gamma\left( 2 - \frac{\Delta_\O2}{2} \right)^2 }{
  \Gamma\left( \frac{4+\ell-\Delta}{4} \right)^2
  }
  \frac{
  \Gamma\left( \frac{\Delta+\ell+4}{4} \right)^2
  \Gamma\left( \frac{\Delta+\ell+2\Delta_\O2-2}{4} \right)
  \Gamma\left( \frac{3+\ell-\Delta}{2} \right)
  \Gamma\left( \frac{2\Delta_\O2-4+\ell-\Delta}{4} \right) }{
  \Gamma\left( \frac{\Delta+\ell+3}{2} \right)
  \Gamma\left( \frac{10+\ell-2\Delta_\O2-\Delta}{4} \right)
  \Gamma\left( \frac{12+\ell+\Delta-2\Delta_\O2}{4} \right)
  \Gamma\left( \frac{\Delta_\O2}{2} \right)^2
} \right|_{poles}\,.
\end{aligned}
\end{equation}
This expression has poles in $\Delta$ whenever the gamma functions in the numerator
diverge, which happens for\footnote{
  There are also poles at $\Delta = 7+\ell+2n$ which come from poles in the
  conformal blocks $g_{l+5,\Delta-5}$, rather than the integral. They do not
  contribute to the dCFT data so we ignore them; for a proper treatment see~\cite{Caron-Huot:2017vep}.
}
\begin{align}
  \frac{2\Delta_\O2-4+\ell-\Delta}{4} = -n
  \qquad \Rightarrow \qquad
  \Delta = 2 \Delta_\O2 - 4 + \ell + 4n\,, \qquad
  n \in \mathbb{Z}_{\ge 0}\,.
\end{align}
Near these poles the gamma function behaves as
\begin{align}
  \Gamma(z)|_{z \to -n} = \frac{(-1)^n}{n! (z+n)} + \dots\,.
\end{align}
with the subleading terms regular as $z \to -n$. The residue at these poles is
thus trivial to calculate. Notice also that while the limit $\Delta_\O2 \to 4$
of the gamma functions diverges above, it is finite at the poles. We obtain
\begin{align}
  \mathsf{C}^t(\Delta,\ell)|_{poles} &=
  \sum_{n \ge 0}
  2^{-2n-1}
  \frac{(\ell+2)(4n+\ell+5)(2n+1)!(2n+\ell+2)!(2n+\ell+3)!}{(4n+1)!! (4n+2\ell+5)!
  (4 + 4n+\ell-\Delta)}\,.
\end{align}
Finally, adding the contribution for the $u$-channel, we can read the OPE
coefficients and reproduce the result~\eqref{eqn:bulkdecompOPE} presented at the beginning of this
section.

\section{Conclusions and outlook}

It is an interesting problem to understand how to perform calculations in a
nonlagrangian theory. For such theories with a large $N$ limit, the conformal
bootstrap offers a systematic approach to calculate correlators perturbatively
in $1/N$, with minimal assumptions, and thus provides a working definition of
these theories.

In this paper we present a case study, the 2-point function of stress
tensor superprimaries $\O2$ in the presence of a surface operator $V$, and adapt and
develop bootstrap techniques to calculate the first subleading contribution at
large $N$ to their correlator $\vev{\O2 \O2 V}$. In doing so we extract dCFT
data characterising the surface operators: We find partial information about the spectrum of
operators in the 2d dCFT associated to $V$, and their interactions with local
bulk operators in the form of the coefficients $a_k$ and $b_{kl}$ entering
respectively the 1-point functions~\eqref{eqn:1pts} and bulk-defect 2-point
functions~\eqref{eqn:bulkdefect2pts}.

The coefficients we calculate are the combinations $b_{m,s}^2$
in~\eqref{eqn:sblockidentityOPE} and~\eqref{eqn:sblocstresstensorOPE}, and $(a
\lambda)_{n,\ell}$ in~\eqref{eqn:bulkdecompOPE}. In the large $N$ limit we expect
degeneracies in the spectrum of operators, and therefore these coefficients to
correspond to averages for all corresponding superblocks of the same
representation. Extracting the individual coefficients $a, b$ from this data
would require lifting the degeneracies; this is not something we attempt here.

The result for the correlator and dCFT data we obtain depends on the choice of
$\cN = (2,0)$ theory and representation for $V$ solely through the anomaly coefficients
$c, d$~\eqref{eqn:anomalycoeffs}, which are known exactly. Our result is valid for
both the $A_{N-1}$ and $D_N$ series of $\cN = (2,0)$ theories at large $N$,
and for any representation so long as $1 \ll d \ll c$.
We emphasize that the correlator we calculate contains information about
defect operators in long supermultiplets not protected by supersymmetry. For
these operators we calculate their anomalous dimension to first order.

The main tool in our analysis is the supersymmetric inversion formula,
presented respectively for the defect and bulk channels
in~\eqref{eqn:susyinversionB} and~\eqref{eqn:bulksusyinvformula}.
These formula are particularly useful at large $N$, since the (double)
discontinuity suppresses the contribution from long operators. In the context of
4-point functions of the stress tensor, this suppression and crossing symmetry
was used to obtain the 1-loop correction to the correlator by ``squaring'' the
tree-level anomalous dimensions of all double-trace
operators~\cite{Aharony:2016dwx,Alday:2017vkk}.  Similarly, at order $1/c^2$, we
expect contributions to the discontinuity from the tree-level anomalous
dimensions of double-trace operators, leading to subleading corrections to the
2-point function of order $d^2/c^2$ and $d/c^2$.

The inversion formula is known to miss contributions to the correlator coming
from defect operators with spins below a certain value $s_*$, so a natural
question is whether the correlator we obtain should be supplemented by
additional superblocks. Clearly the inversion formula misses the contributions
from the defect identity and the displacement supermultiplet, but both of these
supermultiplets are special and their contributions are fixed independently by
kinematics, see~\eqref{eqn:a2} and~\eqref{eqn:disp}. Adding these to our result,
we find that the correlator we obtain also admits a bulk channel decomposition
and so is a nontrivial solution to the crossing symmetry constraints. This
provides substantial evidence that the inversion formula recovers completely the
dynamical part of the correlator at order $d/c$~\eqref{eqn:corrresult}, and so
that our result is complete and unambiguous. We expect that a similar strategy
based on the supersymmetric (as opposed to regular) inversion formula would also
resolve the ambiguities faced in~\cite{Barrat:2021yvp}.

The value of $s_*$ is not known a priori and rather enters as an
assumption on the validity of the inversion formula~\eqref{eqn:sstar}. From the
result~\eqref{eqn:corrresult} we can check that this assumption is verified for
the R-symmetry channel $\cF|_{r=2}$ and the inversion formula converges down to
\textit{negative} spin $s_* = -1$ (this is similar to the case of 4-point
function~\cite{Lemos:2021azv}).  This exceptionally low-value for $s_*$ explains
the surprisingly simple dependence on $z + \bar{z}$ in~\eqref{eqn:corrresult}.
This constrains superblocks containing a primary with $r=2$ to sit on Regge
trajectories, with $s > s_*$ for long superblocks and $s > s_* - 1$ for
$A[1]_{s}$ (shifting $s$ for the superprimary).  For $A[1]_s$, the Regge
trajectory is extended to negative spin by identifying $B[2]$ as the analytic
continuation to $s=-1$ (see Table~\ref{tab:defectsblocks}). We note that even
though the inversion formula does not converge at $s = s_*$, the coefficient of
the displacement multiplet is correctly reproduced by identifying $B[1]$ as the
analytic continuation of $A[1]_s$ to $s=-2$.  This is surprising given that the
supersymmetric inversion formula is oblivious, by construction, to the short
multiplets $B[0], B[1]$ and $A[0]_s$.

Subleading corrections to the correlator are determined by long supermultiplets
acquiring an anomalous dimensions, however the corresponding superblocks do not
contribute to the function $\zeta(z)$, which suggests that $\zeta(z)$ is in fact
\textit{exact}. We prove that this is the case by showing that $\zeta(z)$ is
captured by the chiral algebra subsector identified in~\cite{Beem:2014kka} and
can be calculated exactly using standard techniques from chiral algebras.
This is an exact result for any $ADE$ theory and any representation for
$V$, and it encodes the OPE data of the BPS sector of defect operators. We
emphasize that this approach from chiral algebras does not assume a lagrangian
description for the (2,0) theories and offers a viable alternative to
supersymmetric localization.

At the technical level, the setup we study here is surprisingly simple: We could
obtain all the superconformal blocks explicitly, are able to perform all the
integrals and resummations exactly. We believe that this makes these surface
operators an excellent playground to test and develop analytical bootstrap
methods.

Beyond this technical aspect, the setup we introduce here is interesting because
it makes manifest the relation between the 6d CFT of local operators and the 2d
dCFT of defect operators. In this paper we use our knowledge of the 6d theory
(in particular the existence of a stress tensor) to infer properties of the 2d
dCFT, but it would be interesting to learn something about the 6d theories by
bootstrapping directly the 2d dCFT at large $N$ as initiated in~\cite{Drukker:2020swu}.

The correlator we obtain has a natural interpretation in holography as the
propagator for Kaluza-Klein modes in the graviton supermultiplet in the presence
of M2-branes. It would be interesting to confirm this calculation directly from
supergravity. In particular, when the number of M2-branes $M$ is large, the
M2-branes backreact on the $AdS_7$ geometry and give rise to the bubbling
geometries~\cite{DHoker:2008rje,bachas:2013vza}. A hint of that change in
geometry is that when $d \gg c$, the leading term in $\zeta$ is
$(1-z)^{-2}$, which can be interpreted as the chiral part of the propagator of a
graviton in $AdS_3$.

It would also be interesting to understand the structure of our
result~\eqref{eqn:corrresult} in Mellin space, in analogy with the
simplifications for the 4-point functions of local operators
(see for
instance~\cite{Mack:2009mi,Penedones:2010ue,Rastelli:2017udc,Rastelli:2017ymc}).
Mellin space for defects has recently been introduced
in~\cite{Goncalves:2018fwx}.  Mellin space amplitudes are also interesting for
their flat space limit, and it would be interesting to study the analogous limit
for defects.

Finally, this work sets the basis for further explorations of the 2d dCFT
associated with $V$, and a natural goal for the future is to bootstrap this
correlator at the next order in $c$. This is complicated by the degeneracies at
large $N$ which need to be resolved by considering additional correlators (a
similar problem was recently studied in the context of Wilson
lines~\cite{Ferrero:2021bsb}). We hope to report on it in the near future.

\subsection*{Acknowledgements}

We would like to thank
Lorenzo Bianchi,
Gabriel Bliard,
Nadav Drukker,
Jean-Fran\c{c}ois Fortin,
Pedro Liendo,
Valentina Prilepina and
Pedro Vieira
for enlightening discussions.
MT gratefully acknowledges the support of the Institute for Theoretical and Mathematical
Physics (Lomonosov State University, Moscow) where this project began, and
the Simons Center for Geometry and Physics (Stony Brook University), New York
University, University of Parma and the University of Turin, 
where part of this project was realised.  Research at Perimeter
Institute is supported by the Government of Canada through the Department of
Innovation, Science and Economic Development and by the Province of Ontario
through the Ministry of Research and Innovation.
This work has been supported in part by Istituto Nazionale di Fisica Nucleare (INFN) through the “Gauge and String Theory” (GAST) research project.

\appendix

\section{Bulk channel blocks}
\label{sec:bulkblocks}

In this appendix we present the superconformal blocks $\cG$ arising
in the bulk channel decomposition of the correlator~\eqref{eqn:bulkblockdecomp}.
These superconformal blocks are given by combinations of conformal blocks
$g^{st}$ and $h$ as in~\eqref{eqn:confblockbulkch} and~\eqref{eqn:sblockD20},
and we begin by reviewing the derivation of the latter directly from the OPE in
the limit $z, \bar{z}, \omega \to 1$. The full conformal blocks can then be
recovered from the Casimir equation. Finally we present the complete list of
superconformal blocks that may contribute to the correlator.

\subsection{OPE and normalisation}
\label{sec:bulkOPE}

A straightforward (if cumbersome) approach to calculating the contribution to the correlator
corresponding to the exchange of a given bulk operator is to use the
bulk OPE of $\O2_k \O2_k$. Consider a bulk operator with weights $\Delta,\ell,R$
and nonzero 3-point function ($\ell,R$ are even)
\begin{equation}
\begin{aligned}
  \vev{\O2_k(x_1,y_1) \O2_k(x_2,y_2) \cO_{\Delta,\ell,R}(x_3,y_3,v)}
  =
  \lambda_{k k, \left\{ \Delta,\ell,R \right\}} 
  \frac{|y_{12}|^{2k-R}}{|x_{12}|^{4k-\Delta}}
  \frac{|y_{13}|^{R}}{|x_{13}|^{\Delta}}
  \frac{|y_{23}|^{R}}{|x_{23}|^{\Delta}}
  \frac{(2 x_{12} \cdot v)^\ell}{|x_{12}|^\ell}\,.
\end{aligned}
\end{equation}
The coordinate $y$ encodes the R-symmetry polarisation and is defined
in~\eqref{eqn:projectiveu}; $v$ is a polarisation vector for the spin.
If $\lambda$ is nonzero, the operator $\cO_{\Delta,\ell,R}$ appears in the OPE of
$\O2_k \O2_k$.  Expanding near $x_2 \to x_1$ and similarly for $y_2 \to y_1$
leads to the OPE
\begin{align}
  \O2_k(x_1,y_1) \O2_k(x_2,y_2) \supset
  \lambda_{kk,\left\{ \Delta,\ell,R \right\}}
  \frac{(y_{12}^2)^{\frac{2k-R}{2}}}{(x_{12}^2)^{\frac{4k-\Delta+\ell}{2}}}
  \left[ 1 + \dots \right]
  \cO_{\Delta,\ell,R}(x_1,y_1,x_{12})\,.
  \label{eqn:okope}
\end{align}
The ellipsis contains terms with derivatives in $x_1$ and $y_1$ and
suppressed in the coincident limit. Since $v=x_{12}$ does not satisfy the
condition $v^2 = 0$, to ensure consistency we rewrite $\cO|_{v = x_{12}}$ using
the Todorov operator $D_v$~\cite{Dobrev:1975ru,Dobrev:1977qv} (see
also~\cite{Costa:2011mg}) defined as
\begin{align}
  x \cdot D_v = \left(\frac{q-2}{2} + v \cdot \frac{d}{dv} \right) \left(x \cdot
  \frac{d}{dv} \right) - \frac{1}{2} (x \cdot v) \frac{d^2}{dv \cdot dv}\,,
  \qquad
  \text{here}\ q=6\,.
  \label{eqn:todorov}
\end{align}
This satisfies the identity (with $(n)_\ell \equiv \Gamma(n+\ell)/\Gamma(n)$ the
Pochhammer symbol)
\begin{align}
  \cO_{\Delta,\ell,R}(x_1,v,y_1)|_{v= x_{12}}
  =
  \frac{(2 x_{12} \cdot D_v)^\ell}{(\ell!)(2)_\ell}
  \cO_{\Delta,\ell,R}(x_1,v,y_1)\,,
\end{align}
and ensures tracelessness.  The operator $\cO_{\Delta,\ell,R}$ can have a nonzero
expectation value with $V$ whichs take the form
\begin{align}
  \vev{\cO_{\Delta,\ell,R}(x_1, v, y_1) V}
  =
  a_{\left\{ \Delta,\ell,R \right\}}
  \frac{|y_1^\perp|^R (x_1^\perp \cdot v)^\ell}{|x_1^\perp|^{\Delta+\ell}}\,.
  \label{eqn:obulkvev}
\end{align}
Taking the expectation value of~\eqref{eqn:okope} in the presence of $V$ and
plugging~\eqref{eqn:obulkvev} gives the contribution to the correlator
for the exchange of $\cO_{\Delta,\ell,R}$
\begin{align}
  \vev{ \O2_k(x_1,y_1) \O2_k(x_2,y_2) V}_{\cO_{\Delta,\ell,R}}
  = \lambda_{kk,\left\{ \Delta,\ell,R \right\}} a_{\left\{ \Delta,\ell,R \right\}}
  \frac{(y_{12}^2)^{\frac{2k-R}{2}}}{(x_{12}^2)^{\frac{4k-\Delta+\ell}{2}}}
  \left[ 1 + \dots \right]
  \frac{(2x_{12} \cdot D_v)^\ell}{\ell! (2)_\ell}
  \frac{|y_1^\perp|^R (x_1^\perp \cdot v)^\ell}{|x_1^\perp|^{\Delta+\ell}}\,.
\end{align}
The action of the Todorov operator on the 1-point function can be calculated
exactly using the identity~\cite{Costa:2011mg} ($C_\ell^{(\alpha)}$ are the
Gegenbauer polynomials)
\begin{align}
  (x \cdot D_v)^{\ell} (-2 v \cdot w)^{\ell}
  =
  (\ell!)^2 (x^2 w^2)^{\ell/2}
  C_{\ell}^{(\tfrac{q}{2}-1)} \left( \frac{x \cdot w}{(x^2 w^2)^{1/2}} \right)
  \label{eqn:todorovidentity}
\end{align}
In the present case this implies
\begin{align}
  \nonumber \frac{(2 x_{12} \cdot D_v)^\ell}{\ell! (2)_\ell (x_{12}^2)^{\ell/2}} \frac{(x_1^\perp \cdot
  v)^\ell}{|x_1^\perp|^\ell}
  &=
  \frac{1}{\ell+1}
  C_{\ell}^{(2)}\left( \frac{x_{12} \cdot x_1^\perp}{|x_{12}||x_1^\perp|}
  \right)\,.
\end{align}
In terms of the cross-ratios defined in~\eqref{eqn:crossratios}
we find to leading order
\begin{align}
  g_{\Delta,\ell,R}(z,\bar{z},\omega) =
  \frac{1}{\ell+1}
  \left|(1-z)(1-\bar{z}) \right|^{\frac{\Delta}{2}}
  |1-\omega|^{-R}
  C_{\ell}^{(2)}\left( \frac{2-z-\bar{z}}{2\sqrt{(1-z)(1-\bar{z})}} \right)
   + \dots
   \label{eqn:bulkchannelope}
\end{align}
with subleading terms suppressed in the limit $z,\bar{z},\omega \to 1$. In
particular in the lightcone limit $\bar{z} \to 1$ we recover
\begin{align}
  g_{\Delta,\ell,R}(z,\bar{z},\omega) =
  |1-z|^{\frac{\Delta+\ell}{2}}
  |1-\bar{z}|^{\frac{\Delta-\ell}{2}}
  |1-\omega|^{-R} + \dots\,.
  \label{eqn:bulkblocknormlc}
\end{align}

\subsection{Casimir equation}

A more convenient approach to compute the full conformal blocks is to use
that bulk operators exchanged in the bulk channel OPE transform in
representations of the 6d conformal group and $\sof(5)$ R-symmetry. These
properties can be shown to lead respectively to two constraints satisfied by the
conformal blocks $g_{\Delta,\ell,R}$ in the form of Casimir
equation~\cite{Dolan:2003hv,Billo:2016cpy}
\begin{align}
  \nonumber
  &2 \left[ (z-1)^2 z \partial_z^2
    + (z-1)\left( -(z+3) + \frac{2 z (z-1)}{z-\bar{z}}
    + \frac{2(z-1)}{z \bar{z} -1} \right) \partial_z
    + (z \leftrightarrow \bar{z})
  \right] g_{\Delta,\ell,R}(z,\bar{z},\omega)\\
  &\qquad = C_{2,6}(\Delta,\ell) g_{\Delta,\ell,R}(z,\bar{z},\omega)\,,
  \label{eqn:bulkspconfblockeqn}
\end{align}
and
\begin{align}
  -(1-\omega)^2 \left[ \omega \partial_\omega^2 + \partial_\omega
  \right] g_{\Delta,\ell,R}(z,\bar{z},\omega)
  =
  C_{2,3}(-R,0) g_{\Delta,\ell,R}(z,\bar{z},\omega)\,.
  \label{eqn:bulkrconfblockeqn}
\end{align}
In these equations, $C_{2,d}(\Delta,\ell)$ is the quadratic Casimir of $\sof(d)$
for the representation of dimension $\Delta$ and spin $\ell$
\begin{align}
  C_{2,d}(\Delta,\ell) = \Delta(\Delta-d)+ \ell (\ell+d-2),
\end{align}
while the differential operators on the left are the differential
representations of the same Casimir operators acting on the correlator.

These two equations are separated, and accordingly their solution is given in
terms of the product $g_{\Delta,\ell,R}(z,\bar{z},\omega) =
g^{st}_{\Delta,\ell}(z,\bar{z}) h_R(\omega)$.

The first equation~\eqref{eqn:bulkspconfblockeqn} was observed
in~\cite{Liendo:2019jpu} to reduce to the Casimir equation of the 4d conformal
blocks for 4-point functions of local operators $g^{(a,b)}_{\Delta,\ell}$, with $a
= b = 0$. Following their observation it is easy to verify that a solution to
the Casimir equation is given by
\begin{align}
  g^{st}_{\Delta,\ell} =
  \frac{(1-z)(1-\bar{z})}{1 - z \bar{z}}
  g^{(0,0)}_{\Delta-1,\ell+1}\left( 1-z, 1-\bar{z} \right)\,,
  \label{eqn:gstpartialsol}
\end{align}
with $g^{(a,b)}_{\Delta,\ell}$ given by~\cite{Dolan:2000ut}
\begin{align}
  \label{eqn:gab4d}
  g^{(a,b)}_{\Delta,\ell}(z,\bar{z}) &=
  \frac{z \bar{z}}{z - \bar{z}} \left( 
  k_{\Delta+\ell}(z) k_{\Delta-\ell-2}(\bar{z}) - k_{\Delta+\ell}(\bar{z})
  k_{\Delta-\ell-2}(z)\right)\,,\\
  k_{2h}(z) &= z^{h} {}_2F_1 \left( h-\frac{a}{2},
  h + \frac{b}{2}, 2h; z \right)\,.
  \label{eqn:kbdef}
\end{align}
This solution also has the correct asymptotics required by the OPE
analysis~\eqref{eqn:bulkchannelope}. To see this, expand the blocks first in
$\bar{z} \to 1$ and then $z \to 1$.

The second equation~\eqref{eqn:bulkrconfblockeqn} is closely related to the
equation for Legendre polynomials, and its solution is given by
\begin{align}
  h_R(\omega) &=
  \cN_R 
  \left( \frac{1-\omega}{1+\omega} \right)
  P_{R+1}\left( \frac{1+\omega}{1-\omega} \right)\,.
  \label{eqn:rsymblockpartial}
\end{align}
The normalisation factor $\cN_R$ can be fixed by expanding this block in the
limit $\omega \to 1$ and comparing to~\eqref{eqn:bulkchannelope}. Matching the
normalisation we find
\begin{align}
  \cN_R = \frac{R!(R+1)!}{(2R+1)!}\,.
\end{align}

The result~\eqref{eqn:rsymblockpartial} can be understood by realising the $\sof(5)$ R-symmetry as the
(complexified) conformal group in 3d. From the point of view of kinematics, $V$
specifies a plane (or boundary) in $\bR^3$, and operators in symmetric traceless
representations of rank $R$ give scalar operators with conformal dimension
$-R$. And indeed one can show that~\eqref{eqn:rsymblockpartial} matches the
3d boundary blocks obtained in~\cite{McAvity:1995zd}.

Assembling both results we obtain the blocks given
in~\eqref{eqn:confblockbulkch}.

\subsection{Table of superconformal blocks}
\label{sec:tabsblockbulk}

As reviewed in section~\ref{sec:bulkchannelexpansion}, conformal blocks assemble
in superconformal blocks describing the exchange of all the operators of a given
supermultiplet appearing in the OPE. Here we tabulate the conformal blocks
content of the various superconformal blocks relevant for our analysis.

The superblocks for $\cD[2,0]$ are given in~\eqref{eqn:sblockD20}. For $\cD[4,0]$,
$\cB[2,0]_\ell$ and $\cL[0,0]_{\Delta,\ell}$ they are respectively
\begin{equation}
  \begin{aligned}
    \cG_{\cD[4,0]} =&\ 
    g_{8,0,4}
    - \frac{25}{6237} g_{10,2,2}
    + \frac{1}{58212} g_{12,0,0}\,.
  \end{aligned}
  \label{eqn:sblockD40}
\end{equation}

\begin{equation}
  \begin{aligned}
    \cG_{\cB[2,0]_\ell} =&\ 
    g_{\ell+8,\ell,2}
    - \frac{3}{700} g_{\ell+10,\ell-2,0}
    - \frac{3}{700} g_{\ell+10,\ell+2,0}\\
    &- \frac{(\ell+2)(\ell+9)}{90 (2\ell+9)(2\ell+13)}
    g_{\ell+10,\ell+2,2}
    - g_{\ell+10,\ell+2,4}
    + \frac{3(\ell+4)(\ell+7)}{7000 (2\ell+9)(2\ell+13)}
    g_{\ell+12,\ell,0}\\
    &+\frac{3}{700}
    g_{\ell+12,\ell,2}
    + \frac{(\ell+6)^2(\ell+7)^2}{16 (2\ell+11)(2\ell+13)^2(2\ell+15)}
    g_{\ell+12,\ell+4,2}\\
    &- \frac{3(\ell+6)^2(\ell+7)^2}{11200 (2\ell+11)(2\ell+13)^2(2\ell+15)}
    g_{\ell+14,\ell+2,0}\,.
  \end{aligned}
  \label{eqn:sblockB20}
\end{equation}

\begin{equation}
  \begin{aligned}
    \cG_{\cL[0,0]_{\Delta,\ell=0}} =&\ 
    g_{\Delta,0,0}\\
    &- \frac{\Delta(\Delta+6)}{40(\Delta+1)(\Delta+5)}
    g_{\Delta+2,2,0}\\
    &- g_{\Delta+2,2,2}\\
    &+ \frac{9(\Delta-4)(\Delta-2)(\Delta+4)(\Delta+6)}{8960(\Delta-3)(\Delta-1)(\Delta+3)(\Delta+5)}
    g_{\Delta+4,0,0}\\
    &+ \frac{(\Delta-4)(\Delta+6)}{36 (\Delta-3)(\Delta+5)}
    g_{\Delta+4,0,2}\\
    &+ g_{\Delta+4,0,4}\\
    &+ \frac{(\Delta+4)^2(\Delta+6)^2}{256 (\Delta+3)(\Delta+5)^2(\Delta+7)}
    g_{\Delta+4,4,0}\\
    &- \frac{(\Delta-4) (\Delta+2)(\Delta+4)^2(\Delta+6)^2}{10240
    (\Delta-3)(\Delta+1)(\Delta+3)(\Delta+5)^2(\Delta+7)}
    g_{\Delta+6,2,0}\\
    &- \frac{(\Delta+4)^2(\Delta+6)^2}{256(\Delta+3)(\Delta+5)^2 (\Delta+7)}
    g_{\Delta+6,2,2}\\
    &+ \frac{\Delta^2 (\Delta+2)^2 (\Delta+4)^2
    (\Delta+6)^2}{2^{16}(\Delta-1)(\Delta+1)^2 (\Delta+3)^2 (\Delta+5)^2
    (\Delta+7)}
    g_{\Delta+8,0,0}\,.
  \end{aligned}
  \label{eqn:sblockL00}
\end{equation}

\begin{equation}
  \scalebox{0.8}{
    $
\begin{aligned}
  \cG_{\cL[0,0]_{\Delta,\ell}} =&\ 
  g_{\Delta,\ell,0}\\
  &- \frac{(\Delta-\ell-4)(\Delta-\ell-2)}{40 (\Delta-\ell-3)(\Delta-\ell-1)}
  g_{\Delta+2,\ell-2,0}\\
  &- g_{\Delta+2,\ell-2,2}\\
  &- \frac{(\Delta+\ell)(\Delta+\ell+6)}{40(\Delta+\ell+1)(\Delta+\ell+5)}
  g_{\Delta+2,\ell+2,0}\\
  &- g_{\Delta+2,\ell+2,2}\\
  &+ \frac{(\Delta-\ell)^2(\Delta-\ell+2)^2}{256
  (\Delta-\ell-1)(\Delta-\ell+1)^2(\Delta-\ell+3)}
  g_{\Delta+4,\ell+2,0}\\
  &+ \frac{11 \Delta^4 + 44 \Delta^3 - (22\ell^2 + 88 \ell + 124)\Delta^2 - 4(11 \ell^2 +
  44\ell + 84) \Delta + 11 \ell^4 + 88 \ell^3 + 140 \ell^2 - 144 (\ell+1)}{2240
  (\Delta-\ell-3)(\Delta-\ell+1)(\Delta+\ell+1)(\Delta+\ell+5)}
  g_{\Delta+4,\ell,0}\\
  &+ \frac{\Delta^4 + 4 \Delta^3 - (2 \ell^2 + 8\ell + 19)\Delta^2 - 2 (2
  \ell^2+8\ell+23)\Delta + \ell^4 + 8\ell^3 + 5\ell^2
  -44\ell-24}{36(\Delta-\ell-3)(\Delta-\ell+1)(\Delta+\ell+1)(\Delta+\ell+5)}
  g_{\Delta+4,\ell,2}\\
  &+g_{\Delta+4,\ell,4}\\
  & \frac{\Delta+\ell+4)^2(\Delta+\ell+6)^2}{256
  (\Delta+\ell+3)(\Delta+\ell+5)^2(\Delta+\ell+7)}
  g_{\Delta+4,\ell+4,0}\\
  &- \frac{(\Delta-\ell)^2(\Delta-\ell+2)^2 (\Delta+\ell)
  (\Delta+\ell+6)}{10240(\Delta-\ell-1)(\Delta-\ell+1)^2(\Delta-\ell+3)(\Delta+\ell+1)(\Delta+\ell+5)}
  g_{\Delta+6,\ell-2,0}\\
  &-\frac{(\Delta-\ell)^2(\Delta-\ell+2)^2}{256(\Delta-\ell-1)(\Delta-\ell+1)^2(\Delta-\ell+3)}
  g_{\Delta+6,\ell-2,2}\\
  &-\frac{(\Delta-\ell-4)(\Delta-\ell+2)(\Delta+\ell+4)^2(\Delta+\ell+6)^2}{10240(\Delta-\ell-3)(\Delta-\ell+1)(\Delta+\ell+3)(\Delta+\ell+5)^2(\Delta+\ell+7)}
  g_{\Delta+6,\ell+2,0}\\
  &-
  \frac{(\Delta+\ell+4)^2(\Delta+\ell+6)^2}{256(\Delta+\ell+3)(\Delta+\ell+5)^2(\Delta+\ell+7)}
  g_{\Delta+6,\ell+2,2}\\
  &+\frac{(\Delta-\ell)^2(\Delta-\ell+2)^2(\Delta+\ell+4+)^2(\Delta+\ell+6)^2}{2^{16}(\Delta-\ell-1)(\Delta-\ell+1)^2(\Delta-\ell+3)(\Delta+\ell+3)(\Delta+\ell+5)^2(\Delta+\ell+7)}
  g_{\Delta+8,\ell,0}\,.
\end{aligned}
$
}
  \label{eqn:sblockL00l}
\end{equation}

\section{Defect channel blocks}
\label{sec:defectblocks}

In this appendix we detail the derivation of the defect channel superconformal
blocks. We follow the same strategy as for the bulk channel superconformal
blocks presented in appendix~\ref{sec:bulkblocks}. We start by reviewing the
calculation of the leading behavior of the conformal blocks from the OPE. This fixes
the normalisation and asymptotics of the conformal blocks. We then proceed to
recover the full blocks by using the Casimir equation. Finally we tabulate the
relevant superconformal blocks.

\subsection{OPE and normalisation}

In the presence of the defect $V$, bulk operators admit a defect operator
expansion (dOE)~\eqref{eqn:doe}, which we reproduce here for convenience
\begin{align}
  \O2_k(x,y) V =
  \sum_{\{ \hat{\Delta},s,r \}} b_{k\{\hat{\Delta},s,r\}}
  \frac{|y^\perp|^{k-r}}{|x^\perp|^{2k-\hat{\Delta}}}
  C_{k,\{ \hat{\Delta},s,r \}}
  V[\hat{\cO}_{\hat{\Delta},s,r}(x^\parallel, y^\parallel,v)]\,.
\end{align}
The coefficients $b_{k\{\hat{\Delta},s,r\}}$ are the dCFT data entering the
correlators~\eqref{eqn:bulkdefect2pts}. $C_{k,\{ \hat{\Delta},s,r \}}$ are
differential operators acting on $\hat{\cO}$ and encoding the contributions of
the descendants; they can be calculated by requiring the dOE to reproduce the
correlators~\eqref{eqn:bulkdefect2pts} and to leading order are given by
\begin{align}
  C_{k,\left\{ \hat{\Delta},s,r \right\}}|_1 = (1 + \dots) \frac{(x_1^\perp \cdot
  D_v)^{s}}{|x_1^\perp|^{s}(s!)^2}\,,
  \label{eqn:diffopdefect}
\end{align}
with terms suppressed by $|x_1^\perp|/|x_{12}^\parallel|$, and similarly for $y$.
$D_v$ is the Todorov operator defined in~\eqref{eqn:todorov} (here with $q = 4$).

From the dOE we can directly evaluate the leading contribution to $\cF$ due to the
exchange of a defect operator of weights $\hat{\Delta}, s, r$. Acting twice
on~\eqref{eqn:2pointfunction} and taking the expectation value leads
to an expression for the conformal block $\hat{g}_{\hat{\Delta},s,r}$ of the
form
\begin{align}
  \hat{g}_{\hat{\Delta},s,r} =
  \frac{|x_1^\perp|^{\hat{\Delta}}|x_2^\perp|^{\hat{\Delta}}}{|y_1^\perp|^{r} |y_2^\perp|^{r}}
  C|_1 C|_2
  \vev{V[\hat{\cO}_{\hat{\Delta},s,r}(x_1^\parallel,y_1^\parallel,v_1)
  \hat{\cO}_{\hat{\Delta},s,r}(x_2^\parallel,y_2^\parallel,v_2)]}\,,
\end{align}
where $C|_i$ are the differential operators~\eqref{eqn:diffopdefect} acting on the point $i$.
The action of $D_v$ on the 2-point function is easy to evaluate from the
identity~\eqref{eqn:todorovidentity}, and to leading order we get
\begin{align}
  \hat{g}_{\hat{\Delta},s,r} = 
  \left( \frac{|x_1^\perp||x_2^\perp|}{(x_{12}^\parallel)^2}
  \right)^{\hat{\Delta}}
  \left( \frac{-(y_{12}^\parallel)^2}{|y_1^\perp||y_2^\perp|} \right)^{r}
  U_{s}\left( \frac{x_1^\perp \cdot
  x_2^\perp}{|x_1^\perp||x_2^\perp|} \right) + \dots
  \label{eqn:confblocksasymptotics}
\end{align}
with $U_s(x)$ the Chebyshev polynomials of second kind.
This is a function of the cross-ratios introduced in~\eqref{eqn:crossratios} and
provides the normalisation for the conformal blocks.

\subsection{Casimir equation}
\label{sec:scblocks}

We can derive the full defect channel conformal blocks from a Casimir equation
following~\cite{Dolan:2003hv,Billo:2016cpy}. The defect operators exchanged
in~\eqref{eqn:2pointfunction} transform in the 2d (global) conformal group,
along with $\sof(4)$ rotations tranverse to the plane and $\sof(4)$ residual
R-symmetry. For each of these symmetries there is a corresponding Casimir
equation. These take a nice form in terms of the cross-ratios
\begin{equation}
  \begin{gathered}
  \chi = \frac{(x_{12}^\parallel)^2 + |x_1^\perp|^2 + |x_2^\perp|^2}
  {2 |x_1^\perp| |x_2^\perp|}
  = \frac{1 + z \bar{z}}{2 \sqrt{z \bar{z}}}\,, \qquad
  \cos{\phi} = \frac{x_1^\perp \cdot x_2^\perp}{|x_1^\perp| |x_2^\perp|}
  = \frac{z + \bar{z}}{2 \sqrt{z\bar{z}}}\,,\\
  \psi = \frac{(y_{12}^\parallel)^2 + |y_1^\perp|^2 + |y_2^\perp|^2}
  {2|y_1^\perp| |y_2^\perp|}
  = \frac{\omega + \omega^{-1}}{2}\,.
  \end{gathered}
\end{equation}
They read respectively
\begin{align}
  \left[
    (1-\chi^2) \frac{\partial^2}{\partial \chi^2}
    - 3 \chi \frac{\partial}{\partial \chi}
    + 4\hat{\Delta}(\hat{\Delta}-1)
  \right]
  \hat{g}_{\hat{\Delta},s,r}(\chi,\cos\phi,\psi) &= 0\,,\\
  \left[
    \sin^2{\phi} \frac{\partial^2}{\partial \cos\phi^2}
    - 3 \cos\phi \frac{\partial}{\partial \cos\phi}
    + s(s+2)
  \right]
  \hat{g}_{\hat{\Delta},s,r}(\chi,\cos\phi,\psi) &= 0\,,\\
  \left[
    (1-\psi^2) \frac{\partial^2}{\partial \psi^2}
    - 3 \psi \frac{\partial}{\partial \psi}
    + r(r+2)
  \right]
  \hat{g}_{\hat{\Delta},s,r}(\chi,\cos\phi,\psi) &= 0\,.
  \label{eqn:casimireqns}
\end{align}
The equations are separated and can be solved straightforwardly.
Picking the solution with the right asymptotics, we find respectively
\begin{align}
  \chi^{-\hat{\Delta}} {}_2F_1\left(
  \frac{\hat{\Delta}+1}{2},\frac{\hat{\Delta}}{2};\hat{\Delta};\chi^{-2} \right)\,,
  \qquad
  U_{s}(\cos\phi)\,, \qquad
  U_{r}(\psi)\,.
\end{align}
We note that the solution for $\chi$ matches the conformal block found
in~\cite{Billo:2016cpy} for a plane, and the solution for $\psi$ matches the
boundary block for $\Delta = -r$ and $d=3$ found in~\cite{McAvity:1995zd}.

The conformal block $\hat{g}_{\hat{\Delta},s,r}$ is given by the product of
these 3 blocks, up to an overall normalisation factor.
Taking the limit $\chi \to \infty$ and $\psi \to \infty$ and matching with the
OPE result~\eqref{eqn:confblocksasymptotics} we can fix the normalisation of the
blocks to get
\begin{align}
  \hat{g}_{\hat{\Delta},s,r}(z,\bar{z},\omega) =
  \left( 2\chi \right)^{-\hat{\Delta}}
  {}_2F_1\left( \frac{\hat{\Delta}+1}{2}, \frac{\hat{\Delta}}{2};\hat{\Delta};
  \chi^{-2} \right)
  U_s\left( \cos(\phi) \right)
  U_{r}\left( -\psi \right)\,.
\end{align}
Expressing the cross-ratios in terms of $z, \bar{z}, \omega$ and using the
identity
\begin{align}
  U_r\left( \frac{\omega + \omega^{-1}}{2} \right) = \frac{\omega^{r+1}- \omega^{-(r+1)}}{\omega - \omega^{-1}}\,.
  \label{eqn:chebushevid}
\end{align}
we find the conformal blocks~\eqref{eqn:defectchannelconfbloc}.

\subsection{Table of superconformal blocks}
\label{sec:defectblocktable}

Finally we tabulate the various superconformal blocks appearing in the OPE. Here
we list the conformal blocks content of each superconformal blocks.

The short multiplets $B[r]$ (for $r = 1,2$) have superconformal blocks
\begin{align}
  \hat{\cG}_{B[r]} = 
  \hat{g}_{2r,0,r}
  +\hat{g}_{2r+1,1,r-1}
  + \hat{g}_{2r+2,0,r-2}\,.
  \label{eqn:sblockBr}
\end{align}
Note that for $B[0]$ is the defect identity. The special case $B[1]$ is the
displacement operator supermultiplet and has a shortened superconformal block.
It contains only the blocks $g_{2,0,1}$ and $g_{3,1,0}$, and correspondingly we
can check that the last term in the ansatz above vanishes identically.

The semishort multiplets $A[r]_s$ (defined for $r=0,1$) respectively have superblocks
\begin{align}
  \hat\cG_{A[0]_s} &=
  \hat{g}_{2+s,s,0} + \hat{g}_{3+s,s+1,1} + \hat{g}_{4+s,s+2,0}\,,
  \label{eqn:sblockA0}
\end{align}
and
\begin{align}
  \hat{\cG}_{A[1]_s} &=
  \hat{g}_{4+s,s,1} + \hat{g}_{5+s,s-1,0} + \hat{g}_{5+s,s+1,0} +
  \hat{g}_{5+s,s+1,2} +
  \hat{g}_{6+s,s,1} + \hat{g}_{6+s,s+2,1}
  +\hat{g}_{7+s,s+1,0}\,.
  \label{eqn:sblockA1}
\end{align}
Again for the special case $s=0$ some of these conformal blocks vanish
identically, which reflects the shortening of the multiplet.

Finally the long supermultiplet have superconformal blocks
\begin{equation}
\begin{aligned}
  \hat\cG_{L[s,0]_{\hat{\Delta}}} &=
  \hat{g}_{\hat{\Delta},s,0}
  + \hat{g}_{\hat{\Delta}+1,s+1,1}
  + \hat{g}_{\hat{\Delta}+1,s-1,1}
  + \hat{g}_{\hat{\Delta}+2,s-2,0}
  + \hat{g}_{\hat{\Delta}+2,s,0}\\
  &\phantom{=}\ + \hat{g}_{\hat{\Delta}+2,s+2,0}
  + \hat{g}_{\hat{\Delta}+2,s,2}
  +\hat{g}_{\hat{\Delta}+3,s+1,1}
  + \hat{g}_{\hat{\Delta}+3,s-1,1}
  + \hat{g}_{\hat{\Delta}+4,s,0}\,.
  \label{eqn:sblockL}
\end{aligned}
\end{equation}
Notice that the blocks $\hat{g}$ appear, for our normalization, all with unit coefficient. 

\section{Calculation of $a_2$}
\label{sec:calca2}

We can calculate the coefficient $a_2$ appearing in~\eqref{eqn:1pts} by relating
it to the anomaly coefficients $c,d$. The transformation rules for operators of
the stress tensor multiplet are given in~\cite{Drukker:2020atp}
\begin{align}
\label{eqn:susyTmult}
  \delta T^{\mu\nu} =&\ 
  \frac{1}{2} \varepsilon \gamma^{\rho (\mu} \partial_\rho J^{\nu)}\,,\nonumber\\
  \delta J^\mu =&\ 
  2 \varepsilon \gamma_{\nu} T^{\mu\nu}
  + \frac{2 \alpha_2}{5 \alpha_3} \left( 6 \eta^{\rho \mu}
  \left( \gamma^{\nu \sigma \lambda} + 3 \eta^{\sigma \nu} \gamma^\lambda
  \right)
  - \eta^{\mu\nu} \gamma^{\rho \sigma \lambda} \right)
  \check{\gamma}_I 
  \partial_\nu H_{\rho \sigma \lambda}^I\nonumber\\
  &+\frac{1}{10} \varepsilon
  \left( \gamma^{\mu\nu\rho} -4 \eta^{\mu \rho} \gamma^{\nu} \right)
  \check{\gamma}^{IJ}
  \partial_{\nu} j_{\rho IJ}\,,\nonumber\\
  \delta j^\mu_{IJ} =&\ 
  {-\frac{1}{2}} \varepsilon \check{\gamma}_{IJ} J^\mu
  + \frac{1}{5 \alpha_3} \varepsilon \gamma^{\mu\nu}
  \partial_\nu \check{\gamma}_{[I} \chi_{J]}\,,\nonumber\\
  \delta H^I_{\mu\nu\rho} =&\ 
  \frac{\alpha_3}{8 \alpha_2} \varepsilon \check{\gamma}^I \gamma_{[\mu\nu} J_{\rho]}
  + \frac{1}{120 \alpha_2} \varepsilon \gamma_\sigma \bar{\gamma}_{\mu\nu\rho}
  \partial^\sigma \chi^I\,,\nonumber\\
  \delta \chi^I =&\ 
  \alpha_2 \varepsilon \gamma^{\mu\nu\rho}
  \left( \check{\gamma}^{IJ} + 4 \delta^{IJ} \right)
  H^J_{\mu\nu\rho} 
  + \alpha_3 \varepsilon \gamma_\mu
  \left( \check{\gamma}^{IJK} + 3 \delta^{IJ} \check{\gamma}^{K} \right)
  j^{\mu}_{JK} \nonumber\\
  &+ \frac{1}{\alpha_1} \varepsilon \gamma^\mu \check{\gamma}^{J}
  \partial_\mu O^{IJ}\,,\nonumber\\
  \delta O^{IJ} =&\ 
  \alpha_1 \varepsilon \check{\gamma}^{(I} \chi^{J)}\,.
  \\[-5mm]\nonumber
\end{align}

The constants $\alpha_1, \alpha_2, \alpha_3$ are arbitrary constants that can
absorbed in the normalisation for the operators $\cO, \chi, H$.

The coefficient of the 1-point function of $T^{\mu\nu}$ in the presence of $V$ is
known to be related to the anomaly coefficient
$d$~\cite{Lewkowycz:2014jia,bianchi:2015liz,Drukker:2020atp}. Acting with
supersymmetry, we find that the 1-point function of the superprimary $\O2$ has
coefficient~\cite{Drukker:2020atp}
\begin{align}
  a_2 = \frac{5 \alpha_1 \alpha_3 d}{4 \pi^3}\,.
  \label{eqn:a2fromd}
\end{align}

We can fix the coefficients $\alpha_1 \alpha_3$ by requiring $\O2$ to be normalised as
in~\eqref{eqn:2pts}. A simple way to do so is to compare the 2-point function of
$\O2$ with the 2-point function of R-symmetry currents $j^\mu$, which is fixed
by conformal symmetry to take the form
\begin{align}
  \vev{j^{\mu ij}(x_1) j^{\nu kl}(x_2)}
  =
  C_j \left( \delta^{ik} \delta^{jl} - \delta^{il} \delta^{jk}
  \right)\frac{I_{\mu\nu}(x_{12})}{|x_{12}|^{10}}\,,
  \qquad
  I_{\mu \nu}(x) =
  \delta_{\mu\nu} - 2 \frac{x_\mu x_\nu}{x^2}\,.
\end{align}
The constant $C_j$ is related to the anomaly coefficient
$c$ as~\cite{Manvelyan:2000ef}
\begin{align}
  C_j = \frac{5 c}{2 \pi^6}\,.
\end{align}
Acting twice with supersymmetry, a short calculation shows that
\begin{align}
  \vev{\O2(x_1,u_1) \O2(x_2,u_2)}
  =
  \frac{25 \alpha_1^2 \alpha_3^2 c}{32\pi^6} \frac{(2 u_1 \cdot
    u_{2})^2}{|x_{12}|^8}\,.
\end{align}
Matching with the normalisation of $\O2$~\eqref{eqn:2pts} fixes $(\alpha_1
\alpha_3)^2$, and we should take the negative branch to match the supergravity calculation
\begin{align}
  \alpha_1 \alpha_3 = -\frac{4 \sqrt{2} \pi^3}{5 \sqrt{c}}\,.
\end{align}
Plugging back into~\eqref{eqn:a2fromd} gives the result~\eqref{eqn:a2}.

\bibliographystyle{utphys2}
\bibliography{ref}

\providecommand{\href}[2]{#2}\begingroup\raggedright\begin{thebibliography}{100}\setlength{\parskip}{1pt}\setlength{\itemsep}{0pt
  plus 0.3ex}

\bibitem{maldacena:1997re}
J.~M. Maldacena, ``{The large $N$ limit of superconformal field theories and
  supergravity},'' \href{http://dx.doi.org/10.1023/A:1026654312961,
  10.4310/ATMP.1998.v2.n2.a1}{{\em Int. J. Theor. Phys.} {\bfseries 38} (1999)
  1113--1133}, \href{http://arxiv.org/abs/hep-th/9711200}{{\ttfamily
  hep-th/9711200}}.

\bibitem{Witten:1998xy}
E.~Witten, ``{Baryons and branes in anti-de Sitter space},''
  \href{http://dx.doi.org/10.1088/1126-6708/1998/07/006}{{\em JHEP} {\bfseries
  07} (1998) 006}, \href{http://arxiv.org/abs/hep-th/9805112}{{\ttfamily
  arXiv:hep-th/9805112}}.

\bibitem{Aharony:1998rm}
O.~Aharony, Y.~Oz, and Z.~Yin, ``{M-theory on $AdS_p \times S^{(11-p)}$ and
  superconformal field theories},''
  \href{http://dx.doi.org/10.1016/S0370-2693(98)00508-5}{{\em Phys. Lett. B}
  {\bfseries 430} (1998) 87--93},
  \href{http://arxiv.org/abs/hep-th/9803051}{{\ttfamily arXiv:hep-th/9803051}}.

\bibitem{Ferrara:1973yt}
S.~Ferrara, A.~Grillo, and R.~Gatto, ``{Tensor representations of conformal
  algebra and conformally covariant operator product expansion},''
  \href{http://dx.doi.org/10.1016/0003-4916(73)90446-6}{{\em Annals Phys.}
  {\bfseries 76} (1973) 161--188}.

\bibitem{Ferrara:1973vz}
S.~Ferrara, A.~Grillo, G.~Parisi, and R.~Gatto, ``{Covariant expansion of the
  conformal four-point function},''
  \href{http://dx.doi.org/10.1016/0550-3213(73)90467-7}{{\em Nucl. Phys. B}
  {\bfseries 49} (1972) 77--98}. [Erratum: Nucl.Phys.B 53, 643--643 (1973)].

\bibitem{Polyakov:1974gs}
A.~Polyakov, ``{Nonhamiltonian approach to conformal quantum field theory},''
  {\em Zh. Eksp. Teor. Fiz.} {\bfseries 66} (1974) 23--42.

\bibitem{Rattazzi:2008pe}
R.~Rattazzi, V.~S. Rychkov, E.~Tonni, and A.~Vichi, ``{Bounding scalar operator
  dimensions in 4D CFT},''
  \href{http://dx.doi.org/10.1088/1126-6708/2008/12/031}{{\em JHEP} {\bfseries
  12} (2008) 031}, \href{http://arxiv.org/abs/0807.0004}{{\ttfamily
  arXiv:0807.0004}}.

\bibitem{Beem:2013sza}
C.~Beem, M.~Lemos, P.~Liendo, W.~Peelaers, L.~Rastelli, and B.~C. van Rees,
  ``{Infinite chiral symmetry in four dimensions},''
  \href{http://dx.doi.org/10.1007/s00220-014-2272-x}{{\em Commun. Math. Phys.}
  {\bfseries 336} no.~3, (2015) 1359--1433},
  \href{http://arxiv.org/abs/1312.5344}{{\ttfamily arXiv:1312.5344}}.

\bibitem{Beem:2014kka}
C.~Beem, L.~Rastelli, and B.~C. van Rees, ``{$ \mathcal{W} $ symmetry in six
  dimensions},'' \href{http://dx.doi.org/10.1007/JHEP05(2015)017}{{\em JHEP}
  {\bfseries 05} (2015) 017}, \href{http://arxiv.org/abs/1404.1079}{{\ttfamily
  arXiv:1404.1079}}.

\bibitem{Chester:2018dga}
S.~M. Chester and E.~Perlmutter, ``{M-theory reconstruction from (2,0) CFT and
  the chiral algebra conjecture},''
  \href{http://dx.doi.org/10.1007/JHEP08(2018)116}{{\em JHEP} {\bfseries 08}
  (2018) 116}, \href{http://arxiv.org/abs/1805.00892}{{\ttfamily
  arXiv:1805.00892}}.

\bibitem{Beem:2015aoa}
C.~Beem, M.~Lemos, L.~Rastelli, and B.~C. van Rees, ``{The $(2, 0)$
  superconformal bootstrap},''
  \href{http://dx.doi.org/10.1103/PhysRevD.93.025016}{{\em Phys. Rev. D}
  {\bfseries 93} no.~2, (2016) 025016},
  \href{http://arxiv.org/abs/1507.05637}{{\ttfamily arXiv:1507.05637}}.

\bibitem{Rastelli:2017ymc}
L.~Rastelli and X.~Zhou, ``{Holographic four-point functions in the (2, 0)
  theory},'' \href{http://dx.doi.org/10.1007/JHEP06(2018)087}{{\em JHEP}
  {\bfseries 06} (2018) 087}, \href{http://arxiv.org/abs/1712.02788}{{\ttfamily
  arXiv:1712.02788}}.

\bibitem{Zhou:2017zaw}
X.~Zhou, ``{On superconformal four-point Mellin amplitudes in dimension
  $d>2$},'' \href{http://dx.doi.org/10.1007/JHEP08(2018)187}{{\em JHEP}
  {\bfseries 08} (2018) 187}, \href{http://arxiv.org/abs/1712.02800}{{\ttfamily
  arXiv:1712.02800}}.

\bibitem{Heslop:2017sco}
P.~Heslop and A.~E. Lipstein, ``{M-theory beyond the supergravity
  approximation},'' \href{http://dx.doi.org/10.1007/JHEP02(2018)004}{{\em JHEP}
  {\bfseries 02} (2018) 004}, \href{http://arxiv.org/abs/1712.08570}{{\ttfamily
  arXiv:1712.08570}}.

\bibitem{Alday:2020tgi}
L.~F. Alday, S.~M. Chester, and H.~Raj, ``{6d (2,0) and M-theory at 1-loop},''
  \href{http://dx.doi.org/10.1007/JHEP01(2021)133}{{\em JHEP} {\bfseries 01}
  (2021) 133}, \href{http://arxiv.org/abs/2005.07175}{{\ttfamily
  arXiv:2005.07175}}.

\bibitem{Arutyunov:2002ff}
G.~Arutyunov and E.~Sokatchev, ``{Implications of superconformal symmetry for
  interacting (2,0) tensor multiplets},''
  \href{http://dx.doi.org/10.1016/S0550-3213(02)00359-0}{{\em Nucl. Phys. B}
  {\bfseries 635} (2002) 3--32},
  \href{http://arxiv.org/abs/hep-th/0201145}{{\ttfamily arXiv:hep-th/0201145}}.

\bibitem{Heslop:2004du}
P.~J. Heslop, ``{Aspects of superconformal field theories in six dimensions},''
  \href{http://dx.doi.org/10.1088/1126-6708/2004/07/056}{{\em JHEP} {\bfseries
  07} (2004) 056}, \href{http://arxiv.org/abs/hep-th/0405245}{{\ttfamily
  arXiv:hep-th/0405245}}.

\bibitem{Alday:2020lbp}
L.~F. Alday and X.~Zhou, ``{All tree-level correlators for M-theory on $AdS_7
  \times S^4$},'' \href{http://dx.doi.org/10.1103/PhysRevLett.125.131604}{{\em
  Phys. Rev. Lett.} {\bfseries 125} no.~13, (2020) 131604},
  \href{http://arxiv.org/abs/2006.06653}{{\ttfamily arXiv:2006.06653}}.

\bibitem{Lemos:2021azv}
M.~Lemos, B.~C. van Rees, and X.~Zhao, ``{Regge trajectories for the (2, 0)
  theories},'' \href{http://dx.doi.org/10.1007/JHEP01(2022)022}{{\em JHEP}
  {\bfseries 01} (2022) 022}, \href{http://arxiv.org/abs/2105.13361}{{\ttfamily
  arXiv:2105.13361}}.

\bibitem{Kantor:2022epi}
G.~K\'antor, V.~Niarchos, C.~Papageorgakis, and P.~Richmond, ``{$6D$ (2,0)
  bootstrap with soft-Actor-Critic},''
  \href{http://arxiv.org/abs/2209.02801}{{\ttfamily arXiv:2209.02801}}.

\bibitem{Penedones:2010ue}
J.~Penedones, ``{Writing CFT correlation functions as $AdS$ scattering
  amplitudes},'' \href{http://dx.doi.org/10.1007/JHEP03(2011)025}{{\em JHEP}
  {\bfseries 03} (2011) 025}, \href{http://arxiv.org/abs/1011.1485}{{\ttfamily
  arXiv:1011.1485}}.

\bibitem{Witten:1995zh}
E.~Witten, ``{Some comments on string dynamics},'' in {\em {STRINGS 95: Future
  Perspectives in String Theory}}, pp.~501--523.
\newblock 7, 1995.
\newblock \href{http://arxiv.org/abs/hep-th/9507121}{{\ttfamily
  arXiv:hep-th/9507121}}.

\bibitem{Strominger:1995ac}
A.~Strominger, ``{Open $p$-branes},''
  \href{http://dx.doi.org/10.1016/0370-2693(96)00712-5}{{\em Phys. Lett. B}
  {\bfseries 383} (1996) 44--47},
  \href{http://arxiv.org/abs/hep-th/9512059}{{\ttfamily arXiv:hep-th/9512059}}.

\bibitem{ganor:1996nf}
O.~J. Ganor, ``{Six-dimensional tensionless strings in the large $N$ limit},''
  \href{http://dx.doi.org/10.1016/S0550-3213(96)00702-X}{{\em Nucl. Phys.}
  {\bfseries B489} (1997) 95--121},
  \href{http://arxiv.org/abs/hep-th/9605201}{{\ttfamily hep-th/9605201}}.

\bibitem{Howe:1997ue}
P.~S. Howe, N.~D. Lambert, and P.~C. West, ``{The selfdual string soliton},''
  \href{http://dx.doi.org/10.1016/S0550-3213(97)00750-5}{{\em Nucl. Phys. B}
  {\bfseries 515} (1998) 203--216},
  \href{http://arxiv.org/abs/hep-th/9709014}{{\ttfamily arXiv:hep-th/9709014}}.

\bibitem{Gaiotto:2014kfa}
D.~Gaiotto, A.~Kapustin, N.~Seiberg, and B.~Willett, ``{Generalized global
  symmetries},'' \href{http://dx.doi.org/10.1007/JHEP02(2015)172}{{\em JHEP}
  {\bfseries 02} (2015) 172}, \href{http://arxiv.org/abs/1412.5148}{{\ttfamily
  arXiv:1412.5148}}.

\bibitem{DelZotto:2015isa}
M.~Del~Zotto, J.~J. Heckman, D.~S. Park, and T.~Rudelius, ``{On the defect
  group of a 6d SCFT},''
  \href{http://dx.doi.org/10.1007/s11005-016-0839-5}{{\em Lett. Math. Phys.}
  {\bfseries 106} no.~6, (2016) 765--786},
  \href{http://arxiv.org/abs/1503.04806}{{\ttfamily arXiv:1503.04806}}.

\bibitem{Bhardwaj:2020phs}
L.~Bhardwaj and S.~Sch\"afer-Nameki, ``{Higher-form symmetries of 6d and 5d
  theories},'' \href{http://dx.doi.org/10.1007/JHEP02(2021)159}{{\em JHEP}
  {\bfseries 02} (2021) 159}, \href{http://arxiv.org/abs/2008.09600}{{\ttfamily
  arXiv:2008.09600}}.

\bibitem{Apruzzi:2021mlh}
F.~Apruzzi, L.~Bhardwaj, D.~S.~W. Gould, and S.~Schafer-Nameki, ``{2-Group
  symmetries and their classification in 6d},''
  \href{http://dx.doi.org/10.21468/SciPostPhys.12.3.098}{{\em SciPost Phys.}
  {\bfseries 12} no.~3, (2022) 098},
  \href{http://arxiv.org/abs/2110.14647}{{\ttfamily arXiv:2110.14647}}.

\bibitem{maldacena:1998im}
J.~M. Maldacena, ``{Wilson loops in large $N$ field theories},''
  \href{http://dx.doi.org/10.1103/PhysRevLett.80.4859}{{\em Phys. Rev. Lett.}
  {\bfseries 80} (1998) 4859--4862},
  \href{http://arxiv.org/abs/hep-th/9803002}{{\ttfamily hep-th/9803002}}.

\bibitem{Drukker:2021vyx}
N.~Drukker and M.~Trepanier, ``{M2-doughnuts},''
  \href{http://dx.doi.org/10.1007/JHEP02(2022)071}{{\em JHEP} {\bfseries 02}
  (2022) 071}, \href{http://arxiv.org/abs/2111.09385}{{\ttfamily
  arXiv:2111.09385}}.

\bibitem{Drukker:2022beq}
N.~Drukker and M.~Tr\'epanier, ``{Ironing out the crease},''
  \href{http://dx.doi.org/10.1007/JHEP08(2022)193}{{\em JHEP} {\bfseries 08}
  (2022) 193}, \href{http://arxiv.org/abs/2204.12627}{{\ttfamily
  arXiv:2204.12627}}.

\bibitem{DHoker:2008rje}
E.~D'Hoker, J.~Estes, M.~Gutperle, and D.~Krym, ``{Exact half-BPS flux
  solutions in M-theory II: Global solutions asymptotic to $AdS_7\times
  S^4$},'' \href{http://dx.doi.org/10.1088/1126-6708/2008/12/044}{{\em JHEP}
  {\bfseries 12} (2008) 044},
\href{http://arxiv.org/abs/0810.4647}{{\ttfamily arXiv:0810.4647}}.

\bibitem{bachas:2013vza}
C.~Bachas, E.~D'Hoker, J.~Estes, and D.~Krym, ``{M-theory solutions invariant
  under $D(2,1;\gamma) \oplus D(2,1;\gamma)$},''
  \href{http://dx.doi.org/10.1002/prop.201300039}{{\em Fortsch. Phys.}
  {\bfseries 62} (2014) 207--254},
\href{http://arxiv.org/abs/1312.5477}{{\ttfamily arXiv:1312.5477}}.

\bibitem{Liendo:2012hy}
P.~Liendo, L.~Rastelli, and B.~C. van Rees, ``{The bootstrap program for
  boundary CFT$_d$},'' \href{http://dx.doi.org/10.1007/JHEP07(2013)113}{{\em
  JHEP} {\bfseries 07} (2013) 113},
  \href{http://arxiv.org/abs/1210.4258}{{\ttfamily arXiv:1210.4258}}.

\bibitem{Billo:2016cpy}
M.~Bill\`o, V.~Gon\c{c}alves, E.~Lauria, and M.~Meineri, ``{Defects in
  conformal field theory},''
  \href{http://dx.doi.org/10.1007/JHEP04(2016)091}{{\em JHEP} {\bfseries 04}
  (2016) 091}, \href{http://arxiv.org/abs/1601.02883}{{\ttfamily
  arXiv:1601.02883}}.

\bibitem{Liendo:2016ymz}
P.~Liendo and C.~Meneghelli, ``{Bootstrap equations for $ \mathcal{N} $ = 4 SYM
  with defects},'' \href{http://dx.doi.org/10.1007/JHEP01(2017)122}{{\em JHEP}
  {\bfseries 01} (2017) 122}, \href{http://arxiv.org/abs/1608.05126}{{\ttfamily
  arXiv:1608.05126}}.

\bibitem{Lemos:2017vnx}
M.~Lemos, P.~Liendo, M.~Meineri, and S.~Sarkar, ``{Universality at large
  transverse spin in defect CFT},''
  \href{http://dx.doi.org/10.1007/JHEP09(2018)091}{{\em JHEP} {\bfseries 09}
  (2018) 091}, \href{http://arxiv.org/abs/1712.08185}{{\ttfamily
  arXiv:1712.08185}}.

\bibitem{Caron-Huot:2017vep}
S.~Caron-Huot, ``{Analyticity in spin in conformal theories},''
  \href{http://dx.doi.org/10.1007/JHEP09(2017)078}{{\em JHEP} {\bfseries 09}
  (2017) 078}, \href{http://arxiv.org/abs/1703.00278}{{\ttfamily
  arXiv:1703.00278}}.

\bibitem{Barrat:2021yvp}
J.~Barrat, A.~Gimenez-Grau, and P.~Liendo, ``{Bootstrapping holographic defect
  correlators in $ \mathcal{N} $ = 4 super Yang-Mills},''
  \href{http://dx.doi.org/10.1007/JHEP04(2022)093}{{\em JHEP} {\bfseries 04}
  (2022) 093}, \href{http://arxiv.org/abs/2108.13432}{{\ttfamily
  arXiv:2108.13432}}.

\bibitem{Gimenez-Grau:2021wiv}
A.~Gimenez-Grau and P.~Liendo, ``{Bootstrapping monodromy defects in the
  Wess-Zumino model},'' \href{http://dx.doi.org/10.1007/JHEP05(2022)185}{{\em
  JHEP} {\bfseries 05} (2022) 185},
  \href{http://arxiv.org/abs/2108.05107}{{\ttfamily arXiv:2108.05107}}.

\bibitem{Gimenez-Grau:2022ebb}
A.~Gimenez-Grau, ``{Probing magnetic line defects with two-point functions},''
  \href{http://arxiv.org/abs/2212.02520}{{\ttfamily arXiv:2212.02520}}.

\bibitem{Bianchi:2022sbz}
L.~Bianchi, D.~Bonomi, and E.~de~Sabbata, ``{Analytic bootstrap for the
  localized magnetic field},''
  \href{http://arxiv.org/abs/2212.02524}{{\ttfamily arXiv:2212.02524}}.

\bibitem{Costa:2011mg}
M.~S. Costa, J.~Penedones, D.~Poland, and S.~Rychkov, ``{Spinning conformal
  correlators},'' \href{http://dx.doi.org/10.1007/JHEP11(2011)071}{{\em JHEP}
  {\bfseries 11} (2011) 071}, \href{http://arxiv.org/abs/1107.3554}{{\ttfamily
  arXiv:1107.3554}}.

\bibitem{Witten:1998qj}
E.~Witten, ``{Anti-de Sitter space and holography},''
  \href{http://dx.doi.org/10.4310/ATMP.1998.v2.n2.a2}{{\em Adv. Theor. Math.
  Phys.} {\bfseries 2} (1998) 253--291},
  \href{http://arxiv.org/abs/hep-th/9802150}{{\ttfamily arXiv:hep-th/9802150}}.

\bibitem{Jensen:2018rxu}
K.~Jensen, A.~O'Bannon, B.~Robinson, and R.~Rodgers, ``{From the Weyl anomaly
  to entropy of two-dimensional boundaries and defects},''
  \href{http://dx.doi.org/10.1103/PhysRevLett.122.241602}{{\em Phys. Rev.
  Lett.} {\bfseries 122} no.~24, (2019) 241602},
  \href{http://arxiv.org/abs/1812.08745}{{\ttfamily arXiv:1812.08745}}.

\bibitem{Henningson:1998gx}
M.~Henningson and K.~Skenderis, ``{The holographic Weyl anomaly},''
  \href{http://dx.doi.org/10.1088/1126-6708/1998/07/023}{{\em JHEP} {\bfseries
  07} (1998) 023}, \href{http://arxiv.org/abs/hep-th/9806087}{{\ttfamily
  arXiv:hep-th/9806087}}.

\bibitem{Intriligator:2000eq}
K.~A. Intriligator, ``{Anomaly matching and a Hopf-Wess-Zumino term in 6d
  $\mathcal{N}=(2,0)$ field theories},''
  \href{http://dx.doi.org/10.1016/S0550-3213(00)00148-6}{{\em Nucl. Phys. B}
  {\bfseries 581} (2000) 257--273},
  \href{http://arxiv.org/abs/hep-th/0001205}{{\ttfamily arXiv:hep-th/0001205}}.

\bibitem{Tseytlin:2000sf}
A.~A. Tseytlin, ``{$R^4$ terms in 11 dimensions and conformal anomaly of (2,0)
  theory},'' \href{http://dx.doi.org/10.1016/S0550-3213(00)00380-1}{{\em Nucl.
  Phys. B} {\bfseries 584} (2000) 233--250},
  \href{http://arxiv.org/abs/hep-th/0005072}{{\ttfamily arXiv:hep-th/0005072}}.

\bibitem{Beccaria:2014qea}
M.~Beccaria, G.~Macorini, and A.~A. Tseytlin, ``{Supergravity one-loop
  corrections on AdS$_7$ and AdS$_3$, higher spins and AdS/CFT},''
  \href{http://dx.doi.org/10.1016/j.nuclphysb.2015.01.014}{{\em Nucl. Phys. B}
  {\bfseries 892} (2015) 211--238},
  \href{http://arxiv.org/abs/1412.0489}{{\ttfamily arXiv:1412.0489}}.

\bibitem{Chalabi:2020iie}
A.~Chalabi, A.~O'Bannon, B.~Robinson, and J.~Sisti, ``{Central charges of 2d
  superconformal defects},''
  \href{http://dx.doi.org/10.1007/JHEP05(2020)095}{{\em JHEP} {\bfseries 05}
  (2020) 095}, \href{http://arxiv.org/abs/2003.02857}{{\ttfamily
  arXiv:2003.02857}}.

\bibitem{graham:1999pm}
C.~R. Graham and E.~Witten, ``{Conformal anomaly of submanifold observables in
  $AdS$/CFT correspondence},''
  \href{http://dx.doi.org/10.1016/S0550-3213(99)00055-3}{{\em Nucl. Phys.}
  {\bfseries B546} (1999) 52--64},
  \href{http://arxiv.org/abs/hep-th/9901021}{{\ttfamily hep-th/9901021}}.

\bibitem{Drukker:2020dcz}
N.~Drukker, M.~Probst, and M.~Tr\'epanier, ``{Surface operators in the 6d
  $\mathcal{N} = (2, 0)$ theory},''
  \href{http://dx.doi.org/10.1088/1751-8121/aba1b7}{{\em J. Phys. A} {\bfseries
  53} no.~36, (2020) 365401}, \href{http://arxiv.org/abs/2003.12372}{{\ttfamily
  arXiv:2003.12372}}.

\bibitem{Drukker:2020swu}
N.~Drukker, S.~Giombi, A.~A. Tseytlin, and X.~Zhou, ``{Defect CFT in the 6d
  $(2,0)$ theory from M2 brane dynamics in $AdS_7 \times S^4$},''
  \href{http://dx.doi.org/10.1007/JHEP07(2020)101}{{\em JHEP} {\bfseries 07}
  (2020) 101}, \href{http://arxiv.org/abs/2004.04562}{{\ttfamily
  arXiv:2004.04562}}.

\bibitem{Barrat:2022psm}
J.~Barrat, A.~Gimenez-Grau, and P.~Liendo, ``{A dispersion relation for defect
  CFT},'' \href{http://arxiv.org/abs/2205.09765}{{\ttfamily arXiv:2205.09765}}.

\bibitem{Bianchi:2022ppi}
L.~Bianchi and D.~Bonomi, ``{Conformal dispersion relations for defects and
  boundaries},'' \href{http://arxiv.org/abs/2205.09775}{{\ttfamily
  arXiv:2205.09775}}.

\bibitem{Drukker:2020atp}
N.~Drukker, M.~Probst, and M.~Tr\'epanier, ``{Defect CFT techniques in the 6d
  $\mathcal{N} = (2,0)$ theory},''
  \href{http://dx.doi.org/10.1007/JHEP03(2021)261}{{\em JHEP} {\bfseries 03}
  (2021) 261}, \href{http://arxiv.org/abs/2009.10732}{{\ttfamily
  arXiv:2009.10732}}.

\bibitem{Cordova:2017mhb}
C.~Cordova, D.~Gaiotto, and S.-H. Shao, ``{Surface defects and chiral
  algebras},'' \href{http://dx.doi.org/10.1007/JHEP05(2017)140}{{\em JHEP}
  {\bfseries 05} (2017) 140}, \href{http://arxiv.org/abs/1704.01955}{{\ttfamily
  arXiv:1704.01955}}.

\bibitem{Dolan:2004mu}
F.~A. Dolan, L.~Gallot, and E.~Sokatchev, ``{On four-point functions of 1/2-BPS
  operators in general dimensions},''
  \href{http://dx.doi.org/10.1088/1126-6708/2004/09/056}{{\em JHEP} {\bfseries
  09} (2004) 056}, \href{http://arxiv.org/abs/hep-th/0405180}{{\ttfamily
  arXiv:hep-th/0405180}}.

\bibitem{Liendo:2015cgi}
P.~Liendo, C.~Meneghelli, and V.~Mitev, ``{On correlation functions of BPS
  operators in 3d ${\mathcal{N}}$ = 6 superconformal theories},''
  \href{http://dx.doi.org/10.1007/s00220-016-2715-7}{{\em Commun. Math. Phys.}
  {\bfseries 350} no.~1, (2017) 387--419},
  \href{http://arxiv.org/abs/1512.06072}{{\ttfamily arXiv:1512.06072}}.

\bibitem{Lemos:2016xke}
M.~Lemos, P.~Liendo, C.~Meneghelli, and V.~Mitev, ``{Bootstrapping
  $\mathcal{N}=3$ superconformal theories},''
  \href{http://dx.doi.org/10.1007/JHEP04(2017)032}{{\em JHEP} {\bfseries 04}
  (2017) 032}, \href{http://arxiv.org/abs/1612.01536}{{\ttfamily
  arXiv:1612.01536}}.

\bibitem{Liendo:2018ukf}
P.~Liendo, C.~Meneghelli, and V.~Mitev, ``{Bootstrapping the half-BPS line
  defect},'' \href{http://dx.doi.org/10.1007/JHEP10(2018)077}{{\em JHEP}
  {\bfseries 10} (2018) 077}, \href{http://arxiv.org/abs/1806.01862}{{\ttfamily
  arXiv:1806.01862 [hep-th]}}.

\bibitem{Dolan:2000ut}
F.~A. Dolan and H.~Osborn, ``{Conformal four point functions and the operator
  product expansion},''
  \href{http://dx.doi.org/10.1016/S0550-3213(01)00013-X}{{\em Nucl. Phys. B}
  {\bfseries 599} (2001) 459--496},
  \href{http://arxiv.org/abs/hep-th/0011040}{{\ttfamily arXiv:hep-th/0011040}}.

\bibitem{Isachenkov:2018pef}
M.~Isachenkov, P.~Liendo, Y.~Linke, and V.~Schomerus, ``{Calogero-Sutherland
  approach to defect blocks},''
  \href{http://dx.doi.org/10.1007/JHEP10(2018)204}{{\em JHEP} {\bfseries 10}
  (2018) 204}, \href{http://arxiv.org/abs/1806.09703}{{\ttfamily
  arXiv:1806.09703}}.

\bibitem{Howe:1983fr}
P.~S. Howe, G.~Sierra, and P.~K. Townsend, ``{Supersymmetry in
  six-dimensions},'' \href{http://dx.doi.org/10.1016/0550-3213(83)90582-5}{{\em
  Nucl. Phys. B} {\bfseries 221} (1983) 331--348}.

\bibitem{Eden:2001wg}
B.~Eden, S.~Ferrara, and E.~Sokatchev, ``{(2,0) superconformal OPEs in $d = 6$,
  selection rules and nonrenormalization theorems},''
  \href{http://dx.doi.org/10.1088/1126-6708/2001/11/020}{{\em JHEP} {\bfseries
  11} (2001) 020}, \href{http://arxiv.org/abs/hep-th/0107084}{{\ttfamily
  arXiv:hep-th/0107084}}.

\bibitem{Ferrara:2001uj}
S.~Ferrara and E.~Sokatchev, ``{Universal properties of superconformal OPEs for
  1/2 BPS operators in $3 \le d \le 6$},''
  \href{http://dx.doi.org/10.1088/1367-2630/4/1/302}{{\em New J. Phys.}
  {\bfseries 4} (2002) 2},
  \href{http://arxiv.org/abs/hep-th/0110174}{{\ttfamily arXiv:hep-th/0110174}}.

\bibitem{Gimenez-Grau:2020jrx}
A.~Gimenez-Grau and P.~Liendo, ``{Bootstrapping Coulomb and Higgs branch
  operators},'' \href{http://dx.doi.org/10.1007/JHEP01(2021)175}{{\em JHEP}
  {\bfseries 01} (2021) 175}, \href{http://arxiv.org/abs/2006.01847}{{\ttfamily
  arXiv:2006.01847}}.

\bibitem{Gunaydin:1990ag}
M.~Gunaydin and R.~J. Scalise, ``{Unitary lowest weight representations of the
  noncompact supergroup $OSp(2m^*|2n)$},''
  \href{http://dx.doi.org/10.1063/1.529401}{{\em J. Math. Phys.} {\bfseries 32}
  (1991) 599--606}.

\bibitem{Agmon:2020pde}
N.~B. Agmon and Y.~Wang, ``{Classifying superconformal defects in diverse
  dimensions part I: superconformal lines},''
  \href{http://arxiv.org/abs/2009.06650}{{\ttfamily arXiv:2009.06650}}.

\bibitem{watts1997}
G.~M.~T. Watts, ``$\mathcal{W}$-algebras and their representations,'' in {\em
  Conformal Field Theories and Integrable Models}, Z.~Horv{\'a}th and L.~Palla,
  eds., pp.~55--84.
\newblock Springer Berlin Heidelberg, Berlin, Heidelberg, 1997.

\bibitem{Bouwknegt:1992wg}
P.~Bouwknegt and K.~Schoutens, ``{$\mathcal{W}$ symmetry in conformal field
  theory},'' \href{http://dx.doi.org/10.1016/0370-1573(93)90111-P}{{\em Phys.
  Rept.} {\bfseries 223} (1993) 183--276},
  \href{http://arxiv.org/abs/hep-th/9210010}{{\ttfamily arXiv:hep-th/9210010}}.

\bibitem{Kim:2013nva}
H.-C. Kim, S.~Kim, S.-S. Kim, and K.~Lee, ``{The general M5-brane
  superconformal index},'' \href{http://arxiv.org/abs/1307.7660}{{\ttfamily
  arXiv:1307.7660}}.

\bibitem{Bullimore:2014upa}
M.~Bullimore and H.-C. Kim, ``{The superconformal index of the (2,0) theory
  with defects},'' \href{http://dx.doi.org/10.1007/JHEP05(2015)048}{{\em JHEP}
  {\bfseries 05} (2015) 048}, \href{http://arxiv.org/abs/1412.3872}{{\ttfamily
  arXiv:1412.3872}}.

\bibitem{Campoleoni:2011hg}
A.~Campoleoni, S.~Fredenhagen, and S.~Pfenninger, ``{Asymptotic
  $\mathcal{W}$-symmetries in three-dimensional higher-spin gauge theories},''
  \href{http://dx.doi.org/10.1007/JHEP09(2011)113}{{\em JHEP} {\bfseries 09}
  (2011) 113}, \href{http://arxiv.org/abs/1107.0290}{{\ttfamily
  arXiv:1107.0290}}.

\bibitem{Bastianelli:1999en}
F.~Bastianelli and R.~Zucchini, ``{Three point functions of chiral primary
  operators in $d = 3, \mathcal{N}=8$ and $d = 6, \mathcal{N}=(2,0)$ SCFT at
  large $N$},'' \href{http://dx.doi.org/10.1016/S0370-2693(99)01179-X}{{\em
  Phys. Lett. B} {\bfseries 467} (1999) 61--66},
  \href{http://arxiv.org/abs/hep-th/9907047}{{\ttfamily arXiv:hep-th/9907047}}.

\bibitem{Corrado:1999pi}
R.~Corrado, B.~Florea, and R.~McNees, ``{Correlation functions of operators and
  Wilson surfaces in the $d = 6$, (0,2) theory in the large $N$ limit},''
  \href{http://dx.doi.org/10.1103/PhysRevD.60.085011}{{\em Phys. Rev. D}
  {\bfseries 60} (1999) 085011},
  \href{http://arxiv.org/abs/hep-th/9902153}{{\ttfamily arXiv:hep-th/9902153}}.

\bibitem{Bonetti:2018fqz}
F.~Bonetti, C.~Meneghelli, and L.~Rastelli, ``{VOAs labelled by complex
  reflection groups and 4d SCFTs},''
  \href{http://dx.doi.org/10.1007/JHEP05(2019)155}{{\em JHEP} {\bfseries 05}
  (2019) 155}, \href{http://arxiv.org/abs/1810.03612}{{\ttfamily
  arXiv:1810.03612}}.

\bibitem{Bianchi:2019sxz}
L.~Bianchi and M.~Lemos, ``{Superconformal surfaces in four dimensions},''
  \href{http://dx.doi.org/10.1007/JHEP06(2020)056}{{\em JHEP} {\bfseries 06}
  (2020) 056}, \href{http://arxiv.org/abs/1911.05082}{{\ttfamily
  arXiv:1911.05082}}.

\bibitem{Argyres:2022npi}
P.~C. Argyres, M.~Lotito, and M.~Weaver, ``{Vertex algebra of extended
  operators in 4d $\mathcal{N}=2$ superconformal field theories},''
  \href{http://arxiv.org/abs/2211.04410}{{\ttfamily arXiv:2211.04410}}.

\bibitem{Drukker:2010jp}
N.~Drukker, D.~Gaiotto, and J.~Gomis, ``{The virtue of defects in 4d gauge
  theories and 2d CFTs},''
  \href{http://dx.doi.org/10.1007/JHEP06(2011)025}{{\em JHEP} {\bfseries 06}
  (2011) 025}, \href{http://arxiv.org/abs/1003.1112}{{\ttfamily
  arXiv:1003.1112}}.

\bibitem{Fateev:1987vh}
V.~A. Fateev and A.~B. Zamolodchikov, ``{Conformal quantum field theory models
  in two-dimensions having $\mathbb{Z}_3$ symmetry},''
  \href{http://dx.doi.org/10.1016/0550-3213(87)90166-0}{{\em Nucl. Phys. B}
  {\bfseries 280} (1987) 644--660}.

\bibitem{Afkhami-Jeddi:2017idc}
N.~Afkhami-Jeddi, K.~Colville, T.~Hartman, A.~Maloney, and E.~Perlmutter,
  ``{Constraints on higher spin CFT$_{2}$},''
  \href{http://dx.doi.org/10.1007/JHEP05(2018)092}{{\em JHEP} {\bfseries 05}
  (2018) 092}, \href{http://arxiv.org/abs/1707.07717}{{\ttfamily
  arXiv:1707.07717}}.

\bibitem{Linshaw:2017tvv}
A.~R. Linshaw, ``{Universal two-parameter $\mathcal{W}_{\infty}$-algebra and
  vertex algebras of type $\mathcal{W}(2,3,\dots, N)$},''
  \href{http://dx.doi.org/10.1112/S0010437X20007514}{{\em Compos. Math.}
  {\bfseries 157} no.~1, (2021) 12--82},
  \href{http://arxiv.org/abs/1710.02275}{{\ttfamily arXiv:1710.02275}}.

\bibitem{Fateev:1987zh}
V.~A. Fateev and S.~L. Lukyanov, ``{The models of two-dimensional conformal
  quantum field theory with $\mathbb{Z}_N$ symmetry},''
  \href{http://dx.doi.org/10.1142/S0217751X88000205}{{\em Int. J. Mod. Phys. A}
  {\bfseries 3} (1988) 507}.

\bibitem{Alday:2009qq}
L.~F. Alday, F.~Benini, and Y.~Tachikawa, ``{Liouville/Toda central charges
  from M5-branes},''
  \href{http://dx.doi.org/10.1103/PhysRevLett.105.141601}{{\em Phys. Rev.
  Lett.} {\bfseries 105} (2010) 141601},
  \href{http://arxiv.org/abs/0909.4776}{{\ttfamily arXiv:0909.4776}}.

\bibitem{Cordova:2016cmu}
C.~Cordova and D.~L. Jafferis, ``{Toda theory from six dimensions},''
  \href{http://dx.doi.org/10.1007/JHEP12(2017)106}{{\em JHEP} {\bfseries 12}
  (2017) 106}, \href{http://arxiv.org/abs/1605.03997}{{\ttfamily
  arXiv:1605.03997}}.

\bibitem{Alday:2017vkk}
L.~F. Alday and S.~Caron-Huot, ``{Gravitational $S$-matrix from CFT dispersion
  relations},'' \href{http://dx.doi.org/10.1007/JHEP12(2018)017}{{\em JHEP}
  {\bfseries 12} (2018) 017}, \href{http://arxiv.org/abs/1711.02031}{{\ttfamily
  arXiv:1711.02031}}.

\bibitem{Lauria:2017wav}
E.~Lauria, M.~Meineri, and E.~Trevisani, ``{Radial coordinates for defect
  CFTs},'' \href{http://dx.doi.org/10.1007/JHEP11(2018)148}{{\em JHEP}
  {\bfseries 11} (2018) 148}, \href{http://arxiv.org/abs/1712.07668}{{\ttfamily
  arXiv:1712.07668}}.

\bibitem{Maldacena:2015iua}
J.~Maldacena, D.~Simmons-Duffin, and A.~Zhiboedov, ``{Looking for a bulk
  point},'' \href{http://dx.doi.org/10.1007/JHEP01(2017)013}{{\em JHEP}
  {\bfseries 01} (2017) 013}, \href{http://arxiv.org/abs/1509.03612}{{\ttfamily
  arXiv:1509.03612}}.

\bibitem{Beem:2013qxa}
C.~Beem, L.~Rastelli, and B.~C. van Rees, ``{The $\mathcal N=4$ superconformal
  bootstrap},'' \href{http://dx.doi.org/10.1103/PhysRevLett.111.071601}{{\em
  Phys. Rev. Lett.} {\bfseries 111} (2013) 071601},
  \href{http://arxiv.org/abs/1304.1803}{{\ttfamily arXiv:1304.1803}}.

\bibitem{Liendo:2019jpu}
P.~Liendo, Y.~Linke, and V.~Schomerus, ``{A lorentzian inversion formula for
  defect CFT},'' \href{http://dx.doi.org/10.1007/JHEP08(2020)163}{{\em JHEP}
  {\bfseries 08} (2020) 163}, \href{http://arxiv.org/abs/1903.05222}{{\ttfamily
  arXiv:1903.05222}}.

\bibitem{erdelyi1981higher}
A.~Erd{\'e}lyi and H.~Bateman, {\em Higher transcendental functions}.
\newblock Bateman Manuscript Project. Krieger, 1981.
\newblock \url{https://books.google.ca/books?id=h7UTzQEACAAJ}.

\bibitem{Aharony:2016dwx}
O.~Aharony, L.~F. Alday, A.~Bissi, and E.~Perlmutter, ``{Loops in $AdS$ from
  conformal field theory},''
  \href{http://dx.doi.org/10.1007/JHEP07(2017)036}{{\em JHEP} {\bfseries 07}
  (2017) 036}, \href{http://arxiv.org/abs/1612.03891}{{\ttfamily
  arXiv:1612.03891}}.

\bibitem{Mack:2009mi}
G.~Mack, ``{$d$-independent representation of conformal field theories in $d$
  dimensions via transformation to auxiliary dual resonance models. Scalar
  amplitudes},'' \href{http://arxiv.org/abs/0907.2407}{{\ttfamily
  arXiv:0907.2407}}.

\bibitem{Rastelli:2017udc}
L.~Rastelli and X.~Zhou, ``{How to succeed at holographic correlators without
  really trying},'' \href{http://dx.doi.org/10.1007/JHEP04(2018)014}{{\em JHEP}
  {\bfseries 04} (2018) 014}, \href{http://arxiv.org/abs/1710.05923}{{\ttfamily
  arXiv:1710.05923}}.

\bibitem{Goncalves:2018fwx}
V.~Goncalves and G.~Itsios, ``{A note on defect Mellin amplitudes},''
  \href{http://arxiv.org/abs/1803.06721}{{\ttfamily arXiv:1803.06721}}.

\bibitem{Ferrero:2021bsb}
P.~Ferrero and C.~Meneghelli, ``{Bootstrapping the half-BPS line defect CFT in
  $\mathcal{N}=4$ supersymmetric Yang-Mills theory at strong coupling},''
  \href{http://dx.doi.org/10.1103/PhysRevD.104.L081703}{{\em Phys. Rev. D}
  {\bfseries 104} no.~8, (2021) L081703},
  \href{http://arxiv.org/abs/2103.10440}{{\ttfamily arXiv:2103.10440}}.

\bibitem{Dobrev:1975ru}
V.~K. Dobrev, V.~B. Petkova, S.~G. Petrova, and I.~T. Todorov, ``{Dynamical
  derivation of vacuum operator product expansion in euclidean conformal
  quantum field theory},''
  \href{http://dx.doi.org/10.1103/PhysRevD.13.887}{{\em Phys. Rev. D}
  {\bfseries 13} (1976) 887}.

\bibitem{Dobrev:1977qv}
V.~K. Dobrev, G.~Mack, V.~B. Petkova, S.~G. Petrova, and I.~T. Todorov,
  \href{http://dx.doi.org/10.1007/BFb0009678}{{\em {Harmonic analysis on the
  $n$-dimensional Lorentz group and its application to conformal quantum field
  theory}}}, vol.~63.
\newblock 1977.

\bibitem{Dolan:2003hv}
F.~A. Dolan and H.~Osborn, ``{Conformal partial waves and the operator product
  expansion},'' \href{http://dx.doi.org/10.1016/j.nuclphysb.2003.11.016}{{\em
  Nucl. Phys. B} {\bfseries 678} (2004) 491--507},
  \href{http://arxiv.org/abs/hep-th/0309180}{{\ttfamily arXiv:hep-th/0309180}}.

\bibitem{McAvity:1995zd}
D.~M. McAvity and H.~Osborn, ``{Conformal field theories near a boundary in
  general dimensions},''
  \href{http://dx.doi.org/10.1016/0550-3213(95)00476-9}{{\em Nucl. Phys. B}
  {\bfseries 455} (1995) 522--576},
  \href{http://arxiv.org/abs/cond-mat/9505127}{{\ttfamily
  arXiv:cond-mat/9505127}}.

\bibitem{Lewkowycz:2014jia}
A.~Lewkowycz and E.~Perlmutter, ``{Universality in the geometric dependence of
  R\'enyi entropy},'' \href{http://dx.doi.org/10.1007/JHEP01(2015)080}{{\em
  JHEP} {\bfseries 01} (2015) 080},
\href{http://arxiv.org/abs/1407.8171}{{\ttfamily arXiv:1407.8171}}.

\bibitem{bianchi:2015liz}
L.~Bianchi, M.~Meineri, R.~C. Myers, and M.~Smolkin, ``{R\'enyi entropy and
  conformal defects},'' \href{http://dx.doi.org/10.1007/JHEP07(2016)076}{{\em
  JHEP} {\bfseries 07} (2016) 76},
  \href{http://arxiv.org/abs/1511.06713}{{\ttfamily arXiv:1511.06713}}.

\bibitem{Manvelyan:2000ef}
R.~Manvelyan and A.~C. Petkou, ``{A Note on $R$ currents and trace anomalies in
  the (2,0) tensor multiplet in $d = 6$ $AdS$/CFT correspondence},''
  \href{http://dx.doi.org/10.1016/S0370-2693(00)00568-2}{{\em Phys. Lett. B}
  {\bfseries 483} (2000) 264--270},
  \href{http://arxiv.org/abs/hep-th/0003017}{{\ttfamily arXiv:hep-th/0003017}}.

\end{thebibliography}\endgroup

\end{document}